\documentclass[12pt]{article}
\usepackage[utf8]{inputenc}
\usepackage{geometry}
\usepackage{color}
\usepackage{array}
\usepackage{float}
\usepackage{booktabs}
\usepackage{url}
\usepackage{multirow}
\usepackage{amsmath}
\usepackage{amsthm}
\usepackage{amssymb}
\usepackage{graphicx}
\usepackage{setspace}
\usepackage{placeins}
\usepackage{relsize}
\PassOptionsToPackage{normalem}{ulem}
\usepackage{ulem}
\usepackage[unicode=true]
 {hyperref}

\makeatletter

\DeclareTextSymbolDefault{\textquotedbl}{T1}
\providecommand{\tabularnewline}{\\}

\theoremstyle{plain}
\newtheorem*{thm*}{\protect\theoremname}
\theoremstyle{plain}

\theoremstyle{plain}

\usepackage{amsfonts}

\newtheorem{theorem}{Theorem}

\newtheorem{corollary}[theorem]{Corollary}

\newtheorem{proposition}[theorem]{Proposition}

\usepackage{graphicx}
\usepackage{booktabs}
\usepackage{rotating}
\usepackage{lscape}
\usepackage{longtable}
\usepackage{blindtext}
\usepackage{layouts}
\usepackage{amsmath}
\usepackage{comment}
\usepackage{caption}
\usepackage{footmisc}
\hypersetup{hidelinks}

\makeatother

\providecommand{\lemmaname}{Lemma}
\providecommand{\propositionname}{Proposition}
\providecommand{\theoremname}{Theorem}

\begin{document}
\title{Liquid Democracy or Direct Democracy?\\
One Theoretical Result and Two Experiments\thanks{With the exception of the first author, all others are in alphabetical
order. We thank audiences at numerous academic seminars and conferences,
and in particular Tim Feddersen, Jean-Francois Laslier, Chloe Tergiman, and Richard van Weelden
for comments. We are grateful to Mael Lebreton, Jonathan Nicholas,
Nahuel Salem, and Camilla van Geen for their help with the second
experiment. We thank the Program for Economic Research at Columbia
and the Columbia Experimental Lab for the Social Sciences for their
financial support. We acknowledge computing resources from Columbia
University's Shared Research Computing Facility project, which is
supported by NIH Research Facility Improvement Grant 1G20RR030893-01
and associated funds from the New York State Empire State Development,
Division of Science Technology and Innovation (NYSTAR) Contract C090171,
both awarded April 15, 2010. The experiments have been approved by
Columbia University's IRB. }}
\author{Victoria Mooers\thanks{Columbia University, v.mooers@columbia.edu},
Joseph Campbell\thanks{Columbia University, joseph.campbell@columbia.edu},
Alessandra Casella\thanks{Columbia University, NBER and CEPR, ac186@columbia.edu},
Lucas de Lara\thanks{Columbia University, lpd2122@columbia.edu},
\\
and Dilip Ravindran\thanks{Humboldt University of Berlin, dilip.ravindran@hu-berlin.de}\\
}

\maketitle
\thispagestyle{empty}

\bigskip{}
\bigskip{}
\bigskip{}
\bigskip{}

\bigskip{}
\bigskip{}
\bigskip{}
\bigskip{}
\bigskip{}
\bigskip{}
\bigskip{}
\bigskip{}
\bigskip{}
\bigskip{}
\bigskip{}
\bigskip{}
\bigskip{}
\bigskip{}
\bigskip{}

\bigskip{}

\bigskip{}

\bigskip{}

\pagebreak{}
\begin{abstract}
Proponents of participatory democracy praise Liquid Democracy: decisions
are taken by referendum, but voters delegate their votes freely. When
better informed voters are present and the electorate is finite, we show theoretically that delegation can always strictly increase the probability of a correct decision. However, delegation must be used sparingly because it reduces the information aggregated through voting. In two different
experiments---a tightly controlled lab experiment and a perceptual task run online---we find that subjects choose very high rates of delegation, and the theoretically possible improvements fail to materialize. The experimental evidence favors Direct Democracy, whether with or without abstention. We study the perceptual task, where signals' precisions are not known, both as a test of the robustness of the lab results and as an independent methodological contribution. We argue that tests under ambiguous information are valuable and under-used tools in studying collective decision-making.  

\bigskip{}



\pagebreak{}

\setcounter{page}{1}
\end{abstract}

\section{Introduction}

The widespread sense that party politics is in a crisis has been accompanied
by diverse and increasingly loud proposals for new forms of participatory
democracy. Among them are digital platforms to coordinate
common actions, mini-publics and citizens assemblies, participatory
budgeting, and algorithms supporting online democracy. Among
the latter, Liquid Democracy (LD) has caught the imagination of the young
and the tech-savvy. It advocates a voting system where all decisions
are submitted to referendum, but voters can delegate their votes freely.

LD traces its intellectual roots to Charles Dodgson
(1884) and James Miller (1969) and has been adopted occasionally
for internal decisions by European protest parties---the Swedish and
the German Pirate parties being the most famous examples. Now it finds
vocal support in the tech community, where it aligns both with the
emphasis on a non-hierarchical order and with the use of blockchains
to maintain confidentiality and reliability. LD has
become the governance choice for cryptoworld DAOs (Decentralized
Autonomous Organizations), operating in a variety of fields, from financial investing to art tokens. As documented by Hall and Miyazaki (2024), neither these organizations nor LD are purely intellectual curiosities: several DAOs control billions of dollars and use LD to guide their operations.\footnote{In addition to the 18 DAOs studied by Hall and Miyazaki, see also DEV (https://medium.com/element-finance) and the detailed description of its organizational model. For political applications, see LiquidFeedback (https://liquidfeedback.com/en/), the
Association for Interactive Democracy (https://interaktive-demokratie.org/association.en.html),
or Democracy.Earth (https://democracy.earth/). Google ran an early 3-year
experiment on its internal network, testing LD on minor communal decisions (Hardt and Lopes, 2015).} 


Although the details vary, the common point of different LD implementations is the ease and specificity of delegation: a vote can be delegated freely, and to a different person depending on the issue. According to supporters, LD is superior to representative democracy because representatives can be chosen on the basis of their specific competence on each decision, and is superior to direct democracy because uninformed or uninterested voters can delegate their votes.

The desirability of delegation in LD has typically been studied in common interest environments
with two alternatives and an unknown ``ground truth'' on which experts
have superior information (Kahng, Mackenzie and Procaccia, 2018; Caragiannis and Michas, 2019; Armstrong and Larson, 2021; Halpern et al., 2021, Dhillon et al., 2025; Berinsky et al., 2025). It is the perspective we also adopt in this work.\footnote{A smaller literature focuses on delegation as a means of completing
preference rankings: individuals unable to express a full ranking
over all alternatives proposed delegate to others with whom they agree
on the issues on which they know their own preferences (Christoph and Grossi, 2017; Brill and Talmon, 2018; Harding, 2022).} 

Delegating to better informed experts with preferences that match
our own can seem a trivially beneficial step. But there is a problem: even if the experts are correctly identified, delegation
deprives the electorate of the richness of noisy but abundant information
distributed among all voters. Condorcet's Jury Theorem teaches us
the value of a large electorate with barely accurate but independent
sources of information. In a binary decision, if voters receive independent
signals whose accuracy exceeds 50\%, the percentage of correct signals
exceeds 50\% with probability approaching one as the size of the electorate
becomes very large. If voters vote according to their signals, majority
voting then delivers the correct outcome with probability that approaches
one (Condorcet, 1785). Thus, unless the extent of delegation is modulated
correctly, a smaller number of voters, even if more accurate, may lead to worse decision-making. 

In addition to this basic trade-off between accuracy and variety of information, we should also consider that the benefits LD can provide---overweighing better informed voters and allowing less informed ones not to vote---can be achieved by a more traditional voting system: Direct Democracy with abstention (DD). Here too individuals with less reliable information can refrain from voting, and by that action alone increase the influence of better informed subjects who do cast their vote. We know from McMurray (2013) that, under common interest and in the absence of voting costs,
the simple option of abstention can lead to improvements over all-voting majority rule, and for
reasons that parallel those favoring delegation. The evaluation of LD thus benefits from the comparison to DD. 

We study a canonical common interest model where voters receive independent
signals, conditional on an unknown binary state. The common objective
is to identify the state correctly. Signals vary across individuals in the probability of being
correct (their\emph{ ``precision''}). Experts are publicly identified
and the precision of their signals is known; for all other voters,
signals\textquoteright{} precisions are private information but known
to be weakly lower than the experts\textquoteright . Under LD, every voter chooses whether to cast her vote or delegate, and in this latter case, the vote is randomly assigned to one of the experts. Under DD, each voter chooses whether to vote or to abstain. With both voting rules, the majority of votes cast determines the group decision.

Our first step is to pin down the theoretical properties of the two rules, comparing them to voting by all and to each other, and because we are interested in testing such properties in the lab, we study them in electorates of finite size. Somewhat surprisingly, such theoretical comparisons in this most canonical of voting models were missing from the literature.\footnote{As discussed below, Dhillon et al. (2025), includes results related to ours.} We find that for any continuous distribution of signals' precisions, for any finite size of the
electorate, as long as there is more than a single non-expert, there is always an equilibrium with positive
delegation that strictly dominates\footnote{I.e. such that the group decision has a strictly higher probability of being correct.} majority rule with all voting (MV). However, the same result does not hold for abstention: there exist parameter values  
such that MV strictly dominates any equilibrium with positive abstention. We formalize this observation in a theorem, this study's main theoretical result.

The second part of our work aims at testing the two voting rules experimentally and describes the results of two, quite different, experiments.

The first experiment is designed for the lab and follows the theoretical models very closely. The theorem does not imply that allowing for delegation always weakly dominates allowing for abstention---the ranking of the two rules depends on parameter values. In the lab, we focus on two parametrizations, one favoring delegation and one favoring abstention, but with small quantitative differences. Under both parametrizations, both rules are expected to be superior to majority rule with all voting.\footnote{We implement two different designs, either collecting MV votes as a third rule in the experiment, or simulating it. The results are closely comparable. }

Under LD, under both parameterizations, we find delegation rates that are two to three times the rate predicted by theory for the equilibrium that dominates MV. Abstention rates under DD are instead comparable to the optimal rates. 

In terms of performance, we find that LD underperforms MV in three out of our four cases,\footnote{Two parameterizations and two designs.} while DD always does weakly better. The differences, however, are small and never statistically significant. The lack of significance is to be expected: when voting rules are variations on majority voting, the frequency with which outcomes overlap is very high and statistical significance is elusive. Sharper results require focusing on outcomes that differ across rules, but the number of data then becomes too small. A remedy is using bootstrapping techniques, using experimental data to generate a large number of simulated outcomes and focusing on the frequency of correct outcomes, conditioning on disagreement across rules.\footnote{The use of bootstrapping in analyzing collective choice lab experiments is becoming common. See for example, Bouton et al. 2017, Casella et al. 2022, Reshidi et al., 2025.} The approach yields sharper results: in our case, over 100,000 replications of our 240 experimental group decisions, when MV and LD yield a different outcome, MV is on average correct more than 2/3 of the times; when MV and DD yield a different outcome, on the other hand, the conclusion is reversed: MV is correct just slightly above 1/3 of the times. 

Summarizing, whether looking at strategies or outcomes, we do not find evidence that experimental subjects are able to exploit successfully LD's potential to improve decisions. DD, on the other hand, fares better.  

Why such high rates of delegation? And why the difference with respect
to abstention? In our second experiment, we investigate whether
the results could be due to the experimental design we use in the lab. We test the robustness of Experiment 1's findings in a very
different environment: a perceptual task where individuals do not
have precise information about their and others' perceptual accuracy,
or, in the language of Experiment 1, about the reliability of their
and others' signals. 

There are two reasons for such a test. First, actual voting decisions take
place in an ambiguous world, where voters have ``some sense'' of
how likely to be correct they and the experts are, but such sense
is vague and instinctive. Perceptual tasks capture ambiguity well
and indeed have become part of economists' standard tool-kit when
studying individual decision-making. They are much less common when
testing group decision-making.\footnote{There is an increasing focus on strategic uncertainty. But the question
is different from the lack of basic information about the distributions
of relevant parameters in the population, and even about one's own
parameters (precisions, for us). }
And yet, as realized by a recent theoretical literature (Ellis, 2016; Ryan, 2021; Fabrizi et al., 2021 and 2022), in voting problems, the
complexity of many questions and the small marginal impact of a single
vote make the ambiguity of information a particularly plausible assumption.
Perceptual tasks, with the large and sophisticated literature that
accompanies them, can be a useful tool for social choice scholars. We consider our second experiment a separate methodological contribution, and we hope it may be of interest to scholars studying collective decision-making, beyond the focus of this work on LD and DD.  

In the specific case of delegation and abstention, there is a second
reason for studying a parallel experiment with ambiguous information:
the possibility that the explicit numerical frame of Experiment 1
could influence the results. Consider first LD. A participant told
that the probability that her signal is correct is, for example, 55\%, versus 70\% for an expert,
is naturally induced to compare the two numbers. Choosing to delegate
seems very reasonable: if the decision involved only the participant
and the expert, delegation would always be optimal. Delegation becomes
more dubious only if the participant factors in the behavior of others
and the possibility that they too may delegate. But others' choices
are not made salient by the question of whether to delegate to an
expert. In the experiment with DD, the question is whether to abstain---that is, whether
to leave the decision to all those who vote. 
Abstention invokes the comparison not to the experts' precision
only, but to the precisions of all those participants who do not abstain.
And such precisions are not revealed. The comparison between a precision
of 55\% for oneself and of 70\% for an expert may well be less likely
to trigger an automatic reaction.\footnote{The hypothesis that precise mathematical information becomes counterproductive
when it encourages a wrong mental model finds echos in the literature.
In a recent paper, Esponda et al. (2023) report that subjects given
precise quantitative information in a simple belief updating task
are less able to learn from experience, relative to a control treatment
where such precise information is withheld. The knowledge of the base
parameters induces a persistent mistaken mental model.} 

We implement a classic perceptual task amply used in vision and cognitive
research.\footnote{The Random Dot Kinematogram (RDK) was originally developed to study
the perception of motion under noisy conditions in humans and non-human
primates (e.g. van de Grind et al., 1983). In neuroscience, it has
been used to study the neuronal correlates of motion perception (Newsome
et al., 1989; Britten et al., 1992; Roitman and Shadlen, 2002).} A number of moving dots are displayed for a very short interval (1
second); some move in a coherent direction, either Left or Right in
our binary implementation, others move at random; subjects report
in which direction they think coherent dots are moving. We label experts
ex post as the individuals with recent performance in the highest
quintile, and generate a collective decision by aggregating individual
responses, with the additional option of delegation to the experts
(in the LD treatments) or abstention (in the DD treatments). 

We find results that match closely those observed in Experiment 1.
First, delegation remains much more frequent than abstention. Second, exploiting again our bootstrapping approach,
universal majority voting delivers the highest frequency of correct
group decisions in all treatments; DD is only slightly less efficient,
while LD is dominated by both in all treatments. 
The two experiments, different as they are in
their design, complement each other. They reach the same qualitative
conclusion, strengthening our confidence in its robustness: evaluated
on informational benefits, we do not find evidence in favor of LD. 

Our work then makes three contributions. First, a theoretical result comparing delegation and abstention in finite electorates. Second, an experimental test of the relative effectiveness of LD and DD in aggregating information in a canonical common value environment. Third, we show that experiments on collective decision-making can benefit from using familiar perceptual tasks. In such tasks, information is ambiguous and not conveyed in the precise mathematical format of traditional rigorous lab experiments, increasing their realism and possibly limiting biases.    

Our work is related to three separate literatures. First, to the study
of voting as information aggregation. The informational costs and
benefits of delegation in pure common interest voting problems were
the subject of early studies on the Condorcet Jury Theorem (Margolis,
1976; Grofman et al., 1983; Shapley and Grofman, 1984). These studies
asked important statistical questions but did not focus on equilibrium
behavior. More recent work (Austen-Smith and Banks, 1996; Feddersen and Pesendorfer, 1997; McLennan, 1998; Wit, 1998)
put the analysis of the Condorcet Jury Theorem on solid equilibrium
grounds, but abstracted from the focus on delegation. 
The literature
on proxy voting in finance shares our interest in delegation and information
aggregation and highlights the dangers of overweighing individual voters. Bar-Isaac and Shapiro (2020), for example, study the conditions under
which a voter representing a large block of shares chooses to refrain from voting them all, not to overwhelm the information dispersed among
smaller but independently informed other voters.\footnote{In other related works, Malenko and Malenko
(2019) and Buechel et al. (2023) ask whether for-profit proxy advisors
can induce worse corporate decisions by  
decreasing shareholders\textquoteright{} incentives to acquire independent information.} 


The direct comparison between delegation and abstention, central to
our work, is not the focus of these works. Information aggregation
under abstention has been studied either in models with pure common
interest, like ours (McMurray, 2013), or when partisan voters are
present (as in Feddersen and Pesendorfer, 1996). Morton and Tyran
(2011), Mengel and Rivas (2017), and Battaglini et al. (2010) test experimentally 
the two models, with and without partisan voters, in small committees, but again without the comparison to delegation.

The second strand of related works are studies of LD, mostly in normative
political theory and computer science. Green-Armytage (2015) and Blum
and Zuber (2016) discuss LD's possible advantages,
for both epistemic and egalitarian reasons: decisions are taken by
better informed voters, and LD avoids the creation of a class of semi-permanent
professional representatives. The focus is normative, and these studies
do not analyze strategic incentives. The computer science literature
is instead largely concerned with how LD can work in
practice. It models behavior via a priori algorithms and studies rich
interactions where delegation takes place on networks (Christoff and
Grossi, 2017; Kahng, Mackenzie and Procaccia, 2018; Bloembergen, Grossi
and Lackner, 2019; Caragiannis and Michas, 2019; Berinsky et al., 2025). These authors connect
LD to the social choice tradition, but here too strategic considerations
are absent. An exception is Armstrong and Larson (2021) which discusses
the informational trade-off involved in delegation and focuses on
a Nash equilibrium. The paper retains the algorithmic flavor of this
literature by modeling the delegation choice as sequential; with common interest, complete information, and costly voting, sequential choice results in the equilibrium
superiority of delegation over all-voting majority rule. The theoretical
conclusion is thus similar to ours, but the assumptions driving the
result differ. 

Strategic concerns are at the heart of Dhillon et al. (2025), a recent paper in economics that developed in parallel to ours that also studies information aggregation in liquid democracy. Dhillon et al. study a model where both delegation and abstention are allowed, and partisan voters, \`{a} la Feddersen and Pesendorfer (1996), may be present. Under pure common interest, when partisans are absent, delegation must always weakly improve over abstention alone, but the authors identify a sufficient condition guaranteeing strict improvement: sufficient heterogeneity in the precisions of voters. 
The analysis differs from ours because we allow delegation \textit{or} abstention, but never both, and compare each to majority voting by all. Our model separates cleanly the relative role of delegation and abstention in increasing the weight of better informed voters, and allows us to derive very broad and transparent conditions for improvement over MV. Studying delegation and abstention separately also allows for cleaner equilibrium characterizations which lend our model better to testing with experiments. The cost, however, is the lack of general results in ranking LD and DD. In our model, the two rules' welfare rankings depend on parameters.

All the works cited under this strand of literature are theoretical or discuss numerical simulations. We use instead our theoretical results as foundations for our two experiments.

On the experimental side, Berinsky et al. (2025) tests LD by asking groups to vote on answers to general culture questions. Group members know each other and each individual has the option to select another individual as delegate. The focus of the experiment is the endogenous choice of experts, testing whether delegations go from less to more competent individuals and whether the likelihood of excessive votes concentration is small. The results are encouraging, but the number of data points is small, and LD does not lead to a significantly higher fraction of correct responses, relative to MV. In our experiments, experts are correctly identified exogenously and we too see delegations increasing in the difference of signal precision, in Experiment 1, or of believed competence, in Experiment 2. However, the very high volume of delegations does result in excessive voting weight for the expert(s), weakening LD's performance. Berinsky et al. do not consider the alternative of abstention.\footnote{Delegation in voting is more frequently studied in experiments on representative democracy where individuals are endowed with heterogeneous preferences (see for example, Hamman et al., 2011). Closer to our questions is Kawamura and Vlaseros (2017), which tests experimentally a canonical common interest model with independent private signals. A public statement
by an ``expert'' conveys additional information prior to voting and 
moves participants' prior but there is no actual delegation of votes.} 

Finally, a literature in social psychology studies a question closely related to our second experiment: if a group of individuals
face, individually, a perceptual task but can aggregate
reactions into a group decision, which decision rule for the group
reaches the correct answer most frequently? In Sorkin et al. (1998) a small group of subjects are faced with a signal detection task and
asked whether the display reflects noise only or signal plus noise.
Although the group falls short of normative predictions, simple majority
rule leads to the highest accuracy. Subsequent studies have focused
on communication and confidence (Sorkin et al., 2001; Bahrami et al.,
2012; Silver et al., 2021). Although it seems a natural next step,
we are not aware of similar works that include the possibility of
delegation.

In what follows, we begin by describing the theoretical model (Section
2) and its equilibrium properties (Section 3). We then describe our
first experiment: its treatments and implementation (Section 4), and its results (Section 5). Section 6 discusses the motivation and the design of our second experiment, and Section 7 its results. Section 8 sets the ground for future research,
briefly discussing the effects of relaxing some assumptions of the model. Section 9 concludes. The Appendix collects longer proofs and some
additional experimental findings; further results are in the online Appendix.

\section{The Model}

We study the canonical problem of information aggregation through
voting in a pure common interest problem. $N$ (finite) voters face
an uncertain state of the world $\omega$ and must take a decision
$d$. There are two possible states of the world, $\omega\in\{\omega_{1},\omega_{2}\}$,
and two alternative decisions $d\in\{d_{1},d_{2}\}$. Every voter
$i$'s payoff equals 1 if the decision matches the state of the world
($d=d_{h}$ when $\omega=\omega_{h},\text{ }h=1,2$), and 0 otherwise.
Voters share a common prior $\pi=Pr(\omega_{1})$ and each voter $i$ receives a conditionally
independent signal $s_{i}\in\{s_{1},s_{2}\}$ that
recommends one of the two decisions. We call $q_{i}$ the \emph{precision}
of individual $i$'s signal, or the probability that $i$'s signal
is correct. Precision varies across individuals but is symmetric over
the two possible states of the world: $q_{i}=Pr(s_{i}=s_{1}|\omega_{1})=Pr(s_{i}=s_{2}|\omega_{2})$. 

The group of $N$ voters is composed of $K$ experts and $M$ non-experts: $N=M+K$, with $K$ and $M$ strictly positive integers and $M>1$. Whether any given voter is an expert or a non-expert is commonly known. Every expert $e$ receives a signal of known precision $q_{e} \equiv p \in(1/2,1)$.
The precision of a non-expert $i$'s signal is instead private information:
$q_{i}$ is an independent draw from a commonly known distribution
$F(q)$ everywhere continuous over support $[\underline{q},\overline{q}]$, with $\underline{q}=1/2$ and $\overline{q}=p$. The signals themselves
are also private information, for both experts and non-experts. 

The model deviates only minimally from the classic Condorcet
setup: we add a group of known experts and allow
for asymmetric information about the precision of non-experts' signals. In particular, every non-expert's signal is weakly less precise than any expert's signal. As in the Condorcet model, we rule out communication
among voters. We do it both as a realism check in possible applications
to large elections, and more generally to understand how the options of delegation, in LD, or abstention, in DD, affect canonical results. We assume a fully flat prior: $\pi=Pr(\omega_{1})=1/2$, which simplifies the analysis and again facilitates comparison to the best known Condorcet results. 

Each voter, whether expert or non-expert, holds a single non-divisible
vote. We denote by $EU$ individual ex ante expected payoff, before
the realizations of precisions and signals, noting that $EU$ equals the ex ante
probability that the group reaches the correct decision and is equal for experts and non-experts. We add a subscript
$i$ and write $EU_i$ to indicate interim expected utility, after $i$ is informed
of the signal's precision and realization.


\subsection{Liquid Democracy}

LD allows for delegation. Before the election, each voter receives a signal and is informed of the signal's precision, i.e. observes $q_i$ and $s_i$. 

The voter then chooses whether indeed to
vote, for one or the other of the two options, or whether to delegate
the vote to an expert. All voters, including experts, may choose to delegate. Any delegated vote is assigned randomly, with
equal probability, to any expert (with the exception of the delegator himself, if the delegator is an expert).\footnote{Random assignment
of delegated votes is a natural assumption in the absence of distinguishing characteristics across individuals. It is also desirable because
it tends to balance the accumulation of voting power (for example,
G\H{o}lz et al., 2018, Buechel and Mechtenberg, 2019). In our model,
mixing uniformly when delegating to experts would be an equilibrium
strategy if voters could target individual experts.} 

When counting votes, each individual has a weight equal to the number of votes the individual holds (0 if the voter delegated; 1 for non-experts who did not delegate; and 1 plus the number of delegated votes received, for experts who did not delegate). The
decision receiving more votes is chosen, with ties resolved by a fair coin flip.

In principle, multi-step delegation is a possibility: if delegation targets an expert who has herself chosen delegation, her vote and all votes she was delegated are assigned to another expert. However, if a set of delegation decisions results in a circular delegation flow, we assume that all votes involved in the cycle are cast randomly for either decision, with equal probability.\footnote{Whether votes in the cycle are cast randomly as a block or each is cast independently makes no difference to our results: in our model, under either assumption no delegation cycle occurs in equilibrium. The same conclusion holds if, in case of a cycle, all concerned votes are cast in the wrong direction, as in Berinsky et al. (2025).} 


\subsection{Direct Democracy}

We model DD as majority voting allowing for abstention. The model remains very similar. After voters
learn, privately, the precision and the content of their personal
signal, each decides, simultaneously and independently, whether to
vote or to abstain. Individuals who choose to vote cast a
single vote for one or the other of the two alternatives. The alternative
receiving more votes is chosen. As in the case of delegation, ties are resolved by a coin flip. 

Our model of majority voting with abstention differs under two important aspects from Feddersen and Pesendorfer (1996): there are no partisan voters, and experts are not perfectly informed. We are instead closer to McMurray (2013), but here too there are differences. McMurray does not distinguish experts, but widens the support of the distribution of precisions $F(q)$ to cover the full interval $[1/2,1]$. In addition, because of our experimental aim, we assume that the size of the electorate is known and finite, deviating from McMurray's large Poisson game set-up.    

Note that both delegation (in LD) and abstention (in DD) can improve decision-making because voters with weaker signals can opt out of participation. The two approaches, however, differ: delegation shifts voting power toward experts, whereas abstention amplifies the voices of all those who remain engaged. 

Before studying equilibrium under the two rules, we summarize the common timing of the game. First, nature draws the state $\omega$. Then precisions $q_i$ (degenerate for experts) and signals $s_i$ are drawn. Voter 
$i$ observes $q_i$ and $s_i$ and decides whether to vote for $d_1$, vote for $d_2$, or delegate to an expert (LD)/abstain (DD). Finally, the majority of votes cast determines the collective decision $d$, and payoffs are realized.

\section{Equilibrium }


Within each group of voters, experts and non-experts, the environment is symmetric. We thus focus on equilibria in symmetric strategies within each group. In addition, the signals' precision is independent of the content of the signals. In line with such additional symmetry assumption, we select equilibrium strategies that depend on the
signal's precision, but are symmetric with respect to the signal's
content.  

Under such selection criteria, conditional on voting, voting according to signal is undominated---a result that holds whether
delegation is allowed, or abstention is allowed, or all voters are required
to vote.\footnote{With $q_i$ and $p$ both higher than $1/2$ and symmetric strategies, one's signal can only increase the probability that the corresponding decision is correct. Conditional on voting and on pivotality, voting according to signal is the unique best response.} Moreover, delegation/abstention decisions themselves are not informative about the state, but only affect voting weights. We focus on their role in improving information aggregation by changing the relative powers of voters with signals of different precision.\footnote{To be clear: the symmetry restrictions on strategies are equilibrium selection criteria, not assumptions.}

Thus, we select semi-symmetric Perfect Bayesian equilibria
in undominated strategies where, when voting, voters follow their
signal, delegation or abstention decisions depend on the signal's
precision, and voters of a given category (non-experts or experts) follow
the same strategy, symmetric across signals' content. In what follows, ``equilibrium''
refers to such a notion. We denote the profile of strategies as $\sigma^{R}=\{\sigma_{ne}^{R},\sigma_{e}^{R}\}$, where
$R\in\{LD,DD,MV\}$ refers to delegation (under $LD$), abstention
(under $DD$), or majority rule with voting by all (under $MV$), and the subscript refers to either non-experts ($ne$), or
to experts ($e$). As noted earlier, conditional on voting, voting according to signal is an undominated strategy under all three rules. Our equilibrium notion thus selects a unique equilibrium under MV. Under LD, equilibrium strategies pin down the extent and origin of delegation; under DD, the extent and origin of abstention. 

The following Proposition highlights core features of such strategies: 

\bigskip

\begin{proposition}
\textbf{Equilibria.} \textit{For all $N=K+M$ finite, with $M>1$, and for all $F(q)$ everywhere continuous over $[1/2,p]$, for both
$R=LD$ and $R=DD$, in all semi-symmetric equilibria: (i) All experts vote; (ii) Non-experts' strategies
must be monotonic: there exists equilibrium $\widetilde{q}_{R}$ such that any individual
$i$ with $q_{i}>\widetilde{q}_{R}$ votes and any $i$
with $q_{i}<\widetilde{q}_{R}$ delegates/abstains.}  
\end{proposition}

\bigskip

We leave the proof to the Appendix (see Section \ref{subsec:prop1_proof}). The intuition however, is straightforward. Under LD, all experts vote because delegating to another expert, with equal information precision, wastes a signal without empowering a strictly better one. Under DD, all experts vote because abstaining risks empowering voters with less precise information. As for non-experts, when delegating or abstaining, their information cannot affect the common decision, whereas there is always a strictly positive probability that it might when voting. It follows then that voting must be more advantageous the higher the precision of the voter's information, supporting the equilibrium in monotone threshold strategies described by the Proposition.

The two results in the proposition replicate McMurray (2013)'s findings on abstention in our setting and contribute to the debate on LD. One common concern in the LD literature is the possibility of cycles of delegation. In our model, the likelihood of cycles is reduced by the assumption that only experts can receive delegated votes, but is not eliminated if experts choose to delegate. The Proposition shows that the issue is moot: experts always prefer to vote. In addition, non-strategic models in Computer Science stress that LD's positive effects depend on the correct selection of voters casting votes: they must be the better informed voters.\footnote{See for example the discussion in Berinsky et al. (2025).} The two results in the proposition state that this is indeed a feature of the equilibrium.

Note that if $\widetilde{q}_R \leq 1/2$, all non-experts vote; if $\widetilde{q}_R \geq p$, all non-experts delegate/abstain; if $\widetilde{q}_R \in (1/2,p)$, we say that the equilibrium is interior: non-experts with $q_i<\widetilde{q}_R$ delegate/abstain, and non-experts with $q_i>\widetilde{q}_R$ vote. This implies that if an interior equilibrium exists, there exist non-experts who vote although their information is strictly less precise, and \textit{known} to be strictly less precise, than the experts' information, and do so to maximize the chances of a correct decision. As the threshold $\widetilde{q}_R$ increases, voting power shifts toward better-informed agents, but the total amount of information aggregated in the final decision diminishes. 
The trade-off is at the heart of both systems. 


We are interested in the welfare properties of LD and DD,
relative to each other and to voting by all according to signal under MV.\footnote{As noted earlier, under symmetric strategies, voting according to signal is undominated under MV.} We say that an equilibrium under rule $R$ ``strictly improves" over rule $R'$ if
in equilibrium the ex ante probability of reaching the decision that
matches the state of the world is strictly higher than under any equilibrium of $R'$, or $EU_{R}>EU_{R'}$. We find:

\begin{thm*}
For all $N=K+M$ finite, with $M>1$, and for all $F(q)$ everywhere continuous over $[1/2,p]$:

(i) For any $M$, any $K$, and any $F$, there exists an equilibrium under LD that strictly improves over MV. On the other hand, there exist
$M$, $K$, and $F$ such that MV strictly improves over any equilibrium under DD
with positive probability of abstention. 

(ii) There are $M$, $K$, and $F$ such that there exists an equilibrium under LD that strictly improves over any equilibrium under DD. However, there are $M$, $K$, and $F$ such that there exists an equilibrium under DD that strictly improves over any equilibrium under LD. 
\end{thm*}

In this symmetric Condorcet-type model, majority voting is asymptotically optimal: as $N$ goes
to infinity, no alternative rule can yield a higher probability of a correct decision. The question we study,
and the theorem answers, is how LD and DD compare to 
majority voting and to each other when
the number of voters is finite. We find that LD can dominate MV for any parametrization, but the result does not extend to DD: there exist parameter values for which no equilibrium under DD can strictly improve over MV. As for comparing the two rules to each other, no general result holds: which rule has the potential to dominate the other depends on parameters.   

With finite $N$, voting models are famously sensitive to the details of the parametrization (for example,
whether $N$ is odd or even), and general results are lacking. Comparing LD to voting by all, the theorem gives a surprisingly sharp answer: for any
finite $N$, any ratio of experts and non-experts, and any (well-behaved)
distribution of precisions, there must exist an equilibrium with a
positive probability of delegation that dominates majority voting.\footnote{For future uses of the model, it is worth noting that the result continues to hold if delegation can target non-experts as well.}
The literature on LD typically delivers results on asymptotic equivalence to MV, or, with finite $N$, achieves improvements over MV under specific restrictions. The theorem establishes the general potential for strict improvement over MV with $N$ finite in a canonical model.\footnote{For results under strategic behavior and equilibrium, see Dhillon et al. (2025). Using algorithmic solutions, Berinsky et al. (2025) (building on Kahng et al. (2018)), discuss asymptotic strict improvements over MV if $\underline{q}<1/2$. The assumption, however, is problematic if individuals know their own signal precision and signals are binary. 
We return to this point when we
discuss Experiment 2.} 

The result is less clear-cut for DD, and the difference is interesting in itself. The different conclusions make clear that the existence of experts with superior information does not per se imply the inferiority of voting by all. DD can increase the relative voting weight of the experts through the abstention of less well-informed voters, and yet it may not improve over MV.\footnote{Even under delegation, the possibility of improvement over MV requires a positive probability of voters with close-to-uninformative signals ($\underline{q}=1/2$). For example, if $N$ is odd and there is a single expert, delegation does not dominate MV if $2log[\underline{q}/(1-\underline{q})]>log[p/(1-p)]$. Allowing some probability---even if small---of (almost) uninformed voters seems the better default assumption.}



We prove the theorem in the Appendix (Section 
\ref{subsec:theorem_proof}), but the logic behind result (i) is worth describing. Consider LD first.
We show that when delegation is an option there cannot be an equilibrium where all voters, including those with information barely better than random, choose to vote, thus replicating MV. But in this common interest problem, we know from
McLennan (1998) that if a profile of strategies that maximizes expected
utility exists, it must be an equilibrium. Such a set does exist
in our game, and thus an equilibrium must exist. But then such equilibrium must strictly
improve over MV and must include a positive probability of delegation.

In the case of DD, however,
an equilibrium with voting by all may exist, depending on parameter values, and the simple logic of the
proof cannot be applied. In such cases, whether or not there is an
equilibrium with DD that dominates majority voting needs to
be studied in detail. It is not difficult to find examples where the answer is negative (or examples where the answer is positive). 

The second part of the theorem compares LD and DD directly. We find that whether LD dominates DD, or DD dominates LD depends on parameter values. We report in the Appendix the equilibrium conditions and the expected utility equations for both LD and DD, for arbitrary parameter values (Section \ref{subsec:eqm_conditions}). However, while Proposition 1 characterizes equilibrium strategies, and the Theorem yields broad results on welfare, analytical comparative statics results remain out of reach. The Appendix describes the results of several numerical exercises (Section \ref{simulations}). Below, we discuss regularities uncovered by such exercises that help us build intuition for the parameterizations we use in the lab.

\subsection{Some numerical regularities}\label{sec: numerical regularities}

\begin{figure}[!t]
\begin{centering}
\includegraphics[width=\textwidth]{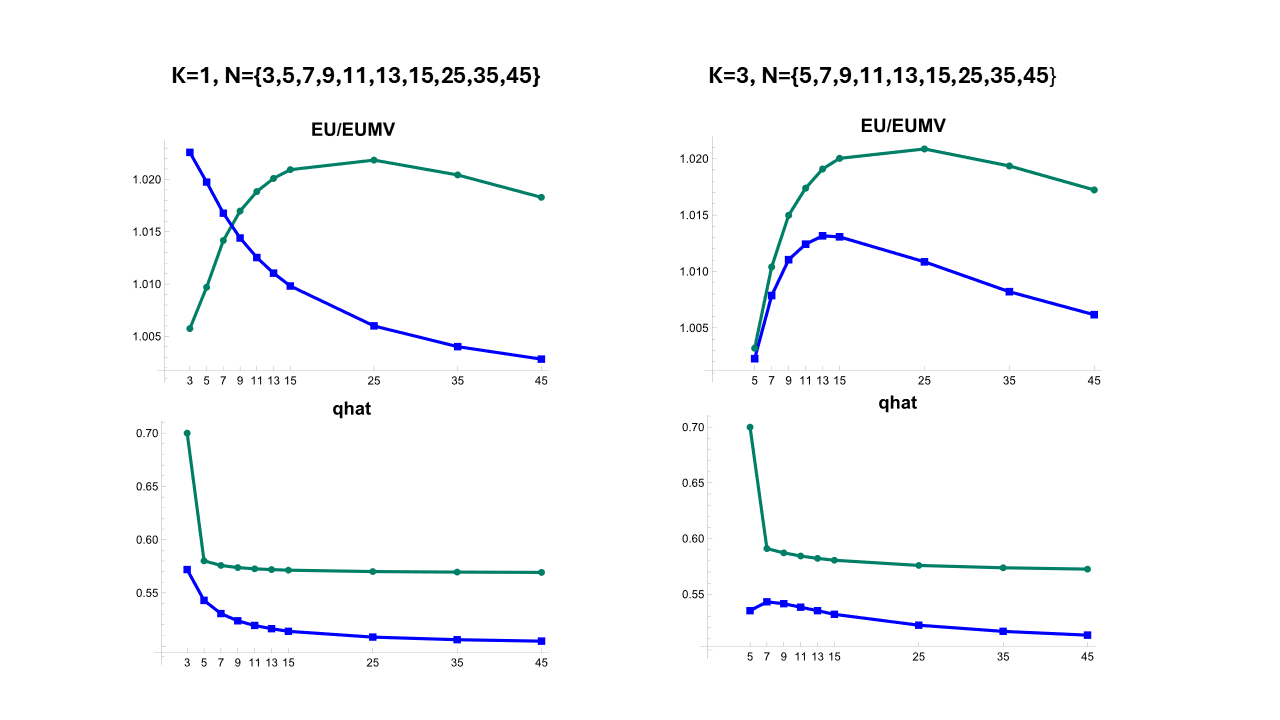}
\caption{\emph{Numerical examples}. Blue is LD, Green is DD. The horizontal axis is $N$. $F$ Uniform over $[0.5,0.7]$, $p=0.7$.}
\label{Fig.SimulationsK1K3.Uniform}
\end{centering}
\end{figure}

Figure \ref{Fig.SimulationsK1K3.Uniform} shows theoretical predictions in the best equilibrium for both LD and DD, under $F$ Uniform, for $K=1$, on the left, and $K=3$ on the right, with values of $N$ ranging from 3 (5 if $K=3$) to 45. LD is represented in Blue, and DD in Green.  

The upper panels depict expected utility, or the probability of the correct outcome under the two rules, relative to MV, as a function of the size of the group. All reported values are above 1, denoting, for these parameter values, improvements over MV for both rules. As the panels show, LD has the potential to dominate DD when the size of the group is small and $K=1$. With a single expert, however, LD's edge over MV declines at larger $N$, as the concentration of votes in the hands of the only expert becomes costly. DD on the contrary initially gains relative to MV as $N$ increases, before the gains eventually decline when $N$ becomes large. Predictably, DD is less sensitive to the number of experts than LD, and increasing $K$ to 3 has only minor effects: the left and right panels are quite similar for DD but not for LD. With $K=3$, LD performance relative to MV mimics DD, improving for intermediate groups sizes, before declining, but is consistently inferior to DD. 

The lower panels report $\widetilde{q}$, the precision threshold in each rule's best equilibrium. When the group is very small (and what ``small" is depends on $K$), there is no interior equilibrium under DD: the best equilibrium requires all non-experts to abstain and $\widetilde{q}_{DD}=p$. At larger group sizes, the optimal threshold is interior and remains quite constant, very slowly declining as $N$ increases. Under LD, there is again a sustained decline at larger $N$, but only after an initial increase if $K=3$. In all cases, under both values of $K$ and for all $N$ we studied, comparing the best equilibria of the two systems, $\widetilde{q}_{DD}>\widetilde{q}_{LD}$: the expected frequency of abstention is higher than the expected frequency of delegation.  

A common finding in all numerical exercises we report here and in the Appendix is the small magnitude of the quantitative improvements over MV: even in the best equilibrium, increases in the frequency of correct decisions from either LD or DD are of the order of 1-3\% regardless of the size of the group and the number of experts. Generating more substantive gains requires assuming that experts are radically better informed than any non-expert. 
What is more surprising, however, is that even in such a case, the relative performance of LD and DD remains very similar.\footnote{To be concrete, consider for example a group of 5 voters with a single expert, $p=0.9$ and $F$ Uniform over $[1/2, \bar{q}]$, with $\bar{q}=0.7$. Then, under the best equilibrium, both LD and DD have expected utility gains of 14\% over MV, but relative to each other their expected performance differs by less than 0.01\%.} We do not pursue this direction, in part because the calibration of experts' v/s non-experts' knowledge is bound to be controversial, but mostly because our focus is on the comparison between LD and DD. 

The numerical results help us put into perspective the model's theoretical predictions at the experimental parameters. 

\section{Experiment 1: Design \label{sec:Experiment-1: Design}}

The game we study in Experiment 1 follows very closely the theoretical
model, with one simplification: in the lab, only non-experts can delegate under LD or abstain under DD. We know from Proposition 1 that in equilibrium the experts always vote; we impose it in the lab to simplify the subjects' task and reduce unnecessary noise. 

The experiment studies four treatments: LD and DD for each of $N=5$ with $K=1$, and $N=15$ with $K=3$. In both cases, the number of experts is fixed at one fifth of the group. Using the number of experts as index, we call the four treatments LD1 and DD1, and LD3 and DD3. In all experiments, we set $p=0.7$, and $F(q)$ Uniform over $[0.5,0.7]$.



Table \ref{table: equilLD} reports detailed theoretical predictions under LD.

\begin{table}[h]
\caption{$p=0.7$, $F(q)$ Uniform over $[0.5,0.7]$}

\begin{centering}
\begin{tabular}{|c|c|c|c|ccccc|c|c|c}
\multicolumn{4}{c}{$LD1:$ $N=5;$ $K=1$} &  &  &  &  & \multicolumn{4}{c}{$LD3:$ $N=15;$ $K=3$}\tabularnewline
\multicolumn{1}{c}{} & \multicolumn{1}{c}{} & \multicolumn{1}{c}{} & \multicolumn{1}{c}{} &  &  &  &  & \multicolumn{1}{c}{} & \multicolumn{1}{c}{} & \multicolumn{1}{c}{} & \tabularnewline
\cline{1-4} \cline{2-4} \cline{3-4} \cline{4-4} \cline{9-12} \cline{10-12} \cline{11-12} \cline{12-12} 
$\widetilde{q}_{LD}$ & $F(\widetilde{q}_{LD})$ & $EU_{LD}$ & $EU_{MV}$ &  &  &  & \multicolumn{1}{c|}{} & $\widetilde{q}_{LD}^{3}$ & $F(\widetilde{q}_{LD}^{3})$ & $EU_{LD}^{3}$ & \multicolumn{1}{c|}{$EU_{MV}^{3}$}\tabularnewline
\cline{1-4} \cline{2-4} \cline{3-4} \cline{4-4} \cline{9-12} \cline{10-12} \cline{11-12} \cline{12-12} 
0.7 & 1 & 0.7 & \multirow{2}{*}{0.717} &  &  &  & \multicolumn{1}{c|}{} & 0.532 & 0.162 & 0.843 & \multicolumn{1}{c|}{0.832}\tabularnewline
\cline{1-3} \cline{2-3} \cline{3-3} \cline{9-12} \cline{10-12} \cline{11-12} \cline{12-12} 
0.543 & 0.215 & 0.731 &  &  &  &  &  & \multicolumn{1}{c}{} & \multicolumn{1}{c}{} & \multicolumn{1}{c}{} & \tabularnewline
\cline{1-4} \cline{2-4} \cline{3-4} \cline{4-4} 
\end{tabular}
\par\end{centering}
\label{table: equilLD}
\end{table}

Treatment LD1 has two symmetric equilibria: an equilibrium where every voter delegates to the expert with probability 1 ($\widetilde{q}_{LD}=p$), and a unique interior equilibrium with $\widetilde{q}_{LD}=0.543$. The first equilibrium is not strict: no individual non-expert is pivotal and delegating one's vote is a weak
best response. The expert alone controls the outcome and the probability of making the correct choice equals $p$ ($EU_{LD}=0.7$), lower than under MV. The interior equilibrium is instead strict and yields a higher probability of the correct outcome, and higher than under MV, in line with the result of the Theorem: $EU_{LD}=0.731$. Note that the interior threshold  $\widetilde{q}$ is low,
and the ex ante probability of delegation is only just above 20\%.\footnote{With a single expert, the uniqueness of the symmetric equilibrium
with interior $\widetilde{q}_{LD}$ can be proven analytically and holds
for arbitrary $N$. Asymmetric equilibria are theoretically possible but require coordination and realistically cannot be engineered in the lab under random rematching and no communication.} 

In LD3, full delegation is not an equilibrium. Intuitively, 
when there are multiple experts and all other non-experts delegate,
voter $i$ can be pivotal only if the experts disagree among themselves. The disagreement reduces the attraction of delegation.
The unique equilibrium has interior $\widetilde{q}_{LD}$, and equilibrium delegation remains rare: the expected frequency of individual delegation falls to 16\%. The interior equilibrium again dominates MV. However, with the increase in $N$, the effect of the Condorcet Jury Theorem becomes pronounced and the quantitative improvement declines.  


Table \ref{table:equilMVA} shows the equilibria under DD.

\begin{table}[h]
\caption{$p=0.7$, $F(q)$ Uniform over $[0.5,0.7]$}

\begin{centering}
\begin{tabular}{|c|c|c|c|cccc|c|c|c|c|}
\multicolumn{4}{c}{$DD1:$ $N=5;$$K=1$} &  &  &  & \multicolumn{1}{c}{} & \multicolumn{4}{c}{$DD3:$ $N=15;$$K=3$}\tabularnewline
\multicolumn{1}{c}{} & \multicolumn{1}{c}{} & \multicolumn{1}{c}{} & \multicolumn{1}{c}{} &  &  &  & \multicolumn{1}{c}{} & \multicolumn{1}{c}{} & \multicolumn{1}{c}{} & \multicolumn{1}{c}{} & \multicolumn{1}{c}{}\tabularnewline
\cline{1-4} \cline{2-4} \cline{3-4} \cline{4-4} \cline{9-12} \cline{10-12} \cline{11-12} \cline{12-12} 
$\widetilde{q}_{DD}$ & $F(\widetilde{q}_{DD})$ & $EU_{DD}$ & $EU_{MV}$ &  &  &  &  & $\widetilde{q}_{DD}^{3}$ & $F(\widetilde{q}_{DD}^{3})$ & $EU_{DD}^{3}$ & $EU_{MV}^{3}$\tabularnewline
\cline{1-4} \cline{2-4} \cline{3-4} \cline{4-4} \cline{9-12} \cline{10-12} \cline{11-12} \cline{12-12} 
0.7 & 1 & 0.7 & \multirow{3}{*}{0.717} &  &  &  &  & 0.7 & 1 & 0.784 & \multirow{3}{*}{0.832}\tabularnewline
\cline{1-3} \cline{2-3} \cline{3-3} \cline{9-11} \cline{10-11} \cline{11-11} 
0.580 & 0.40 & 0.724 &  &  &  &  &  & 0.581 & 0.40 & 0.849 & \tabularnewline
\cline{1-3} \cline{2-3} \cline{3-3} \cline{9-11} \cline{10-11} \cline{11-11} 
0.5 & 0 & 0.717 &  &  &  &  &  & 0.5 & 0 & 0.832 & \tabularnewline
\cline{1-4} \cline{2-4} \cline{3-4} \cline{4-4} \cline{9-12} \cline{10-12} \cline{11-12} \cline{12-12} 
\end{tabular}
\par\end{centering}
\label{table:equilMVA} 
\end{table}

For both group sizes, under DD, there are three symmetric equilibria. Two
are boundary equilibria, with either zero ($\widetilde{q}_{DD}=0.5)$
or full ($\widetilde{q}_{DD}=0.7$) abstention; one is an interior
equilibrium and is unique among interior equilibria.\footnote{The existence of the boundary equilibria depends on $N$ and $K$ odd.} As in the case of LD, under DD as well the interior equilibrium dominates MV, and this is true for both group sizes.

The interior equilibrium threshold $\widetilde{q}_{DD}$ remains nearly constant around $0.58$ for both $N=5$ and $N=15$, and in both cases is higher than $\widetilde{q}_{LD}$. With an expected frequency of 40\%, it implies a larger expected number of abstentions than delegations. For example, under the interior equilibrium and rounding up to integers, when $N=15$, we expect 2 non-experts to delegate under LD, but 5 non-experts to abstain under DD. 

The choice of parameters does not stack the experiment in favor of either LD or DD. Comparing the best equilibria, LD dominates DD at $N=5,$ but the reverse holds at $N=15$. In addition, as noted earlier, expected improvements over MV are quantitatively minor for both rules and both parametrizations, even in the best equilibria. In the lab, we should expect the two rules (and MV) to perform very similarly. The experiment will tell us how the three rules rank in the data.\footnote{An important question concerns the robustness of the two voting rules to strategic mistakes. In the online Appendix in Section \ref{subsec:robustness}, we study the experimental treatments supposing that the threshold $\widetilde{q}$ is chosen incorrectly but, for each rule, remains symmetric across all non-experts. For both group sizes, we find that strategic mistakes of this type are more costly for LD than for DD, in terms of both the frequency and the expected size of losses relative to MV. However, the magnitude of the losses remains small.}




We ran the experiment online, using the Zoom
videoconferencing software. Participants were recruited from the Columbia
Experimental Laboratory for the Social Sciences (CELSS)\textquoteright{}
ORSEE website.\footnote{Greiner (2015). CELSS' ORSEE subjects are primarily undergraduate
students at Columbia University or Barnard College.} They received instructions and communicated with the experimenters
via Zoom, and accessed the experiment interface on their personal
computer's web browser. The experiment was programmed in oTree and,
with the exception of a more visual style for the instructions, developed
very similarly to an in-person experiment. 


Participants were asked to vote on the correct selection of a box
containing a prize, out of two possible choices, a green box and a
purple box. The computer selected the winning box putting equal probability
on either; conditionally on the computer's random choice, participants
then received a message suggesting a color and were told the probability
that the message was accurate.\footnote{To limit decimal digits, the precision of the signal was drawn uniformly
from a discrete distribution with bins of size 0.01. When comparing
the experimental results to the theory, below, we compute equilibria
using the corresponding discrete distribution of precisions. The differences
from using a continuous distribution are minute.} The same screen also informed them of whether or not they were an
expert (for that round).

We ran two different designs, which we index here by 1 and 2. Under Design 1, after learning their type and precision, participants were asked to vote for
one of the two boxes, if experts, or, if non-experts, to either choose
one of the boxes or delegate their vote to an expert (in the LD treatments), or abstain (in the DD treatments). We ran 10 sessions of Design 1, with a total of 150 subjects. 

Under Design 1, we did not ask subjects who chose to delegate or abstain how they would have voted otherwise, and thus we did not collect votes under MV.\footnote{The rationale is that asking inexperienced subjects to choose a box immediately after receiving an informative signal risks triggering random behavior.} To verify the robustness of the results, we later ran a second design. In Design 2, participants were informed that, after voting, the computer could decide, with equal probability, whether or not to require voting by all. Thus, if a participant chose to delegate or abstain, in the following screen we also collected their vote in case the computer required it. Once voting for the round was concluded, the computer randomly chose how votes would be counted (summing all individual votes, or allowing delegation/abstention). Thus, under Design 2 we collect all individual votes. We ran 10 sessions of Design 2, again with a total of 150 subjects.    

Under both designs, after each round, all participants
received the same feedback, reporting where the prize was and how many
votes each box received. In Design 1, the feedback also included how the experts voted, and how many non-experts
delegated their vote or abstained. In Design 2, participants were told whether the computer forced voting by all; if not, information on delegations or abstentions was conveyed, as in Design 1. Under both designs, across rounds, expert/non-expert identities were re-assigned randomly, under the constraint that groups of 5 voters had a single expert, and a group of 15 had three; if the session involved multiple groups, they were re-formed randomly. A copy of the instructions is reproduced in online \href{https://vmooers.github.io/assets/pdf/LD.ExperimentalInstructions.pdf}{Appendix C}.

Under Design 1, participants played 20 rounds each of two treatments (40 rounds in total), balancing order and
treatment composition across sessions, as reported in 
Table \ref{tab:Experiment-1:-Experimental}. Under Design 2, only about half of the rounds resulted in delegation/abstention and thus, to allow for similar learning, each session had 40 rounds of a single treatment (one of LD1, LD3, DD1, DD3, plus collecting votes under MV). We report the sessions of Design 2 in Table \ref{tab:Experiment-1:-Experimental-2}. Under each design, we have data for 240 rounds for LD1 and for DD1, and 120 rounds for LD3 and for DD3.

\begin{table}[H]
\centering{}\caption{\emph{Experiment 1: Experimental Design 1\label{tab:Experiment-1:-Experimental}}}
\begin{tabular}{|c|c|c|c|c|}
\hline 
Sessions & Treatments & Rounds & Subjects & Groups\tabularnewline
\hline 
\hline 
1a & LD1, LD3 & 20, 20 & 15 & 3, 1\tabularnewline
\hline 
1b & LD3, LD1 & 20, 20 & 15 & 1, 3\tabularnewline
\hline 
2a & DD1, DD3 & 20, 20 & 15 & 3, 1\tabularnewline
\hline 
2b & DD3, DD1 & 20, 20 & 15 & 1, 3\tabularnewline
\hline 
3a, 3a' & LD3, DD3 & 20, 20 & 15 & 1, 1\tabularnewline
\hline 
3b, 3b' & DD3, LD3 & 20, 20 & 15 & 1, 1\tabularnewline
\hline 
4a & LD1, DD1 & 20, 20 & 15 & 3, 3\tabularnewline
\hline 
4b & DD1, LD1 & 20, 20 & 15 & 3, 3\tabularnewline
\hline 
\end{tabular}
\end{table}

\begin{table}[H]
\centering{}\caption{\emph{Experiment 1: Experimental Design 2\label{tab:Experiment-1:-Experimental-2}}}
\begin{tabular}{|c|c|c|c|c|}
\hline 
Sessions & Treatments & Rounds & Subjects & Groups\tabularnewline
\hline 
\hline 
5, 5' & LD1 & 40 & 15 & 3 \tabularnewline
\hline 
6, 6', 6'' & LD3 & 40 & 15 & 1 \tabularnewline
\hline 
7, 7' & DD1 & 40 & 15 & 3 \tabularnewline
\hline 
8, 8', 8'' & DD3 & 40 & 15 & 1 \tabularnewline
\hline 
\end{tabular}
\end{table}

Under both designs, each session lasted about
90 minutes with average earnings of \$25.50, including a show-up fee of \$5. 

\section{Experiment 1: Results}

The theory leads us to focus on four main questions. At the individual subject level, we ask: (1) Are individual strategies in the lab monotonic? (2) Do experimental subjects use thresholds close to what theory predicts? (3) Is the observed frequency of abstention higher than the frequency of delegation, as expected? At the group level, the main question concerns the outcome: (4) How does the frequency of correct decisions in the lab compare under LD and DD, relative to each other and to majority voting by all?

\subsection{Individual behavior}

Under both voting systems, equilibrium voting strategies are monotonic
in precision (if non-expert $i$ votes at precision $q(i)$, then
$i$ votes at all $q'(i)>q(i)$). As we document more precisely in Appendix Section \ref{subsec:appendix_monotonicity},
individual behavior in the experiment is mostly monotonic, with small and comparable rates of monotonicity violations under 
LD and DD for both designs. Averaging over the two designs and the two rules, approximately 50\% of subjects have no violations at all, and the fraction reaches 80\% and above if we limit attention to the last 10 rounds of each treatment.\footnote{In Design 1, just below 60\% of subjects have no violations at all under LD, and just above 60\% under DD; in Design 2, 40\% of subjects have no violations at all under LD, and just above 40\% under DD. (Recall that in Design 2, subjects play twice as many rounds of a given treatment.) In Design 1 the results are invariant to the size of the group, while in Design 2 monotonicity violations are somewhat more common in the larger groups.}

 

\begin{figure}[!t]
\begin{centering}
\includegraphics[width=0.6\textwidth]{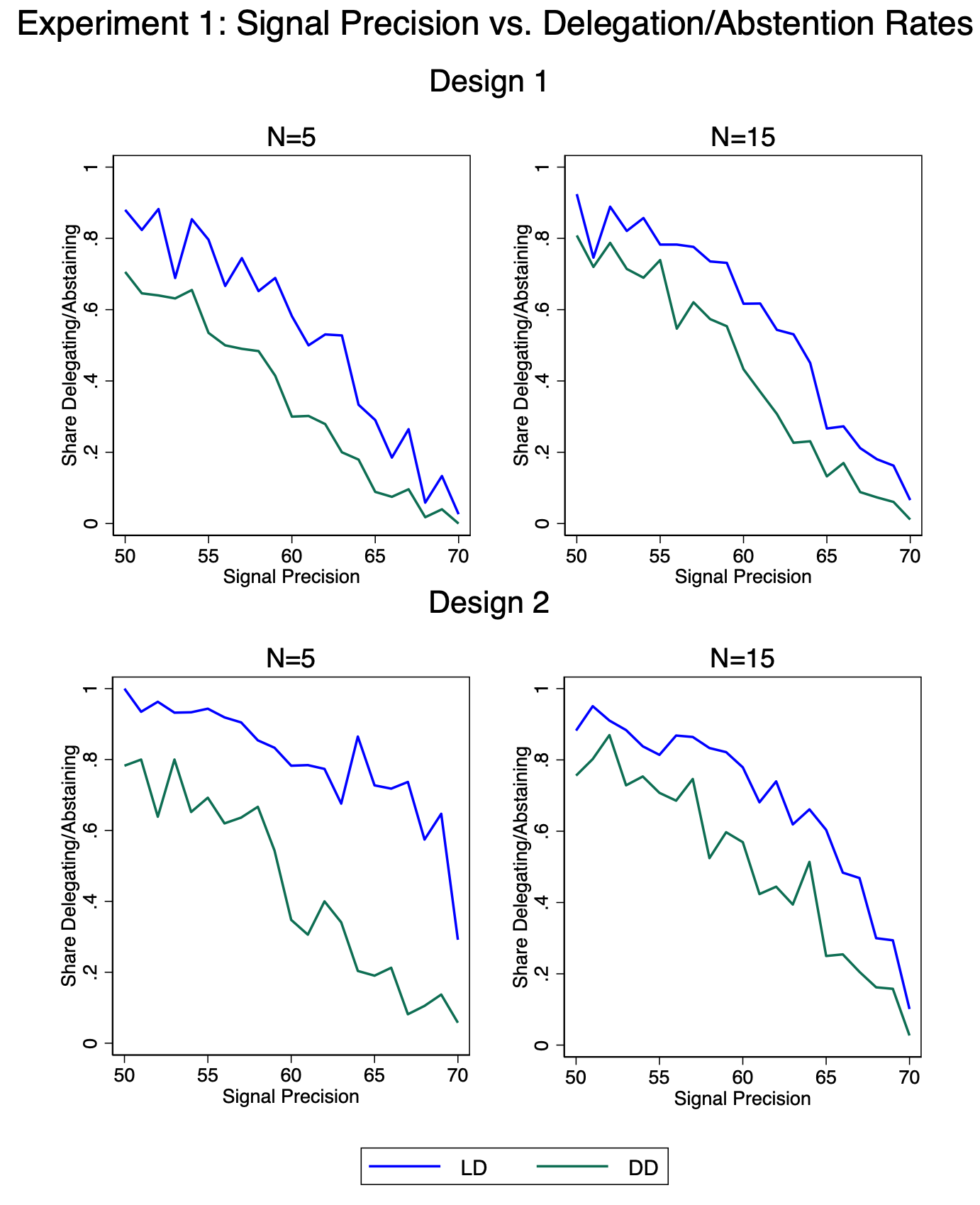}
\par\end{centering}
\caption{\emph{Delegation and abstention frequencies by signal precision.} }

\label{Fig.ind_freq.rates}
\end{figure}

The data convey the subjects' sensitivity to the precision of the
signals. Figure \ref{Fig.ind_freq.rates} reports
the empirical frequencies of delegation or abstention as functions
of individual precision. In all treatments, the frequencies decline
at higher signal precision. Note however the differences across the two rules:
for any precision, the delegation rate is higher than the abstention
rate, for both group sizes and in both designs.

\begin{figure}[!t]
\begin{centering}
\includegraphics[width=0.6\textwidth]{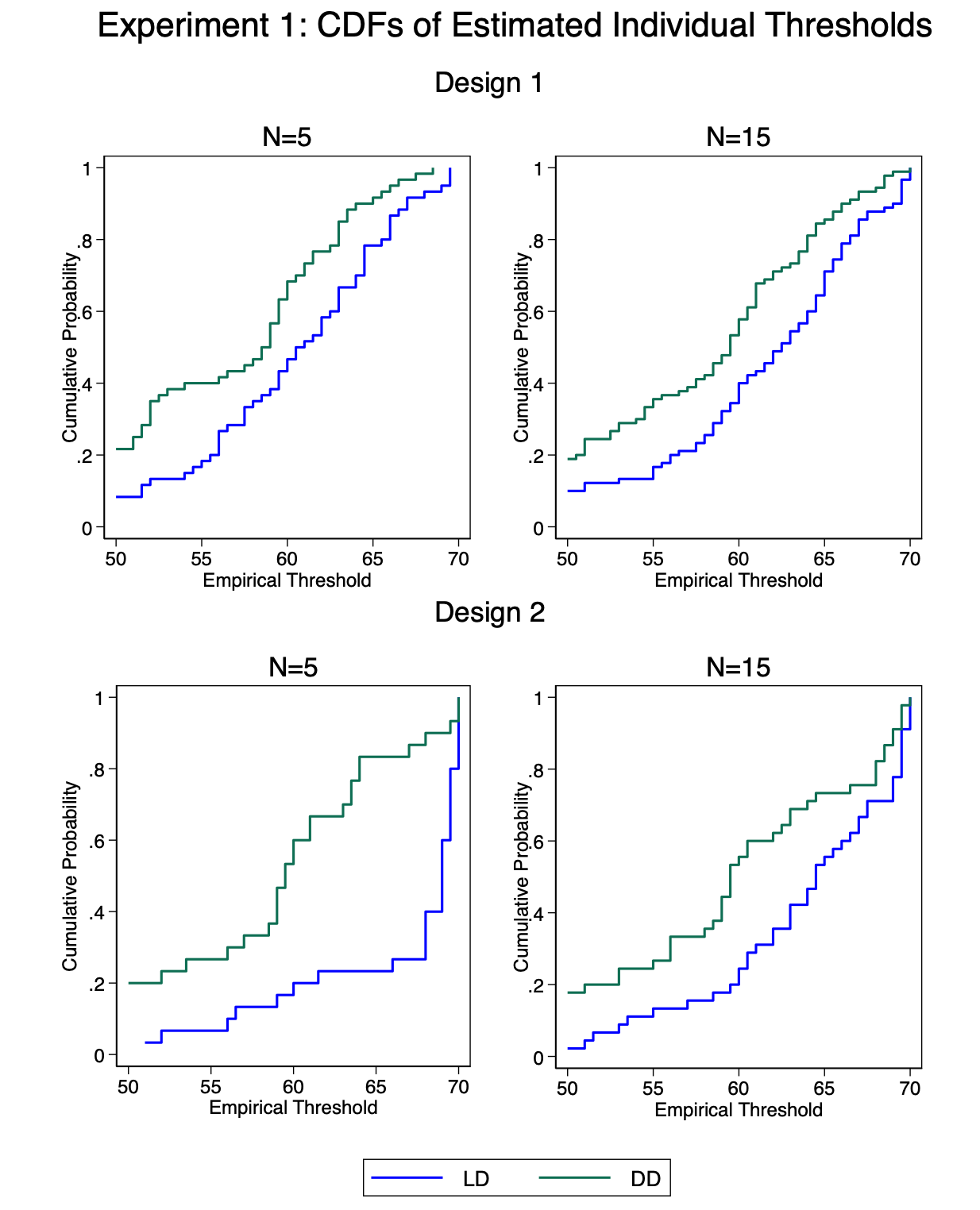}
\par\end{centering}
\caption{\emph{CDFs of estimated individual voting thresholds.} }

\label{Fig.ind_freq.cdfs}
\end{figure}



The higher propensity towards delegation, relative to abstention,
is confirmed by Figure \ref{Fig.ind_freq.cdfs}, where we plot the the cumulative distribution functions of estimated
individual voting thresholds.\footnote{Exploiting monotonicity, we can estimate individual voting thresholds
$\widetilde{q}_{i,LD}$ (or $\widetilde{q}_{i,DD}$) that, for each
subject, minimize the frequency of monotonicity violations. See Appendix Section \ref{subsec:appendix_monotonicity} for details.} For both group sizes and in both designs, the LD distribution, in blue, first order stochastically
dominates the DD distribution, in green: at any precision, the fraction of thresholds estimated to be below the precision is lower under LD. Thresholds below the precision value correspond to subjects voting at that precision, implying that fewer subjects are voting under LD than under DD, or, equivalently, that more subjects are delegating than abstaining.


Two-sample Kolmogorov-Smirnov
tests adjusted for discreteness confirm the visual impression: for
both group sizes and both designs, the probability that the two samples of thresholds,
for LD and for DD, are drawn from the same distribution is very low.
(Across both designs: for $N=5$, $p<0.035$, and for $N=15$, $p<0.011$.)

The gradual decline in delegation or abstention as precision increases
(in Figure \ref{Fig.ind_freq.rates}) and the dispersion
in estimated thresholds (in Figure \ref{Fig.ind_freq.cdfs}) are typical of similar
experiments (for example, Levine and Palfrey, 2007; Morton and Tyran,
2011). The data cast some doubt on the focus on symmetric equilibria,
but each individual's precision draws are relatively few,\footnote{As noted, 16 rounds on average are played as non-expert in Design 1, and 32 rounds as non-expert in Design 2.}
unlikely to span the whole support of precisions, and heterogeneous
across subjects. Some dispersion
in estimated thresholds will result by construction. 

Regressions of individual voting behavior on signal quality, controlling
for round and, in Design 1, for treatment order effects, confirm the sensitivity to
precision and the higher propensity towards delegation. Table \ref{tab:Determinants-of-delegation-1}
reports linear probability regressions for LD1 and DD1 ($N=5$) and
for LD3 and DD3 ($N=15$).\footnote{The first two columns refer to Design 1, in which subjects participated in two treatments in each session. \textquotedblleft Second\textquotedblright{} indicates that the treatment
appeared second in the session. \textquotedblleft Mixed\textquotedblright{}
indicates that both an LD treatment and a DD treatment appeared
in the session. In both columns, the excluded case is DD played as
first treatment in DD-only sessions.} As expected, the propensity to abstain or delegate responds negatively
to higher precision of the signal, across both group sizes and both designs.\footnote{In Design 2, for $N=5$, the coefficient on the interaction between LD and signal precision is positive and highly significant, indicating that delegation falls less with precision than abstention does---a little more than half as quickly. For $N=15$ under both designs, and for Design 1 $N=5$, the coefficients on this interaction are small and not significant.} 
The coefficient of the LD dummy is positive in all cases and generally highly significant.\footnote {It is marginally significant in Design 2 for LD3. In Design 1, with multiple treatments per session, there is some evidence of order effects, but
the net effect of delegation remains always positive. 
In Appendix Section \ref{subsec:appendix_additional_empirical}, we report probit estimates of Table \ref{tab:Determinants-of-delegation-1} as well as between-subject regressions run on first treatments only for Design 1, with unchanged results.}

\begin{table}[p]
\begin{singlespace}
\centering{}%
\resizebox{\textwidth}{!}{
\begin{tabular}{l>{\centering}p{2.8cm}>{\centering}p{2.8cm}>{\centering}p{2.8cm}>{\centering}p{2.8cm}}
\multicolumn{5}{c}{Experiment 1: Frequency of Delegation or Abstention.}\tabularnewline
\midrule 
 & \multicolumn{2}{c}{\rule{0pt}{1em}Design 1} & \multicolumn{2}{c}{Design 2} \\
 \cline{2-3} \cline{4-5}
 & \rule{0pt}{1em}(1) & (2) & (3) & (4)\tabularnewline
 & N=5 & N=15 & N=5 & N=15\tabularnewline
\midrule 
LD & 0.328{*}{*}{*} & 0.208{*}{*} & 0.206{*}{*} & 0.192{*} \tabularnewline
 & (0.073) & (0.071) & (0.041) & (0.095) \tabularnewline
\vspace{6pt} & {[}0.006{]} & {[}0.022{]} & {[}0.015{]} & {[}0.099{]} \tabularnewline
Signal Precision & -0.777{*}{*}{*} & -0.861{*}{*}{*} & -0.811{*}{*}{*} & -0.767{*}{*}{*} \tabularnewline
 & (0.080) & (0.047) & (0.027) & (0.039) \tabularnewline
\vspace{6pt} & {[}0.000{]} & {[}0.000{]} & {[}0.000{]} & {[}0.000{]} \tabularnewline
LD {*} Signal Precision & -0.078 & 0.011 & 0.342{*}{*}{*} & 0.063 \tabularnewline
 & (0.059) & (0.037) & (0.056) & (0.081) \tabularnewline
\vspace{6pt} & {[}0.248{]} & {[}0.783{]} & {[}0.009{]} & {[}0.472{]} \tabularnewline
Round & 0.031 & 0.078 & 0.092 & 0.077{*} \tabularnewline
 & (0.057) & (0.056) & (0.085) & (0.033) \tabularnewline
\vspace{6pt} & {[}0.613{]} & {[}0.208{]} & {[}0.362{]} & {[}0.066{]} \tabularnewline
LD {*} Round & -0.112 & -0.051 & -0.032 & -0.085 \tabularnewline
 & (0.100) & (0.067) & (0.104) & (0.076) \tabularnewline
\vspace{6pt} & {[}0.314{]} & {[}0.475{]} & {[}0.780{]} & {[}0.319{]} \tabularnewline
Second & 0.154{*}{*}{*} & -0.096{*}{*} & & \tabularnewline
 & (0.010) & (0.035) & & \tabularnewline
\vspace{6pt} & {[}0.000{]} & {[}0.029{]} & & \tabularnewline
LD {*} Second & -0.090 & 0.037 & & \tabularnewline
 & (0.050) & (0.037) & & \tabularnewline
\vspace{6pt} & {[}0.134{]} & {[}0.349{]} & & \tabularnewline
Second {*} Mixed & -0.129{*}{*}{*} & 0.078{*}{*}{*} & & \tabularnewline
 & (0.002) & (0.006) & & \tabularnewline
\vspace{6pt} & {[}0.000{]} & {[}0.000{]} & & \tabularnewline
LD {*} Second {*} Mixed & -0.025{*}{*}{*} & -0.166{*}{*} & & \tabularnewline
 & (0.006) & (0.055) & & \tabularnewline
\vspace{6pt} & {[}0.008{]} & {[}0.019{]} & & \tabularnewline
Constant & 0.675{*}{*}{*} & 0.832{*}{*}{*} & 0.794{*}{*}{*} & 0.847{*}{*}{*} \tabularnewline
 & (0.048) & (0.069) & (0.008) & (0.093) \tabularnewline
\vspace{6pt} & {[}0.000{]} & {[}0.000{]} & {[}0.000{]} & {[}0.000{]} \tabularnewline
\midrule 
Observations & 1,920 & 2,880 & 1,920 & 2,880 \tabularnewline
R-squared & 0.309 & 0.309 & 0.323 & 0.249 \tabularnewline
\midrule
\multicolumn{5}{l}{{*}{*}{*} p\textless 0.01, {*}{*} p\textless 0.05, {*} p\textless 0.1}\tabularnewline
\end{tabular}
}
\caption{\emph{Determinants of delegation and abstention}. Linear probability
models. Standard errors in parentheses, clustered at the session level.
P-values in brackets. Delegation/abstention is measured as a binary
0-1 subject decision. The values for signal precision and round have
been scaled to be between 0 and 1. \label{tab:Determinants-of-delegation-1}}
\end{singlespace}
\end{table}

\subsection{Aggregate results}

\subsubsection{Frequency of delegation and abstention}

\begin{figure}[!t]
\begin{centering}
\includegraphics[scale=0.16]{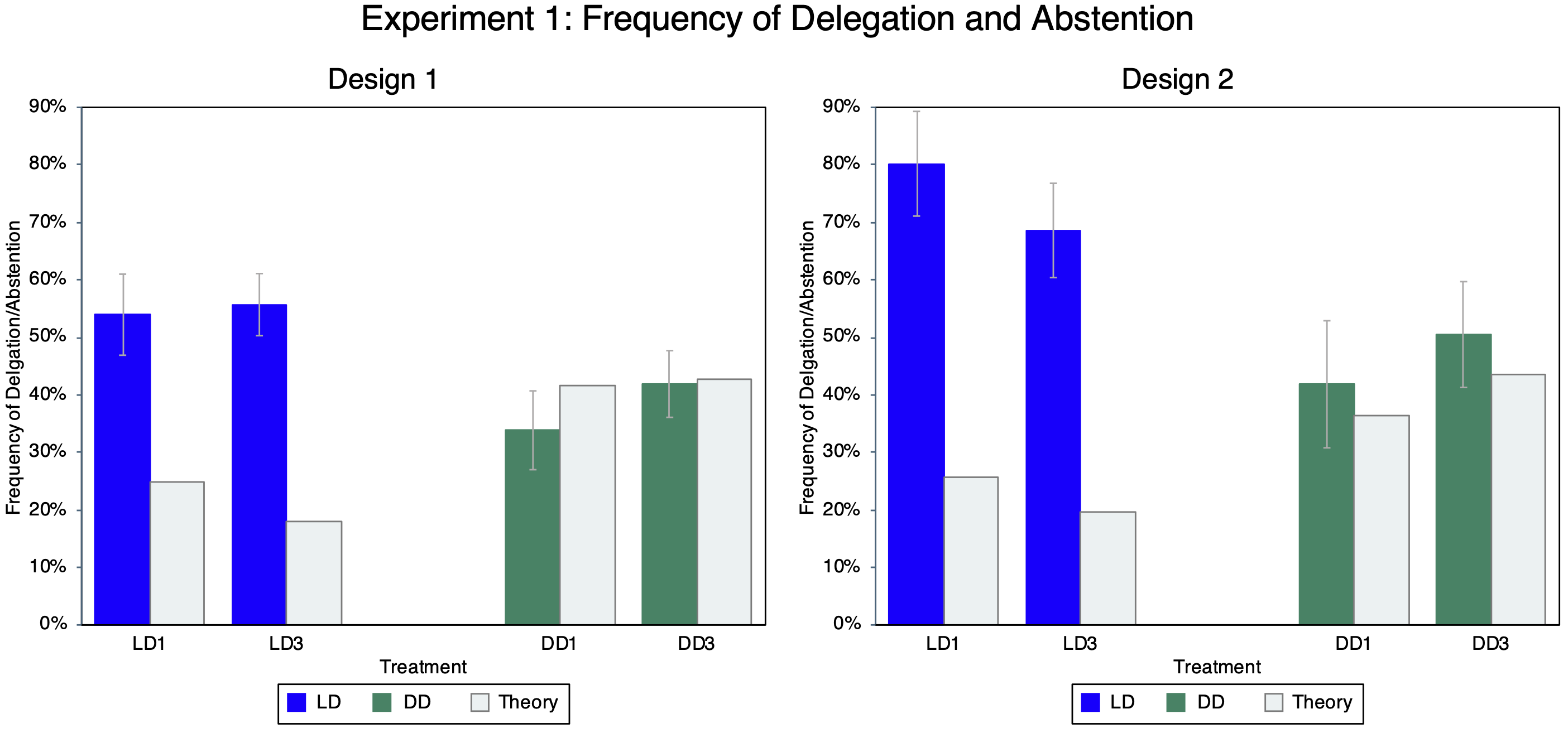}
\par\end{centering}
\caption{\textit{Aggregate frequency of delegation and abstention}. Confidence
intervals are calculated from standard errors clustered at\textbf{
}the individual level. \label{fig:Exp1.delegation.abstention} }
\end{figure}

Figure \ref{fig:Exp1.delegation.abstention} reports the
aggregate frequencies of delegation (in blue) and abstention (in green)
in the data, and according to the predictions of the interior equilibrium,
given realized precision draws in the experiment (in grey). Columns
on the left refer to LD treatments; columns on the right to DD. The
95\% confidence intervals are calculated from standard errors clustered at the individual level.\footnote{Under Design 1, standard errors can alternatively be clustered at the session level and results are unchanged. Under Design 2, sessions per treatment are too few for clustering at the session level.}

Delegation rates in the experiment are between two and three-and-a-half times
what theory predicts for the interior equilibrium. 
Abstention rates, on the other hand, are comparable to the predictions. 
The conclusion is robust to disaggregating by session, or to considering only the 10 final rounds. The interior equilibrium is the relevant reference because it is under such an equilibrium that both LD and DD outperform MV. In the experiment, subjects approximate it closely under DD, but overdelegate substantially under LD. Observed delegation rates suggest that subjects are not using LD effectively.

Leaving aside efficiency considerations, a separate question is whether the high delegation rates we see reflect different equilibria. Asymmetric equilibria, when they exist, are highly improbable in the lab because they require coordination that random group rematching across rounds makes practically unfeasible. A (weak) symmetric equilibrium with all non-experts delegating exists for LD1, but the number of subjects who always delegate in the experiment under LD1 is very small: 3 out of 60 in Design 1, and 4 out of 30 in Design 2. As Figure \ref{fig:Exp1.delegation.abstention} shows, a $100\%$ delegation rate is outside the LD1 confidence interval for both designs.\footnote{Symmetric equilibria at the boundaries (with non-experts either all voting, or all abstaining) exist for DD1 and DD3, where however the data are better explained by the interior equilibria.} Note also, in passing, that the high frequency of delegation cannot be attributed
to participants best responding to others' experimental choices: too high of delegation rates by others depress own optimal delegation rates.

\subsubsection{Frequency of correct choice}


Beyond regularities of delegation and abstention, the real variable
of interest is the frequency with which the voting system leads to
the correct choice. 

\begin{figure}[h]
\centering{}\includegraphics[scale=.16]{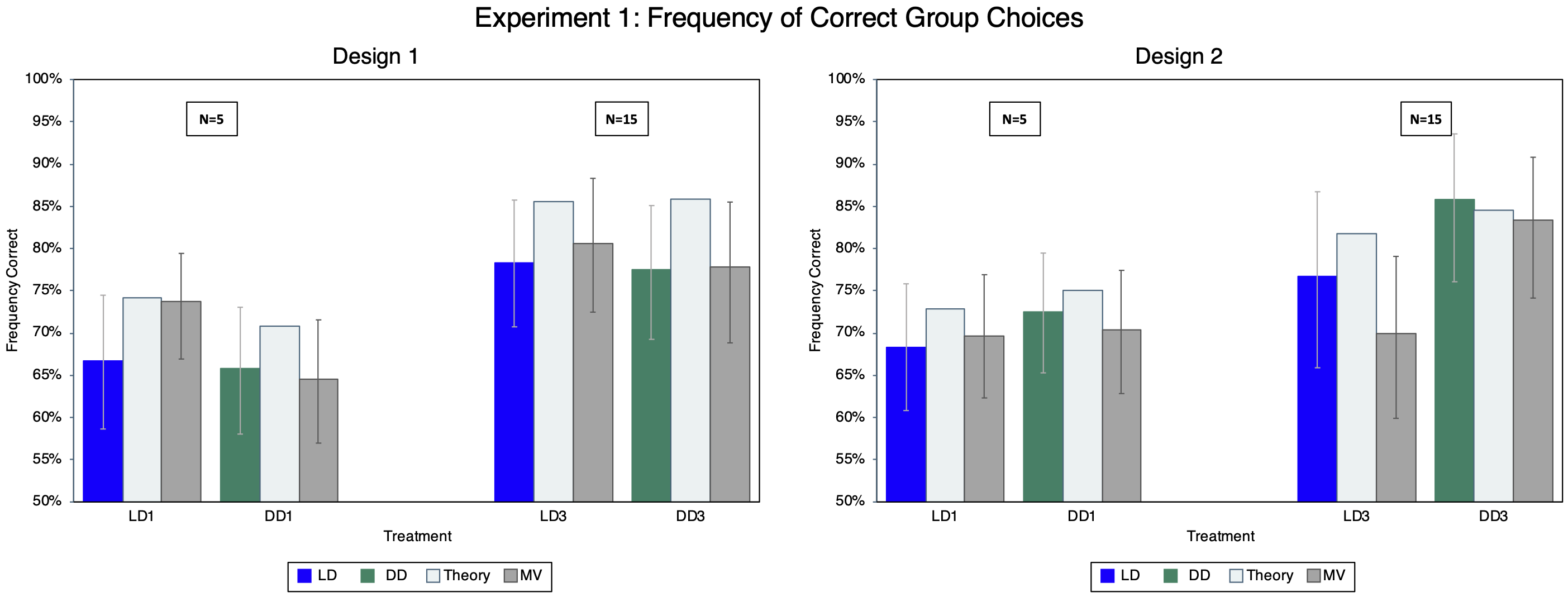}\caption{\textit{Frequency of correct outcomes.} Confidence
intervals are calculated from bootstrapping, using 100,000 simulated
data sets and allowing for within-subject correlation. }\label{fig:Exp1.Correct choices}
\end{figure}

Figure \ref{fig:Exp1.Correct choices} reports the frequency of correct choices in the experimental
data and compares them to the theoretical interior equilibrium (in light grey) and
to MV (in darker grey).\footnote{In Design 2, MV data are as collected; in Design 1, the
outcomes under MV for subjects who delegated or abstained are constructed allowing for a positive probability of voting
against signal. In the figure, such a probability equals the frequency
observed in the treatment, but the results are robust to plausible alternatives. To account for
the randomness in MV data and then for consistency, all 95\% confidence
intervals are calculated from bootstrapping, using 100,000 simulated
data sets. Each subject is drawn with a full set of choices (20 in Design 1, 40 in Design 2), thus
allowing for within-subject correlations.} The vertical axis is the frequency
of correct outcomes over the full data set for the corresponding treatment. 


The figure shows that, with a single exception, both LD and DD underperform relative to their best theoretical equilibrium, with underperformance somewhat more pronounced for LD. With one exception, LD tends to underperform also relative to MV, while DD is equivalent or slightly superior to MV in all cases.

This said, we cannot draw convincing conclusions from Figure \ref{fig:Exp1.Correct choices}. As the figure shows, none of the qualitative differences described above is statistically significant. Collective choice experiments yield few data on outcomes, relative to data on individual actions. Here, in addition, both LD and DD are variations of MV, and very frequently result in the same group decision. Across the two designs, more than $ 70\%$ of outcomes are identical under LD and MV, and more than $84\%$ under DD and MV. To evaluate the two rules' potential to improve over MV, we should compare instances when outcomes differ. But restricting the data sample to those cases only leaves us with too few data points.    

To overcome this difficulty, we use bootstrapping methods to simulate a large number of elections in a population for which our data are representative. By simulating many elections, conditioning on different outcomes becomes feasible.

The procedure we implement allows for correlation across an individual subject's
multiple decisions, and uses randomization to generate the correct
balance of experts and non-experts. For each voting system and group
size, we generate outcomes by drawing subjects, with replacement, and matching them
randomly into groups. To allow for correlation at the individual level, we draw each subject with their full set of decisions: 20 rounds in Design 1, and 40 rounds in Design 2. We then study the outcomes corresponding to
100,000 replications of the experiment for each treatment, using the
population of subjects for that voting rule and group size. Thus, in each of the two designs,
we analyze 100,000 replications of 240 decisions for LD1 and DD1,
and 120 decisions for LD3 and DD3. We describe the procedure in more
detail in Online Appendix Section \ref{subsec:online_app_bootstrapping}.

\begin{figure}[!htbp]
\begin{centering}
\includegraphics[scale=0.44]{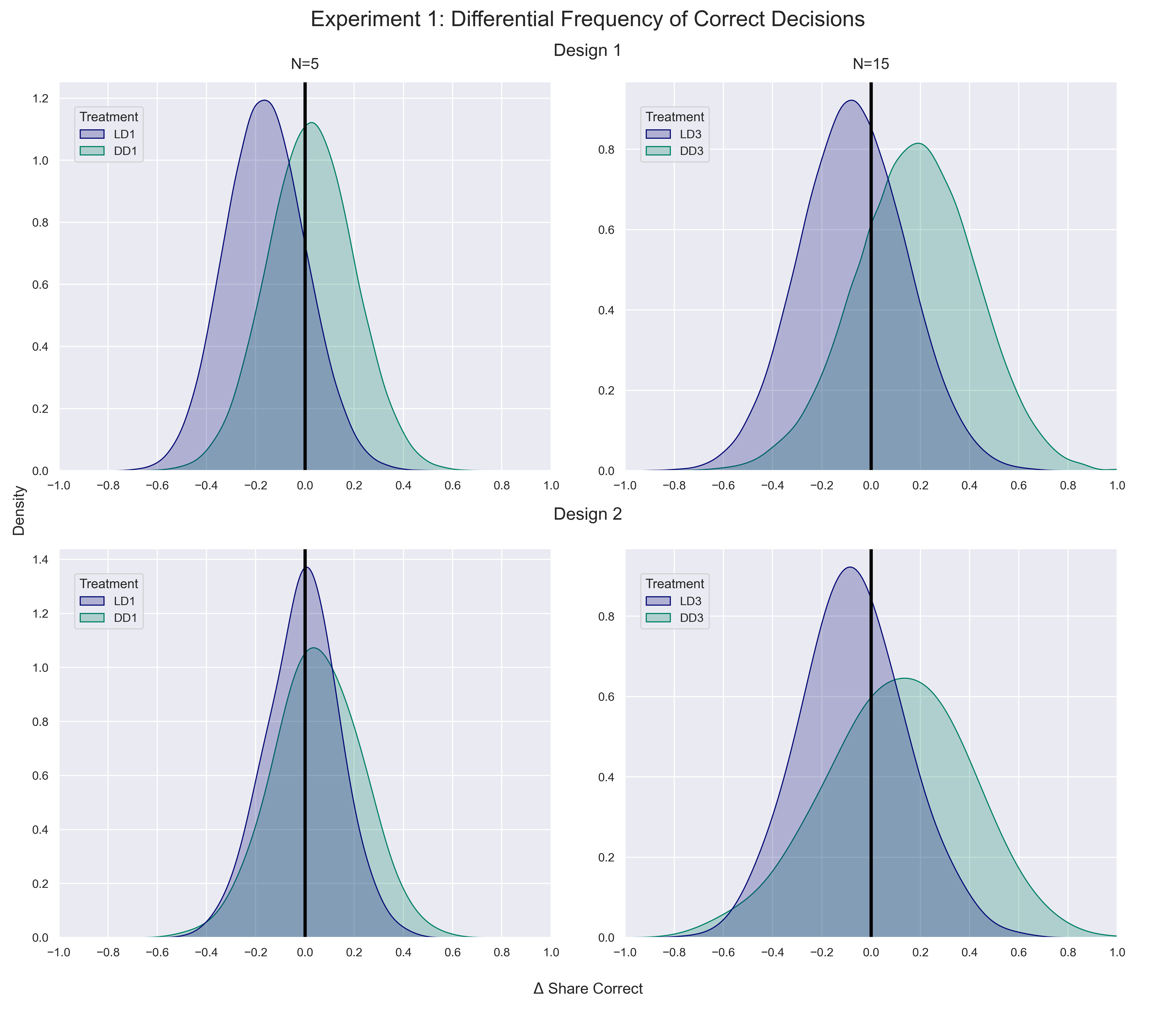}
\par\end{centering}
\caption{\textit{Differential frequency of correct decisions, relative to MV,
conditional on different outcomes}. Distributions over 100,000 bootstrap
replications of each of 240 elections under LD1 or DD1, and 120 elections under LD3 or DD3. \label{fig:Exp1.differential correct.sim}}
\end{figure}

Figure \ref{fig:Exp1.differential correct.sim}
shows the distributions of the differential frequency of correct decisions
between the voting systems we are studying and MV, for each group
size, conditioning on outcomes being different. The upper panels correspond to data from Design 1; the lower panels to data from Design 2. 

Consider for
example LD1. For each of the 100,000 simulations, we focus on the
subset of elections $E_{LD1}$ such that LD1 and MV reach a different
outcome. Call $\gamma_{LD1}(E_{LD1})$ $(\gamma_{MV}(E_{LD1}))$ the
frequency with which LD1 (MV) is correct over subset $E_{LD1}$, a
variable that ranges from 0 to 1. We are interested in $\gamma_{LD1}(E_{LD1})-\gamma_{MV}(E_{LD1})$,
where, by construction, $\gamma_{MV}(E_{LD1})=1-\gamma_{LD1}(E_{LD1})$.
Hence $\gamma_{LD1}(E_{LD1})-\gamma_{MV}(E_{LD1})=2\gamma_{LD1}(E_{LD1})-1$.
Our measure then ranges from 1---when, conditional on disagreement,
LD1 always reaches the correct outcome and MV the incorrect outcome---to
$-1$, when the opposite holds; a value of zero indicates that the
two rules are correct with equal frequency, conditioning on disagreement. The panels on the left of Figure \ref{fig:Exp1.differential correct.sim}
plot, in blue, the distribution of such variable over the 100,000 replications. The equivalent distribution for DD is plotted in the
same panels in green. The panels on the right report the results for groups
of size 15.\footnote{Averaging over all replications, the share of elections in which the
outcome differs from MV in Design 1 is 
23\% for LD1, 15\% for DD1, 20\% for LD3, and 15\% for DD3; in Design 2, it is 
28\% for LD1, 17\% for DD1, 24\% for LD3, and 14\% for DD3.
These differ from the analogous shares in the experimental data by 6pp or less in each case.}

Relative to the zero point indicated by the vertical black line, the blue distribution is shifted to the left, distinctly so for both LD1 and LD3 in Design 1 and for LD3 in Design 2; barely so for LD1 under Design 2. In all cases, when LD and MV differ, the correct decision is more likely to be the one reached by MV. The reverse is true for DD: the green distribution is shifted to the right in all cases, although most noticeably for DD3 in both designs. When DD and MV differ, the correct decision is more likely to be the one reached by DD. 
Averaging across both designs, the probability that MV is superior to LD, conditional on disagreement---i.e., the mass left of zero---is 70\% for N=5 and 67\% for N=15. The same measure yields the opposite conclusion for DD: conditional on disagreement, the probability that MV is superior to DD is 44\% for N=5 and 31\% for N=15, less than 50\% for both group sizes.\footnote{The distributions are largely similar across the two designs, with LD1 as the exception. In Design 1, 85\% of LD1's mass is left of zero, while in Design 2 the corresponding fraction is 55\%.} 

The distributions are also informative about the gap in the probability of being correct, relative to MV. For N=5, the mode occurs on average at $-8\%$ for LD but at $5\%$ for DD. This tells us that over the 100,000 replications of the 240 decisions, conditional on disagreement, we expect a most likely frequency of correct decisions of about $46\%$ for LD1, against $54\%$ for MV (or about 110 correct decisions under LD1, and 130 under MV), and about $52\%$ for DD1, against $48\%$ for MV (or about 125 correct decisions under DD1, and 115 under MV). This difference is larger for N=15, where the mode occurs on average at $-5\%$ for LD but at $14\%$ for DD.

Our first experiment, then, offers no evidence of 
LD's ability to improve decision-making over MV. Especially in the larger group, we see LD dominated by MV. DD, on the other hand, appears consistently comparable or superior to MV. By this measure, we see DD perform better than LD, even under a parametrization that should favor the latter rule ($N=5$).

Why does LD underperform, relative to the potential identified by the theory? Could these results be artifacts of the lab? We analyze this question in our second experiment. 

\section{Experiment 2: The Random Dot Kinematogram}

The laboratory designs we employed are fully canonical. And yet, could they be biasing individuals choices? 
There are three possible concerns. First, conveying information about one's own and the experts' precision in numerical values emphasizes the bilateral relationship between the subject and the expert, at the expense of equilibrium effects that reflect the behavior of the electorate as a whole.

Second, in our data, we observe a non-negligible fraction of votes against signal---10\% of all votes.\footnote{9.8\% in Design 1 (where subjects cast 3,752 votes when not delegating or abstaining) and 10.7\% in Design 2 (where a choice was collected for all 6,000 decisions). Those numbers are comparable to those found in similar experiments. Guarnaschelli et al.
 (2000), Goeree and Yariv (2011), and Mengel and Rivas (2017), for example, report frequencies of
 voting against signal of 6-10\% for juries voting under simple majority and pure common interest, in the absence of communication.} Could such votes too stem from the numerical format in which precision
is conveyed? Votes against signal occur mostly at low signal precision. Thinking that a signal barely more likely to be right than wrong should be disobeyed about half the time is an incorrect but plausible
thought. Outside the lab, on the other hand, individuals are unlikely to vote against their best estimate of the superior alternative.

Finally, by imposing that all signals are correct with probability
larger than 50\% our formulation could be biasing the analysis in
favor of MV. The assumption is natural in a model with binary choices
and known own precision, and is routinely made in experiments: conveying
information with less than random accuracy would require adding a
second level of noise, confusing participants. But is it a plausible
assumption? 

The ambiguity of a perceptual task gives us a simple entry
into a more realistic set-up.\footnote{In a still current analysis of the Condorcet Jury theorem, suggestively
titled ``A Note on Incompetence'', Margolis (1976) discussed the
tension between the asymptotic efficiency promised by the theorem
and political reality. What if, over some questions and for some voters,
information is actually correct with probability \emph{lower} than
$1/2$? As we learnt after conducting the present study, Margolis
went on to advocate understanding judgment, including judgment in
voting and political reasoning, through the lens of pattern recognition,
starting with perception biases (Margolis, 1987).} 

With these arguments in mind, we decided to run a second experiment, where the
information about the precision of the signals is ambiguous. The experiment
consists of a perceptual task---the Random Dot Kinematogram (RDK)---where
individual signals correspond to the accuracy of individual perceptions,
but neither own nor others' accuracies are described or known in precise
probabilistic terms. Because the task may be unfamiliar, we describe
it in some detail.\footnote{Additional information is in online Appendix Section \ref{subsec:online_app_rdk}, 
reproduced for convenience at the end of the current file. Experimental
instructions are online at \uline{\href{https://vmooers.github.io/assets/pdf/LD.ExperimentalInstructions.pdf}{Appendix C}}.}

We ran the experiment via Cloud Research, with prescreening of subjects
recruited from Amazon Mechanical Turk. We used three electorate sizes:
$N=5$ and $N=15$, as in Experiment 1, but also $N=125$, i.e.
a larger size than we could run in the lab or conveniently on Zoom.
In our implementation, 300 moving dots appear in each subject's screen
for one second only; a small fraction of them (dependent on treatment)
moves in a coherent horizontal direction, either Left or Right with
equal ex ante probability; the rest move randomly
and independently, and thus can each move in any direction over 360
degrees. The ``coherence'' of a task is the fraction of dots moving
in the coherent horizontal direction. After one second, the image
disappears and each participant reports whether the perceived coherent
direction was Left or Right. We describe the precise parameters in online Appendix Section \ref{subsec:online_app_rdk} (the size and color of the dots, the movements per
frame, the random process for the dots moving randomly, etc.), but
it should be clear that our experiment does not aim at measuring perception
per se---for example, we cannot control the ambient light, screen
size, or contrast of the monitors our subjects use. Our focus remains
on collective decision-making.\footnote{Heer and Bostock (2010) and Woods et al. (2015) report on the replication
successes and challenges of conducting research on perceptual stimuli
online.} 

We divide the experiment into two parts. Each part is preceded by
a few practice tasks and is subdivided into six blocks, each block
consisting of 20 tasks of equal coherence. Part 1 plays the role of
extended training: subjects are rewarded on the basis of their individual
accuracy only and experience decreasing rates of coherence across
blocks, reaching the same coherence used in Part 2 for the final two
blocks. At the end of Part 1, each subject is informed of her fraction
of correct answers in each block.

In Part 2, each task has both an individual component (``Choose the
coherent direction''), and a subsequent group decision with the possibility
of delegation (under LD), or abstention (under DD). (``You said
Left. Do you want to Vote or to Delegate (Abstain)?'').\footnote{Experiment 2 proceeds similarly to Design 2 in Experiment 1, with the difference that the preferred coherent direction (the preferred choice) is elicited before asking about delegation or abstention. In Experiment 1, the order of the questions is reversed because there is an external signal, and inserting some distance between signal and individual choice helps to avoid confusing the subjects.} When delegation
is chosen, the vote is assigned randomly to an ``expert'', that
is, one of the participants whose accuracy is in the top 20\% of the
group over the last 2 blocks (40 tasks). Experts are not allowed to
delegate (under LD) or to abstain (under DD). In line with
Experiment 1, groups of 5 have 1 expert, and groups of 15 have 3.
The group of 125 has 25 experts, and, following our standard notation,
we denote the two treatments with the larger group by LD25 and DD25.
The group decision corresponds to the majority of votes cast, and
individuals are rewarded both for their individual accuracy and for
the accuracy of the group. As in Part 1, feedback about average individual
accuracy in each block is provided at the end of Part 2.\footnote{Feedback over group accuracy cannot be provided because it depends
on choices made by others and is calculated ex post. Recall that participants
are online and come to the experiment at different times. Participants
were randomly divided into groups after completing the experiment,
and assigned to the same group for the whole experiment.} In Part 2, coherence is kept constant across all blocks. We chose
its value according to two main criteria: the task should not be so
difficult that subjects are discouraged and act randomly, and should
not be so easy that MV accuracy, especially in the large group, leaves
effectively no room for possible improvement. Based on the results
of two preliminary pilots, we fixed coherence in Part 2 at 5\% for
electorates of sizes 5 and 15, and at 3\% for the electorate of size
125. The task is objectively hard, as the reader can verify at the
following link: \url{https://blogs.cuit.columbia.edu/ac186/files/2022/05/rdk-video.gif}

The experiment used the RDK plugin in jsPsych (Rajananda et al., 2018)
and was hosted on cognition.run. For each of LD and DD, we recruited
60 subjects divided into 12 groups for the $N=5$ treatment and 90
subjects divided into 6 groups for $N=15$ (thus replicating the corresponding
number of subjects and groups in Experiment 1), and an additional
125 subjects for the largest group. There were then 275 subjects for
each of LD and DD, or 550 in total. The group size and the relevant
number of experts were always made public. The experiment lasted about
20 minutes. Subjects earned \$1 for participation and a bonus proportional
to the number of correct responses, for a total average compensation
of \$4.92, or just below \$15 an hour.

Relative to Experiment 1, Experiment 2 differs under several dimensions.
The most important and our core motivation for running the experiment is that choices are made under ambiguity, and neither one's own precision
nor the difference in precision between oneself and the experts are
known. In line with psychologists' use of the task, we view differences in accuracy as due to innate differences in perception.\footnote{Note that the exposure to the stimulus
is extremely brief---one second---and although a subject can withdraw
attention, above a low threshold it seems practically very difficult to increase accuracy through increased effort. As for the impact of effort on the delegation/abstention decision,
recall that subjects will have already expressed an incentivized individual choice. In addition,
here too, our focus is the comparison between LD and DD,
both of which allow for not voting.} Second, not only are probabilities not known, but neither we, the
experimenters, nor the subjects know others' beliefs. We cannot formulate
a tight theoretical prediction to which the data can be compared. Third, contrary to Experiment 1, subjects do not receive feed-back after each task. This is typical of perceptual experiments, where the very rapid flow of the tasks is designed to elicit innate perception skills, but also aligns for us with the mechanics of the online experiment.\footnote{Giving feedback on the accuracy of the group decision in Part 2 would require the simultaneous presence of multiple subjects and greatly complicate logistics and programming.} Finally, the subjects are drawn from pre-screened Amazon Mechanical Turk subjects and likely to be more diverse than the students recruited for Experiment 1.     

\section{Experiment 2: Results}

\subsection{Accuracy}

\begin{figure}[h]
\begin{centering}
\includegraphics[scale=0.5]{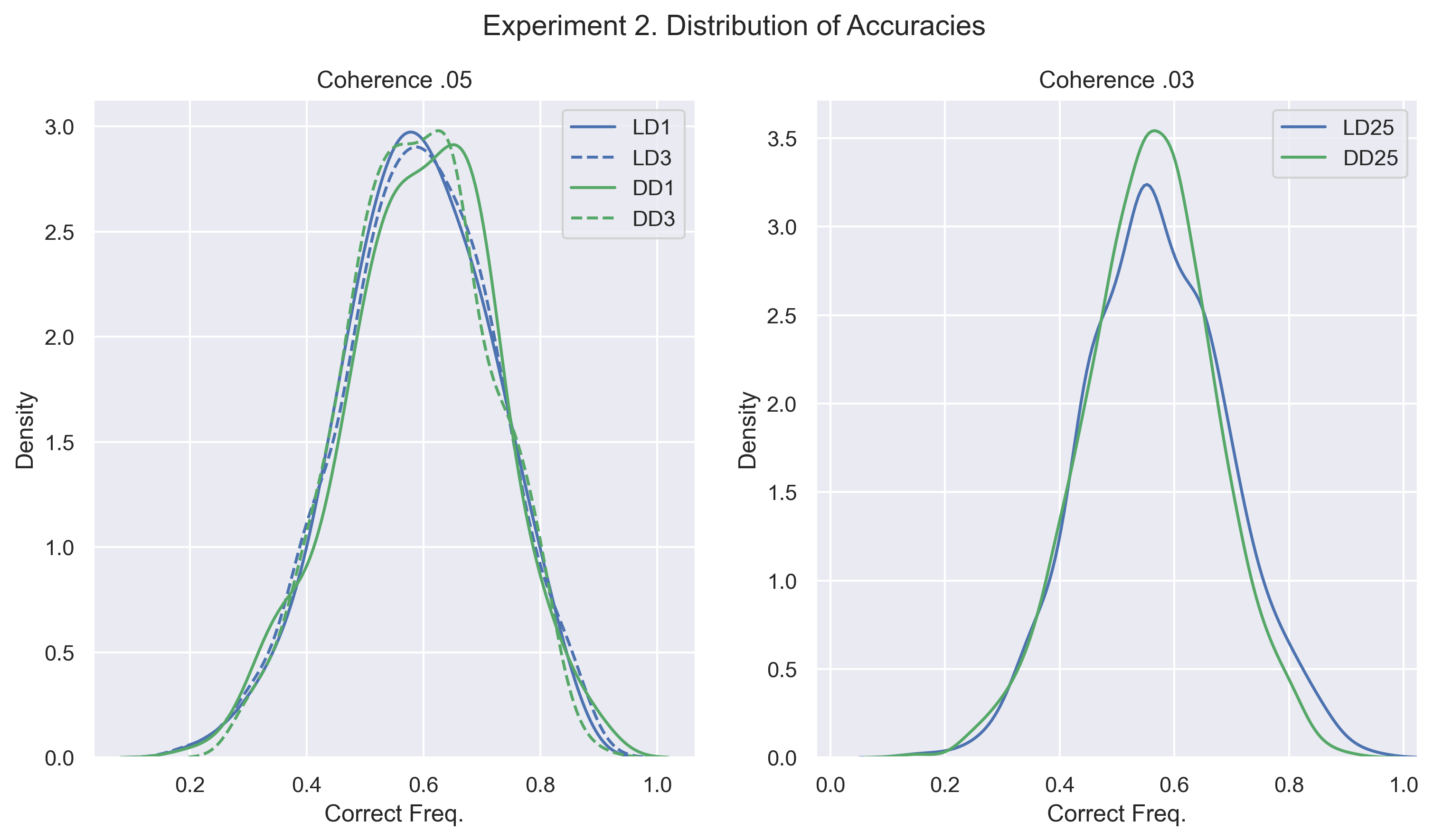}
\par\end{centering}
\centering{}\caption{\textit{Accuracies per block per subject. Distributions}. A block consists
of 20 tasks.\label{fig:Accuracies-per-bloc-per-subject} }
\end{figure}

We define an individual's accuracy as the fraction of correct responses.
Figure \ref{fig:Accuracies-per-bloc-per-subject} reports the distributions
of accuracy in Part 2 calculated over each of the 6 blocks for each
subject, that is, averaged over 20 tasks. The two panels correspond to the
two levels of coherence used in the experiment (0.05 on the left;
0.03 on the right).

For given coherence, the distributions are very similar across treatments.
In all cases, the spread in the distribution of accuracies is large,
ranging from about 25\% all the way to 95\%. Mean accuracy over all
participants is 59\% in treatments with 0.05 coherence, and 56\% in
treatments with 0.03 coherence. 

Experts' accuracy is higher than non-experts': experts' average accuracy over all blocks in which they are labeled as experts is 63\% (v/s 58\% for non-experts)
in treatments with 0.05 coherence, and 59\% (v/s 55.5\%) in treatments
with 0.03 coherence.\footnote{In regressions of individual accuracy, the coefficient of the expert
dummy is positive and strongly significant in all cases. Across all treatments, the share of individual blocks in which experts have a higher accuracy than non-experts is almost 75\%.} 

The frequency of blocks with accuracy below 50\% is non-negligible
(18\% for coherence of 0.05, just above 23\% for coherence of 0.03)
and, surprisingly, persists when we aggregate over a larger number
of tasks. Averaging at the subject level over all 120 tasks, 9\% of
subjects have accuracy below randomness with coherence 0.05, and 12\%
with coherence 0.03.\footnote{Individual subjects' accuracies show high variability across blocks,
evidence of random noise in perceiving and recording the stimulus
in the brain, as formalized in psychophysics research. } If we want to study voting and information aggregation when information
may be faulty, perceptual tasks can provide a very useful tool. 

\subsection{Frequency of delegation and abstention}

\begin{figure}[h]
\begin{centering}
\includegraphics[scale=0.45]{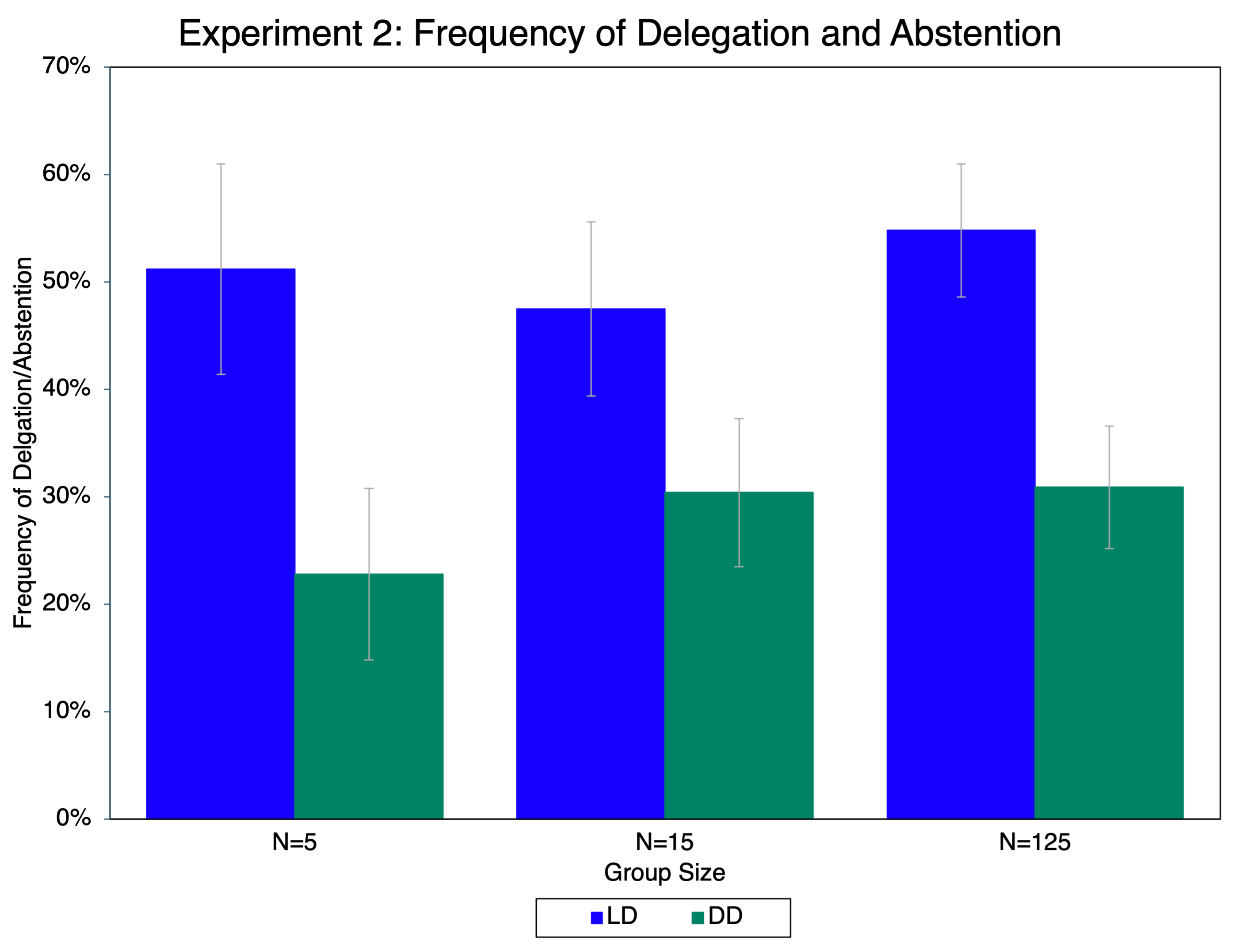}
\par\end{centering}
\caption{\textit{Aggregate frequency of delegation and abstention (non-experts).
}\textit{\emph{95\% confidence intervals calculated from standard
errors clustered at the individual level. \label{fig:Exp2-Freq-Del-Abs}}}\textit{ }}
\end{figure}

Absent information on subjects' beliefs, we do
not have theoretical predictions for LD and DD. We
can however compare the data from the two rules, under the plausible assumption,
supported by Figure \ref{fig:Accuracies-per-bloc-per-subject}, that
accuracies and beliefs about accuracies are comparable across the
LD and DD samples. Figure \ref{fig:Exp2-Freq-Del-Abs} plots the
frequencies of delegation and abstention for each group size.\footnote{In line with experts' obligation to vote and for possible comparison to Experiment 1, the figure reports the frequencies for non-experts only. It remains almost identical if frequencies are calculated over
the full sample.} The 95\% confidence intervals are calculated from standard errors
clustered at the individual level.

If we cannot talk rigorously of overdelegation, we can nevertheless see
that in Experiment 2, delegation remains much more common than abstention,
for all group sizes. In groups of 5, where the disparity is largest,
delegation is more than twice as frequent; in groups of 15, where
we see the least disparity, delegation is still 60\% more common.
The decline in coherence, from $N=5$ or $15$ to $N=125$, has small
effects on the data. 

The higher frequency of delegation is confirmed in the regressions
reported in Table \ref{Table:Exp2.Del.or.Abs.LPM}. The unit of analysis
is the block at the individual subject level (hence 6 blocks per subject),
with data grouped by coherence level. The regressions reported below
confirm the results of the figure: in all treatments, delegation is
significantly more frequent than abstention. In Experiment 2, accuracy
is at best a very weak predictor of participation in voting, never
significant at conventional levels, confirming the high uncertainty
in subjects' evaluation of their own accuracy.\footnote{Corresponding probit regressions are reported in Appendix Table \ref{Table:Exp2.Del.or.Abs.Probit}.}

\begin{table}[ph]
\begin{centering}
\begin{tabular}{>{\raggedright}p{3.1cm}>{\centering}p{3.1cm}>{\centering}p{3.1cm}}
\multicolumn{3}{c}{Experiment 2: Frequency of Delegation or Abstention.}\tabularnewline
\midrule 
 & (1) & (2)\tabularnewline
 & N=5 \& N=15 & N=125\tabularnewline
\midrule 
Accuracy & -0.122 & 0.004\tabularnewline
 & (0.084) & (0.102)\tabularnewline
 & {[}0.146{]} & {[}0.970{]}\tabularnewline
 &  & \tabularnewline
LD & 0.226{*}{*}{*} & 0.223{*}{*}{*}\tabularnewline
 & (0.038) & (0.042)\tabularnewline
 & {[}0.000{]} & {[}0.000{]}\tabularnewline
 &  & \tabularnewline
N=15 & 0.004 & \tabularnewline
 & (0.038) & \tabularnewline
 & {[}0.908{]} & \tabularnewline
 &  & \tabularnewline
Keys: {[}E{]}{[}Y{]} & -0.009 & 0.029\tabularnewline
 & (0.038) & (0.041)\tabularnewline
 & {[}0.816{]} & {[}0.484{]}\tabularnewline
 &  & \tabularnewline
Block & 0.001 & 0.010\tabularnewline
 & (0.012) & (0.013)\tabularnewline
 & {[}0.965{]} & {[}0.459{]}\tabularnewline
 &  & \tabularnewline
Constant & 0.343{*}{*}{*} & 0.295{*}{*}{*}\tabularnewline
 & (0.066) & (0.067)\tabularnewline
 & {[}0.000{]} & {[}0.000{]}\tabularnewline
 &  & \tabularnewline
\midrule 
Observations & 1,800 & 1,500\tabularnewline
R-squared & 0.100 & 0.097\tabularnewline
\midrule
\multicolumn{3}{l}{{*}{*}{*} p\textless 0.01, {*}{*} p\textless 0.05, {*} p\textless 0.1}\tabularnewline
\end{tabular}
\par\end{centering}
\centering{}\caption{\textit{Frequency of delegation or abstention}. Linear probability
models. Standard errors are clustered at the individual level. P-values
in brackets. Delegation/abstention is measured as the share of rounds
in a given block in which a subject chose to delegate/abstain (with
a range from 0 to 1). Accuracy is the share of rounds in the block
that subject answered correctly. Subjects randomly use either keys
{[}V{]} and {[}N{]} or {[}E{]} and {[}Y{]} to choose whether to vote or not;
a dummy for being assigned {[}E{]}{[}Y{]} is included. The values
for block have been scaled to be between 0 and 1; the coefficient for
\textquotedblleft block\textquotedblright{} thus indicates the effect
of going from the first to last block. In both columns, DD (abstention) is the excluded case.  \label{Table:Exp2.Del.or.Abs.LPM}}
\end{table}

\subsection{Frequency of correct outcomes}

\begin{figure}[h]
\begin{centering}
\includegraphics[scale=0.15]{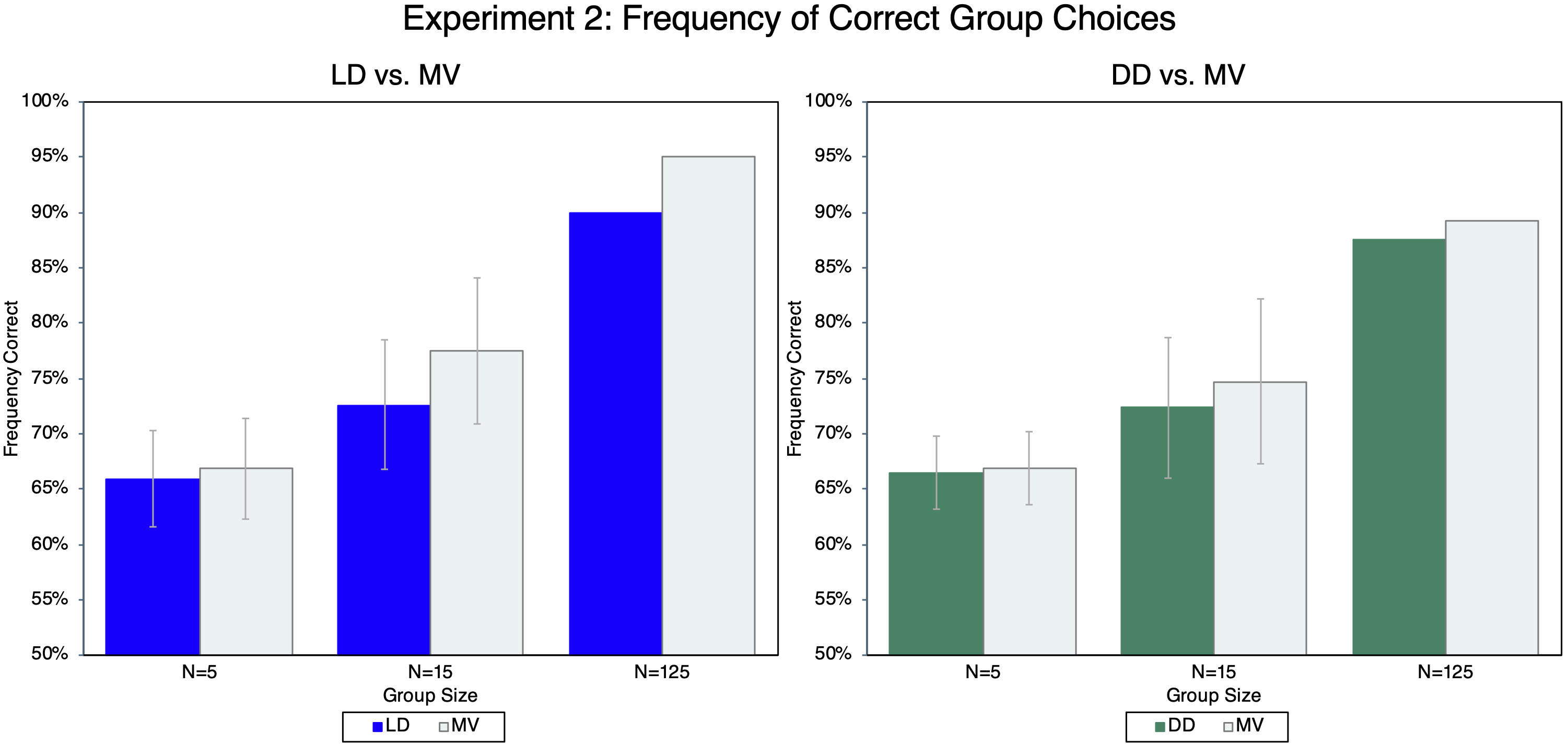}
\par\end{centering}
\begin{centering}
\caption{\emph{Frequency of correct outcomes. }95\% confidence intervals are
calculated from standard errors clustered at the group level. There
is no confidence interval for $N=125$ because there is a single group.\label{fig:Exp2.CorrectOutcomes}\emph{ }}
\par\end{centering}
\end{figure}

How well do LD and DD do in Experiment 2, relative to each other and to MV? Figure \ref{fig:Exp2.CorrectOutcomes}
reports the frequency of correct group decisions, aggregated over
all groups and tasks for given treatment. 

As expected, for all three rules, the fraction of correct decisions
increases with the size of the group, ranging from about 65\% at $N=5$
to 90--95\% at $N=125$. The aggregation of independent signals remains
very powerful, even in the presence of weak accuracies, and MV consistently outperforms the other two rules.  Relative to
MV, DD performs somewhat better than LD.

As in the case of Experiment 1, the relative performance of LD and
DD can be evaluated more precisely only when conditioned on disagreement
with MV.\footnote{The fraction of outcomes that coincide with MV outcomes ranges from
a minimum of 75\% under LD1 to a maximum of 92\% under DD25. } We again bootstrap the data and replicate Part 2 of the experiment 100,000 times, generating a large sample of decisions over which the voting results differ, as we did with the data from Experiment 1.\footnote{For each group size, we replicate Part 2 of the experiment 100,000 times, and for each replication we select those decisions over which the rule we consider (LD or DD) differs from MV and calculate the differential fraction of correct decisions, relative to MV. Each group faces 120 collective decisions (6 blocks of 20 tasks each); there are 12 groups when $N=5$, or a total of 1,440 decisions; 6 groups when $N=15$, or 720 decisions, and a single group completing 120 decisions when $N=125$. Thus each of the 100,000 data points of the bootstrapping exercise reports the differential frequency of correct decisions over 1,440 elections when $N=5$, 720 decisions when $N=15$, and 120 decisions when $N=125$.} 

\begin{figure}[h]
\begin{centering}
\includegraphics[scale=0.35]{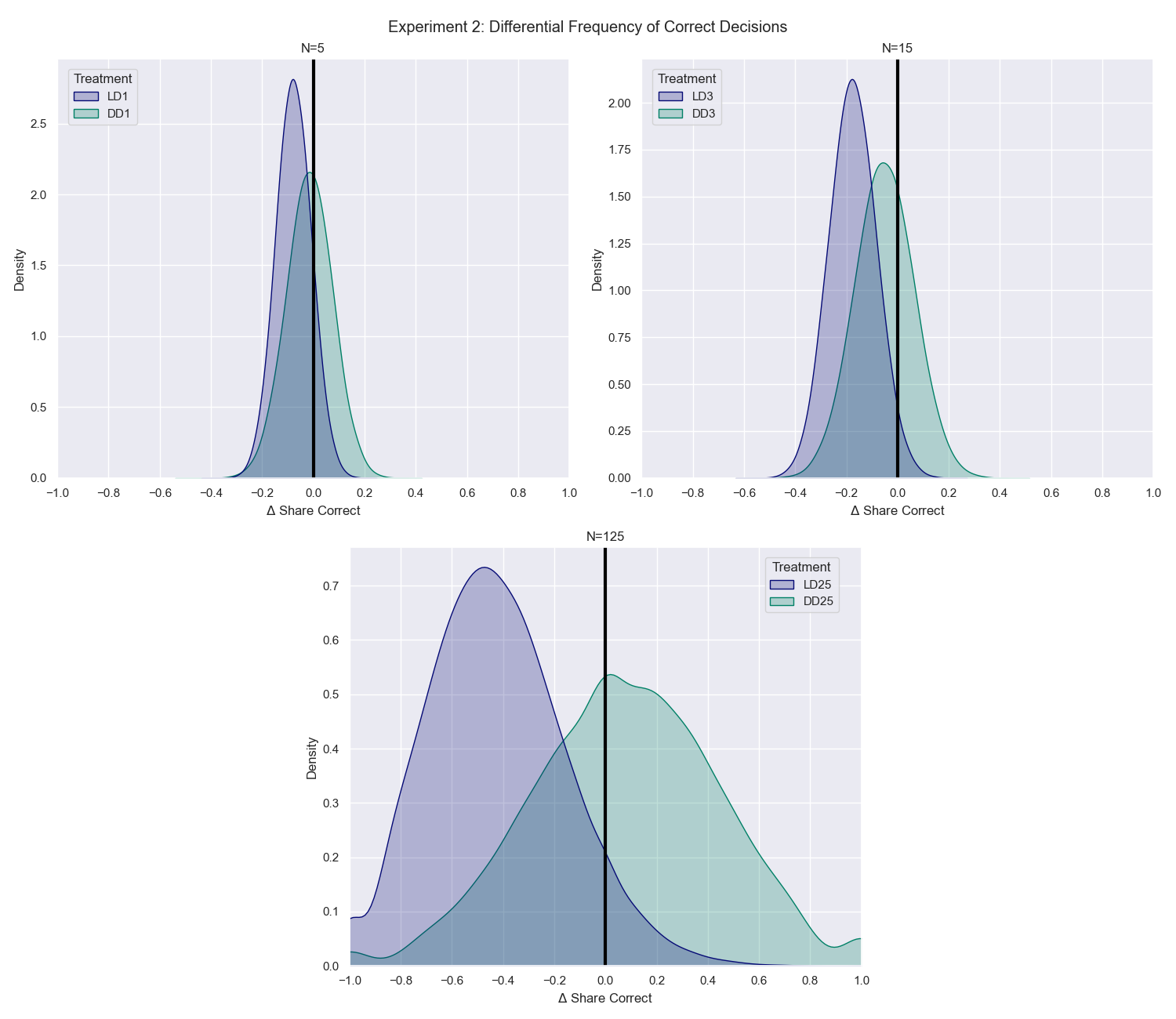}
\par\end{centering}
\caption{\textit{Differential frequency of correct decisions, relative to MV,
conditional on different outcomes}. Distributions over 100,000 bootstrap
replications. \label{fig:Exp2.differential.correct} }
\end{figure}

Figure \ref{fig:Exp2.differential.correct} reports the results of the
100,000 simulations, with three panels corresponding, in order, to
$N=5$$,$ $N=15$, and the single large group at $N=125$. As in
Figure \ref{fig:Exp1.differential correct.sim}, we plot the distributions
of the differential frequency of correct decisions under LD (in blue)
or DD (in green), relative to MV. Recall that, if the distribution
is skewed to the left of the vertical line at zero, then conditional
on disagreement, the correct decision is more likely to be the one
reached by MV; and vice-versa if the distribution is skewed to the
right. 

In all three panels, the blue mass is shifted to the left: in Experiment
2, according to these simulation results, LD again underperforms. Conditional
on disagreement, the share of simulated experiments in which MV is
more likely than LD to yield the correct outcome is 87\% for LD1,
97\% for LD3, and 95\% for LD25. DD (green in the figure) fares better:
the corresponding numbers are 58\% for DD1, 69\% for DD3, and 48\%
for DD25, when DD is just barely more likely to be correct than MV. 

For $N=5$ and $N=15$, the distributions are much more concentrated than in Figure \ref{fig:Exp1.differential correct.sim} because the number of decisions in each of the 100,000 replications is much higher. It still remains true that the shapes of the distributions tell us the frequencies with which
the two systems are correct, relative to MV. When $N=5$ for example, the mode of the blue distribution is again at $-8\%$, implying that over the 100,000 simulations
the most likely result is a frequency of correct decisions of $46\%$ for LD1 versus $54\%$ for MV. On the other hand, the mode of the green distribution is at $-1\%$, such that the most likely result is a frequency of correct decisions of $49.5\%$ for DD1 versus $50.5\%$ for MV. 


In perceptual research, the focus is the immediate, apparently unconscious
reaction to the stimulus, and the experimental design minimizes interference
with such a process. With this in mind, we chose not to ask participants
for their beliefs within each task. We did, however, add two summary
questions at the end of the experiment. We asked: ``On average, what
percentage of trials in the second part do you think you got right?''
and ``On average, what percentage of trials in the second part do
you think the experts got right?'' With these two questions only,
barely incentivized,\footnote{We rewarded replies with 25 cents if the answer was within 5\% of
the observed percentage. For the second question, ``the experts'' referred to the group the participant was assigned
to.} our information on beliefs can only be very noisy. Yet, we learnt
three lessons. First, in all treatments, beliefs about own accuracy
track actual accuracy surprisingly well: for treatments with 0.05
coherence, average believed accuracy is 58\% (v/s 59\% for average
realized accuracy); with coherence 0.03, average believed accuracy
is 55\% (v/s 56\% for average realized accuracy). Second, beliefs
about experts\textquoteright{} accuracy are inflated by about 15\%
at both coherence levels (71\% v/s 63\% at 0.05 coherence, and 70\%
v/s 59\% at 0.03 coherence). Third, while, as we saw, accuracy per se is
a very weak predictor of choosing to vote, beliefs about own accuracy relative to the experts are instead strong predictors of voting behavior.
Beliefs, however, cannot explain the difference in the frequency of
delegation and abstention: controlling for beliefs, delegation remains
substantially and significantly higher than abstention.\footnote{See the regressions in Appendix Table \ref{tab:Beliefs-and-the} in the online Appendix, where we also plot distributions
of errors in beliefs in Appendix Figure \ref{fig:Difference-between-actual}. Beliefs about experts track well average accuracy in the top quintile of the \textit{current} block, but conditioning the status of expert on ex post accuracy while voting occurs before the state of the world is revealed is logically inconsistent.}

\section{Discussion}

We conclude the discussion of our results with two questions. Both
are introduced here as possible guidelines for future research. The
first question asks the reasons for the consistently
high frequency of delegation, in absolute terms and relative to abstention. The second question addresses the maintained
assumptions of our model. 


\subsection{Why is delegation so frequent, and more frequent than abstention? }\label{subsec:disc-del-vs-abs}

Why were participants, in both experiments, drawn to delegation more than to abstention?
One plausible answer is that abstention, but not delegation,
suffers from moral stigma: not taking a position can be seen as cowardly,
as opposed to actively delegating to a more knowledgeable expert.
Note however two arguments to the contrary. First, the explanation
suggests too little abstention, as opposed to our finding of excessive
delegation. Second, in our experiments the moral argument is weak:
without either costs of voting or costs of information acquisition,
abstention does not deliver any private gain. Nor is voting with weak
information virtuous: it worsens everybody's prospects.\footnote{In an environment similar to ours, Morton and Tyran (2011) find excessive
abstention. Mengel and Rivas (2017) introduce asymmetric priors and
in their complex environment find too little abstention, but also repeated
evidence of random behavior. } 

A different conjecture focuses on low confidence in other non-expert
voters. If a participant believes that too many poorly informed non-experts
are voting, she will wish to increase the power of the experts and
delegate, but, on the contrary, will refrain from abstaining. The
logic, whose validity we verified theoretically under our experimental parameters,
is intuitively plausible and suggests an interesting contrast between
delegation and abstention. It is challenged, however, by the feed-back
participants receive in Experiment 1: after voting, each participant
is informed of how many other votes are delegated under LD (and how
many are cast under DD). Even lacking information about other non-experts'
precisions, if voting by others is a concern, such feedback should
be salient, and the misperception that too many others are voting
should be corrected: the probability of delegating/abstaining should
respond to the number of other subjects delegating/abstaining in the
previous round. As we show in the online Appendix in Section \ref{sec:online_app_discussion} and Appendix Tables \ref{tab:response_to_prev_round1} and \ref{tab:response_to_prev_round2}, we find very little evidence of this.\footnote{With the sole exception of DD1 in Design 2, we find no significant correlation between the probability of delegating/abstaining and the number of other subjects delegating/abstaining in the previous round. In DD1, Design 2, the abstention rate is marginally significantly increasing in the number of abstainers in the previous round.}

Our favorite hypothesis is that understanding the cost of excessive vote concentration under delegation is hard, while no such cost accompanies abstention. We propose two possible mistakes. The first, a form of ``correlation neglect'', does not fully account for the loss of signal variety that comes with delegation. It is not unreasonable for delegation decisions to respond primarily to differences in known or believed precision of information relative to the experts, resulting in excessive concentration in voting power---this is what we can state rigorously in Experiment 1, relative to LD's best equilibrium. Abstention makes such concentration less likely because, conditional on voting, all voters cast the same number of votes.\footnote{Consider the following thought experiment. Abstracting from strategic considerations, a single individual is asked to allocate a number of votes among computer-experts with different, known precisions. (Computer-experts always vote according to their signal.) Will the individual concentrate the votes inefficiently in the hands of the higher precision experts only? If instead the individual is asked to choose which of the experts would vote and which ones would abstain, will voting power be more diffuse?}  

A second possible mistake arises when the individual voter fails to factor in the actions of the other voters. We denote it by ``equilibrium neglect'': we propose that under LD, each individual voter focuses on the pairwise relationship between herself and the expert, neglecting the other voters. Under DD, the likelihood of such a distortion is reduced because abstention automatically invokes the behavior of others: ``If I abstain, who votes?'' 

We can detect some evidence in favor of this conjecture in our data. If an experimental subject focuses on her own delegation to the expert only, then delegating is a plausible choice at any $q_i<p$, including in the near neighborhood of $p$.\footnote{Indeed, it is optimal for all $q_i<p$ if the number of experts is odd.} On the other hand, under DD, abstention at high $q_i$ is less compelling. As we show in Appendix Tables \ref{tab:high_precision_reg_n5} and \ref{tab:high_precision_reg_n15}, conditioning on $q_i\geq 0.65$, delegation rates at high precisions in Experiment 1 amount to over 37\% in LD1, v/s 9\% in DD1, and to 27\% in LD3 v/s 13\% in DD3 (averaging over the two designs). All differences are highly significant and robust to controlling for signal precision, round, and, in Design 1, treatment order. Recall that while an equilibrium with full delegation exists for LD1, such equilibrium also exists for DD1, and, importantly does not exist for LD3 (while it exists for DD3).    
The Computer Science literature in particular has long been concerned about the possibility of excessive concentration of votes under LD. Our equilibrium analysis shows that fully rational voters can avoid it by keeping delegation rates low, but the experimental results appear to support the literature's caveats. We offer the two conjectures above as grounds for future inquiry.

\subsection{Are the advantages of LD undermined by the model's simplifying assumptions?}

Is LD handicapped, relative to DD, by the model's maintained assumptions?

We discuss
here very briefly three possible extensions: (1) introducing private
values; (2) making information costly; (3) allowing for correlation
in signals. In online Appendix Section \ref{sec:three_simple_models}, we study the three extensions in a simple example: a 3-voter environment, with a single expert with superior information and two equally informed non-experts. None of the extensions unambiguously favors either LD or DD. The reasons provide some useful insights. 

Consider first introducing private values. Suppose that voters can be of two types, favoring either matching or mis-matching the decision to the state. The expert's type is known; the non-experts' are not. Delegation to the expert can then be valuable not only because of the expert's superior information, but also because of the alignment of private interests. But note that by the same token a voter who is misaligned with the expert prefers DD: under DD, aligned voters' incentive is to abstain and non-aligned voters retain some decision power. In our simple example, ex ante utility, before the realization of types, is identical under LD and DD. The example is special but the intuition is valuable: private values need not favor LD because abstention under DD is a supple means of giving some weight to both aligned and misaligned preferences.\footnote{Note that
multiple experts would not solve the problem: with possibly different private preferences, experts will be more likely to disagree, reducing the value of their expertise.}  

Similar considerations are relevant if we return to a pure common interest environment and relax a different assumption. Suppose each non-expert can independently invest, at some cost, in information as precise as the expert's.\footnote{The impact of different voting procedures on the acquisition of costly
information has been studied in a number of influential papers. Among
others, see Persico (2004) and Martinelli (2006) for theory; Grosser
and Sebauer (2016), Elbittar et al. (2020), Bhattacharya et al. (2017),
and Mechtenberg and Tyran (2019), for experiments. To our knowledge,
none has studied delegation, either alone or in comparison to abstention.} Delegation to the expert  makes it possible to achieve, vicariously, more precise information while saving on information costs. Free-riding on the expert, however, also means reduced aggregate information in the system, relative to DD. There
are savings in information costs, but more faulty decisions. Examples
show that the latter can dominate the former, leading to lower ex
ante utility under LD.

We can argue along similar lines about the effect of introducing correlation
in the signals of non-expert voters: suppose such signals are fully correlated with some known probability (while the expert's signal is conditionally independent). Does LD then outperform DD? The answer remains ambiguous. We find that both DD and LD can support equilibria with the same probability of reaching a correct outcome and that the best equilibria under both rules coincide. In this case, too, we
have no ground to believe that the model's original assumptions are stacked against LD.

\section{Conclusions}

Liquid Democracy is a computer-mediated voting system where all
decisions are subject to popular referendum but voters can delegate
their votes freely. The option of delegation to better informed experts
seems intuitively valuable, and indeed, theory shows that if experts
are correctly identified, in finite electorates delegation can always improve the chances of reaching
the correct decision. However, delegation must be used sparingly because,
by reducing the number of independent voices, it also reduces the
aggregate amount of information expressed by the electorate. This
paper reports the results of two different experiments that measure
participants' propensity to delegate their vote and compare the decisions
of the group when delegation is possible, when abstention is possible,
as in Direct Democracy with abstention, and under majority voting by all. In line with Condorcet's message,
we find that in both experiments majority voting by all leads to
the highest frequency of correct outcomes, even in small groups; abstention
is closely comparable, while delegation appears inferior to both, even in
our simplified world where experts are indeed better informed. 

Together with the weaker performance of Liquid Democracy, in our experiments we see participants delegating with very high frequency, much more frequently than they abstain and two to three times more frequently than optimal in the experiment for which we have precise theoretical predictions. The high rate of delegation makes the quality of decision-making highly dependent on the accuracy of the experts and vulnerable to small downward noise. Abstention, in practice a form of delegation to all those who cast
their votes, is a more robust system, exactly because it aggregates
the information of more voters.

Our results are obtained in experimental environments where experts
are identified exogenously and are objectively better informed than non-experts. And yet, even in this favorable environment, our tentative
conclusion is that, on informational grounds alone, the arguments
in favor of Liquid Democracy should be considered with caution.



The paper also makes a methodological contribution. We match a canonical,
fully controlled lab experiment on voting with a perceptual experiment
where information is ambiguous: participants do not know the probabilities
with which either their own or the experts' information is correct.
We use such a design because there are plausible concerns that the
precise mathematical framing of the canonical experiment may affect
the results. In addition, the ambiguity of the information in the
perceptual experiment seems closer to realistic conditions of voting
on political decisions. We find that the results of the first experiment
replicate closely in the second, and we consider the robustness of the
conclusions a central contribution of our study. Beyond the specific
results, one of our goals is to stress that perceptual experiments
are a very useful tool for the study of group decision-making under
ambiguity, especially in conjunction with the precise and explicit
information structure of traditional lab experiments. Social psychologists
have been studying them for decades; economists interested
in social choice should add them to their tool box as well.

\pagebreak

\pagebreak

\appendix

\section{Appendix}

\setcounter{figure}{0} \renewcommand{\thefigure}{A.\arabic{figure}}
\setcounter{table}{0} \renewcommand{\thetable}{A.\arabic{table}}

\subsection{Theoretical results}

\subsubsection{The proof of Proposition 1}\label{subsec:prop1_proof}

\begin{proposition}
\textbf{Equilibria.} \textit{For all $N=K+M$ finite, with $M>1$, and for all $F(q)$ everywhere continuous over $[1/2,p]$, for both
$R=LD$ and $R=DD$, in all semi-symmetric equilibria: (i) All experts vote; (ii) Non-experts' strategies
must be monotonic: there exists equilibrium $\widetilde{q}_{R}$ such that any individual
$i$ with $q_{i}>\widetilde{q}_{R}$ votes and any $i$
with $q_{i}<\widetilde{q}_{R}$ delegates/abstains.}  
\end{proposition}

\begin{proof}
Note that with all votes cast according to signal, strategic choices are limited to voting or delegating/abstaining. We study the two rules in order. 

I. Consider first the model under DD, where abstention is possible. With all votes cast according to signal, strategic choices are limited
to voting or abstaining. We begin by proving (i): all experts vote. We divide the proof in three steps. 

(1) Suppose first that all non-experts abstain (i.e. $\widetilde{q}_{DD}\geq p)$.
We are focussing on semi-symmetric equilibria; hence either all experts abstain, or all vote, or all mix between voting and abstaining.
(a) If no other expert votes, expert $i$ is pivotal, and with $p>1/2$ strictly prefers voting to a coin toss. (b) Suppose then that all other experts vote.
If $K-1$ is even, $i$ is pivotal when the other experts' votes are split into two equal sides. But with all experts having equal precision, the side that $i$'s vote would make the majority is more likely to be correct. Hence $i$ strictly prefers voting to abstaining and leaving the decision to a coin toss. If $K-1$ is odd, $i$'s vote is pivotal when it induces a tie, and thus a coin-toss choice. Hence the two sides are equally likely to be correct and $i$ is indifferent between voting or abstaining. (c) Finally, suppose then that all other
experts vote with some probability $s$. If, ignoring $i$, the realized number  of expert votes is even, $i$ prefers to vote; if it is odd, $i$ is indifferent. Hence if all other experts randomize, $i$ strictly prefers to vote. Hence there cannot be a symmetric equilibrium mixed strategy. If all non-experts abstain, the only symmetric equilibrium strategy
is for all experts to vote.\footnote{Note that even if $K$ is even (and thus $K-1$ is odd), all experts
abstaining cannot be an equilibrium (because voting is then strictly profitable) and neither can randomizing (by the argument above).} 

(2) Suppose now that all non-experts vote (i.e. $\widetilde{q}_{DD}\leq\underline{q})$.
Suppose first that all other experts vote as well. Expert $i$'s vote is pivotal if it either breaks (if $N-1$ is even) or causes (if $N-1$ is odd) a tie. In either case, each of the two opposite sides can
consist of different numbers of experts and non-experts, but, by the symmetric structure of our environment and the independent signal draws, the average precision of the two opposite sides is expected to be equal, and thus equal to the average precision of all individuals who are voting. Note that it must be inferior to the precision of the expert,
$p$. It then follows that, conditional on expert $i$ being pivotal, the side favored by $i$ is strictly more likely to be correct, and thus, whether $N-1$ is even or odd, it is strictly better for $i$ to vote. In fact, the conclusion does not depend on all other experts voting---it continues to hold if some or all other experts abstain. In all cases, the expected average precision of the two opposite sides,
absent $i$'s vote, is equal and inferior to $p$. Hence it is strictly better for $i$ to vote. If all non-experts vote, the only symmetric equilibrium strategy is for all experts to vote.

(3) Finally suppose that non-experts' optimal voting threshold is interior: $\widetilde{q}_{DD}\in(\underline{q},p)$. Higher
precision non-experts vote, while lower-precision non-experts abstain. The average precision of the voting non-experts depends on $\widetilde{q}_{DD}$,
but the argument above continues to hold identically. Again, the only symmetric equilibrium strategy is for all experts to vote. 

We can now prove claim (ii), again for model DD: In all semi-symmetric equilibria, non-experts' strategies must be monotonic threshold strategies. 

We denote by $EU_{i}^{DD}(\sigma_{i}^{DD},\sigma_{-i}^{DD})$
$i'$s interim expected utility from strategy $\sigma_{i}^{DD}$ when other voters' strategies are $\sigma_{-i}^{DD}$ and
$\sigma\in\{a,v\}$, where $v$ stands for casting one's vote, and
$a$ for abstaining. Given $q_{i}$ symmetric across
states and strategies symmetric relative to the content
of the signals, $EU_{i}^{DD}(a,\sigma_{-i}^{DD})$ does not depend on
$q_{i}$ because $i$ is not voting. On the other hand, $EU_{i}^{DD}(v_{i},\sigma_{-i}^{DD};q_{i})$,
$i$'s interim expected utility from voting, does depend on $q_{i}:$

\begin{align*}
EU_{i}^{DD}(v_{i},\sigma_{-i}^{DD};q_{i}) & =q_{i}Pr(d_{k}\text{ wins }|\omega=\omega_{k},i\text{ votes for }d_{k},\sigma_{-i}^{DD})+\\
 & +(1-q_{i})Pr(d_{k'}\text{ wins }|\omega=\omega_{k'},i\text{ votes for }d_{k},\sigma_{-i}^{DD})
\end{align*}
for $k'\neq k.$ From the symmetry of the prior, information structures,
and others' strategies, this expression is the same for $k=1,2$ (i.e.
for $i$ receiving either signal realization). The same symmetry conditions
imply:
\[
Pr(d_{k}\text{ wins }|\omega=\omega_{k},i\text{ votes for }d_{k},\sigma_{-i}^{DD})>Pr(d_{k'}\text{ wins }|\omega=\omega_{k'},i\text{ votes for }d_{k},\sigma_{-i}^{DD})
\]
where the inequality holds strictly because there is always a positive
probability that $i$ is pivotal. The strict inequality implies that
$EU_{i}^{DD}(v_{i},\sigma_{-i}^{DD};q_{i})$ is strictly increasing
in $q_{i}$. Hence non-expert $i$ must vote for all $q_{i}$ such
that $EU_{i}^{DD}(v_{i},\sigma_{-i}^{DD};q_{i})>EU_{i}^{DD}(a_{i},\sigma_{-i}^{DD})$,
and abstain for all $q_{i}$ such that $EU_{i}^{DD}(v_{i},\sigma_{-i}^{DD};q_{i})<EU_{i}^{DD}(a_{i},\sigma_{-i}^{DD})$. Defining $\widetilde{q}_{DD}$ such that:

\[
EU_{i}^{DD}(a,\sigma_{-i}^{*DD})=EU_{i}^{DD}(v_{i},\sigma_{-i}^{*DD};\widetilde{q}_{DD}),
\]
where $\sigma^{*DD}$ denotes equilibrium strategies
under DD, the Lemma then follows. Note that the proof does not depend on result (i), i.e. on experts never abstaining. For any sincere strategy by the experts, a non-expert who votes has a strictly positive probability of being pivotal, and thus an incentive to vote that is strictly monotonic in $q_i$. Result (ii) then follows. 
\medskip

II. Consider now the model under LD, where delegation is possible. Notice first that the proof of result (ii) remains identical under LD. Under LD, define strategy $\sigma$ as $\sigma \in \{d, v \}$ where $d$ stands for delegating. As in the case of abstaining under DD, independently of experts' strategies, $EU_{i}^{LD}(d_{i},\sigma_{-i}^{LD})$ does not depend on $q_i$, while $EU_{i}^{LD}(v_{i},\sigma_{-i}^{LD};q_{i})$ is monotonically increasing in $q_i$. Claim (ii) then follows.

Consider then claim (i): all experts vote. The result is to be expected given that delegating for an expert implies no increase in precision and a reduction in the signals aggregated in the group decision. The proof, however, is not immediate because each expert holds a random number of votes delegated by non-experts, and thus delegation breaks the link between the number of votes cast in a given direction and the corresponding number of signals.

We begin by noting that pivotality here is defined relative to delegation: we say that voter $j$ is pivotal if $j$'s choice of whether to vote (sincerely) or delegate changes the group's decision. If the delegated vote(s) go to an expert who agrees with $j$, the choice is irrelevant. Hence we condition on scenarios in which $j$'s votes, if delegated, go to an expert who disagrees with $j$.  

If no-one delegates or if $K=1$, the problem is effectively identical to the case of abstention: whenever pivotal, given $p>q_i$ any expert benefits from voting. The result is less transparent when votes are delegated and $K>1$. Consider the perspective of expert $e_i$ and suppose for now that all non-experts delegate and no other expert does. Expert $e_i$ prefers to vote if:
\[
Pr(s_i=\omega) Pr(piv_i|s_i=\omega)>Pr(s_i\neq \omega) Pr(piv_i|s_i \neq \omega)
\]

Call $a$ the number of experts, excluding $e_i$ whose signal agrees with $e_i$'s signal, with $a={0,...,K-1}$. Call $X_a$ the total number of votes they hold, and $X_{K-1-a}$ the total number of votes held by the $K-1-a$ experts whose signal differs from $s_i$. Call $y_i$ the number of votes held by $e_i$. Note that $a$ is a random variable; so is $y_i$ and, for given $a$ and $y_i$, so is $X_a$ (while $X_{K-1-a}=K+M-y_i-X_a$). Note for that any $a$, $e_i$ is pivotal if:
\[
y_i+X_a>X_{K-1-a}\text{ \textit{and} } X_a<y_i+X_{K-1-a}
\]
or:
\[
y_i>|X_a-X_{K-1-a}|
\]

Because delegated votes are allocated to experts randomly, with equal probability, we can exploit the symmetry of the problem, and write:
\[
Pr(piv_i|s_i=\omega)=\sum_{a=0}^{K-2} \binom{K-1}{a}p^a(1-p)^{K-1-a}Pr\left(y_i>|X_{K-1-a}-X_a|\right)
\]
and
\[
Pr(piv_i|s_i \neq \omega)=\sum_{a=0}^{K-2} \binom{K-1}{a}(1-p)^a p^{K-1-a}Pr\left(y_i>|X_{K-1-a}-X_a|\right)
\]  

Note that, by symmetry of the terms around the midpoint of the summation:

\begin{equation*}
\begin{aligned}
\sum_{a=1}^{K-2} \binom{K-1}{a} & p^a(1-p)^{K-1-a}Pr\left(y_i>|X_{K-1-a}-X_a|\right)= \\
\sum_{a=1}^{K-2} \binom{K-1}{a} & (1-p)^a p^{K-1-a}Pr\left(y_i>|X_{K-1-a}-X_a|\right)
\end{aligned}
\end{equation*}

Hence, separating the terms corresponding to $a=0$, we can write:
\begin{equation*}
\begin{aligned}
Pr(piv_i|s_i \neq \omega)= & Pr(piv_i|s_i= \omega)-(1-p)^{K-1}Pr\left(y_i>X_{K-1}\right)+p^{K-1}Pr\left(y_i>X_{K-1}\right) \\
= & Pr(piv_i|s_i= \omega)+\left(p^{K-1}-(1-p)^{K-1} \right) Pr\left(y_i>X_{K-1}\right) 
\end{aligned}
\end{equation*}

Given $Pr(s_i=\omega)=p$, it then follows that expert $e_i$ prefers to vote if:
\begin{equation*}
p Pr(piv_i|s_i=\omega)>(1-p) \left[ Pr(piv_i|s_i=\omega)+\left( p^{K-1}-(1-p)^{K-1} \right) Pr\left(y_i>X_{K-1}\right) \right] 
\end{equation*}

or:
\begin{equation*}
\left (p-(1-p) \right) Pr(piv_i|s_i=\omega)>(1-p) \left(p^{K-1}-(1-p)^{K-1} \right) Pr\left(y_i>X_{K-1}\right) 
\end{equation*} 
 
With $p \in(1/2,1)$, the term $\left(p^{K-1}-(1-p)^{K-1} \right)$ is positive and declining in $K$ (strictly so for $K>2$) for all  and thus:
\[
p^{K-1}-(1-p)^{K-1} \leq p-(1-p)
\]
It follows that a sufficient condition, ensuring that $e_i$ prefers to vote is then:
\[
Pr(piv_i|s_i=\omega)>(1-p) Pr\left(y_i>X_{K-1}\right) 
 \]

Note that:
\[
\sum_{a=0}^{K-2} \binom{K-1}{a}p^a(1-p)^{K-1-a}=1-p^{K-1}
\]
and
\[ 
Pr\left(y_i>|X_{K-1-a}-X_a|\right) >  Pr\left(y_i>X_{K-1}\right) \text{ for all $a>0$}.
\]
Hence:
\[
Pr(piv_i|s_i=\omega)> (1-p^{K-1})Pr\left(y_i>X_{K-1}\right) \geq (1-p) Pr\left(y_i>X_{K-1}\right)
\]
which establishes the result. If all non-experts delegate, and the other experts all vote, then expert $e_i$ prefers to vote.

Suppose now that some non-experts have precision realizations above $\widetilde{q}_{LD}$ and vote. Consider again the problem from the perspective of expert $e_i$, and continue to suppose that the other experts all vote. Note that non-experts who vote can only cast a single vote each, and in their case then the number of votes equals the number of voters, and thus of signals. Recall that there are $M$ non-experts. Call $m$ the number of non-experts who vote, $z$ the number of non-experts who agree with $e_i$, and $\mu_v$ the expected precision of the non-experts who vote ($\mu_v = \int_{\widetilde{q}_{LD}}^{\bar{q}} q dF(q)$). Then:
\begin{equation*}
\begin{aligned}
& Pr(piv|s_i=\omega)= \left[ \sum_{m=0}^M \left(1-F(\widetilde{q}_{LD})^m F(\widetilde{q}_{LD} \right)^{M-m} \right] \\
&\left \{\sum_{a=0}^{K-2} \sum_{z=0}^m \binom{K-2}{a} \binom{m}{z} p^a (1-p)^{K-1-a} \mu_v^z (1-\mu_v)^{m-z}
Pr \left(y_i>|X_{K-1-a}+(m-z)-(X_a+z)| \right) \right\}  
\end{aligned}
\end{equation*}
and
\begin{equation*}
\begin{aligned}
& Pr(piv|s_i \neq\omega)= \left[ \sum_{m=0}^M \left(1-F(\widetilde{q}_{LD})^m F(\widetilde{q}_{LD} \right)^{M-m} \right] \\
&\left \{\sum_{a=0}^{K-2} \sum_{z=0}^m \binom{K-2}{a} \binom{m}{z}(1- p)^a p^{K-1-a} (1-\mu_v)^z \mu_v^{m-z}
Pr \left(y_i>|X_{K-1-a}+(m-z)-(X_a+z)| \right) \right\}  
\end{aligned}
\end{equation*}

Although the equations are more cumbersome, the intuition remains close to the discussion above. Because of the symmetry of the model and the random allocations of delegated votes, the only terms that differ between $Pr(piv|s_i=\omega)$ and $ Pr(piv|s_i \neq \omega)$ are those corresponding to $a=0$. Thus we can write:
\begin{equation*}
\begin{aligned}
& Pr(piv|s_i=\omega)= \left[ \sum_{m=0}^M \left(1-F(\widetilde{q}_{LD})^m F(\widetilde{q}_{LD} \right)^{M-m} \right] \\
&\left \{\sum_{z=0}^m \binom{m}{z} (1-p)^{K-1-} \mu_v^z (1-\mu_v)^{m-z}
Pr \left(y_i>|X_{K-1}+(m-z)-z| \right) \right\} + Pr(piv|a>0)  
\end{aligned}
\end{equation*}
and
\begin{equation*}
\begin{aligned}
& Pr(piv|s_i \neq \omega)= \left[ \sum_{m=0}^M \left(1-F(\widetilde{q}_{LD})^m F(\widetilde{q}_{LD} \right)^{M-m} \right] \\
&\left \{\sum_{z=0}^m \binom{m}{z} p^{K-1-} (1-\mu_v)^z \mu_v^{m-z}
Pr \left(y_i>|X_{K-1}+(m-z)-z| \right) \right\} + Pr(piv|a>0)
\end{aligned}
\end{equation*}

We can establish that experts $e_i$ perfers to vote for any realization of $m$, and thus we can ignore the term in square brackets. Repeating exactly the same strategy followed earlier, for given $m$, voting is preferable if:
\[
( p-(1-p) )C > (1-p) [p^{K-1} D - (1-p)^{K-1} F]
\]
where:
\begin{equation*}
\begin{aligned}
&C=\sum_{a=0}^{K-2} \sum_{z=0}^m \binom{K-2}{a} \binom{m}{z} p^a (1-p)^{K-1-a} \mu_v^z (1-\mu_v)^{m-z} \\
& \left[Pr \left(y_i>|X_{K-1-a}+(m-z)-(X_a+z)| \right)\right] \\
&D=\sum_{z=0}^m \binom{m}{z} (1-\mu_v)^z \mu_v^{m-z} Pr\left(y_i>|X_{K-1}+(m-z)-z| \right)  \\
&F=\sum_{z=0}^m \binom{m}{z} (\mu_v)^z (1-\mu_v)^{m-z} Pr \left(y_i>|X_{K-1}+(m-z)-z| \right) \\
\end{aligned}
\end{equation*}

Note that:
\[
\sum_{z=0}^m \binom{m}{z} (1-\mu_v)^z \mu_v^{m-z} =\sum_{z=0}^m \binom{m}{z} (\mu_v)^z (1-\mu_v)^{m-z}
\]
but, with $\mu_v>0.5$, the left-hand side is larger at lower values of $z$, and the right-hand side at higher values. Because $ Pr \left(y_i>|X_{K-1}+(m-z)-z| \right)$ is higher at higher value of $z$, it follows that $F>D$.  Thus:
\[
(p-(1-p)) C > (1-p) (p^{K-1}-(1-p)^{K-1}) D
 \]
is a sufficient condition for $e_i$ preferring to vote. Since we know $(p-(1-p)) \geq (p^{K-1}-(1-p)^{K-1})$, the result follows if:
\[
C>(1-p)D
\]
or:
\begin{equation*}
\begin{aligned}
&\sum_{a=0}^{K-2} \sum_{z=0}^m \binom{K-2}{a} \binom{m}{z} p^a (1-p)^{K-1-a} \mu_v^z (1-\mu_v)^{m-z} \\
& \left[Pr \left(y_i>|X_{K-1-a}+(m-z)-(X_a+z)| \right)\right] > \\ 
&(1-p) \left[ \sum_{z=0}^m \binom{m}{z} (1-\mu_v)^z \mu_v^{m-z} Pr\left(y_i>|X_{K-1}+(m-z)-z| \right)\right] 
\end{aligned}
\end{equation*}
 
The same logic applied earlier can be followed here. As before:
\[
\sum_{a=0}^{K-2} \binom{K-1}{a}p^a(1-p)^{K-1-a}=1-p^{K-1}
\]
and
\[ 
Pr\left(y_i>|X_{K-1-a}-X_a|\right) >  Pr\left(y_i>X_{K-1}\right) \text{ for all $a>0$}.
\]
Hence:
\begin{equation*}
\begin{aligned}
&\sum_{a=0}^{K-2} \sum_{z=0}^m \binom{K-2}{a} \binom{m}{z} p^a (1-p)^{K-1-a} \mu_v^z (1-\mu_v)^{m-z} \\
& \left[Pr \left(y_i>|X_{K-1-a}+(m-z)-(X_a+z)| \right)\right] > \\ 
&(1-p^{K-1})Pr\left(y_i>|X_{K-1}+(m-z)-z| \right) >\\
&(1-p) \left[ \sum_{z=0}^m \binom{m}{z} (1-\mu_v)^z \mu_v^{m-z} Pr\left(y_i>|X_{K-1}+(m-z)-z| \right)\right] 
\end{aligned}
\end{equation*}
establishing the result. 

Finally, suppose now that other experts delegate with positive probability. From $e_i$'s perspective the reasoning is unchanged if, upon delegation, $e_i$'s votes do not enter into a cycle. If instead $e_i$'s votes do enter into a cycle, then all votes in the cycle, including $e_i$'s votes, are cast randomly. Whether they are cast individually or in block, this must be inferior to $e_i$ casting her votes informatively. Note that if there is any probability of other experts delegating, this reasoning breaks any possible indifference. There cannot be an equilibrium where any expert votes with positive probability.

\end{proof}

\subsubsection{The proof of the theorem}\label{subsec:theorem_proof}

\begin{thm*}
For all $N=K+M$ finite, with $M>1$, and for all $F(q)$ everywhere continuous over $[1/2,p]$, the following results hold:

(i) For any $M$, any $K$, and any $F$, there exists an equilibrium under LD that strictly improves over MV. On the other hand, there exist
$M$, $K$, and $F$ such that MV strictly improves over any equilibrium under DD
with positive probability of abstention. 

(ii) There are $M$, $K$, and $F$ such that there exists an equilibrium under LD that strictly improves over any equilibrium under DD. However, there are $M$, $K$, and $F$ such that there exists an equilibrium under DD that strictly improves over any equilibrium under LD. 
\end{thm*}

\begin{proof}

The proof proceeds in three logical steps.
(1) An equilibrium exists. It exists because a maximal strategy profile exists and by McLennan (1998) in a pure common interest game any maximal strategy profile is an equilibrium.  
(2) If $\widetilde{q}=0.5$, that is, MV, is not an equilibrium, then it cannot be maximal in EU (since otherwise it would be an equilibrium, again by McLennan) 
(3) Hence $\sigma^{R*}$ must do strictly better. If $R=LD/DD$, by (1) and (2) above $\sigma^{R*}$ must include a strictly positive probability of delegation/abstention.

We begin by proving step (1). With votes cast according to signal, strategic choices
are limited to delegation/abstention. By Proposition 1, in all semi-symmetric equilibria, if any exist, all experts vote, and non-experts' strategies are summarized by a precision threshold $\widetilde{q}_{R}$ such that non-expert $i$ votes (according to signal) if $q_{i}>\widetilde{q}_{R}$, and delegates/abstains if $q_{i}<\widetilde{q}_{R}$. Thus a profile of strategies $\sigma^{R}=\{\sigma_{ne}^{R},\sigma_{e}^{R}\}$ can be written as 
$\sigma_{ne}^{R}=\sigma_{ne}^{R}\{\widetilde{q}_{R}\}$, and $\sigma_{e}^{R}=\{v\}$.  

Although the set of strategies $\sigma^{R}$ is infinite, it is compact.
Because $EU^{R}$ is a continuous function of $\sigma^{R}$, we can
then apply Weierstrass\textquoteright s Theorem: there exists a profile $\sigma^{R*} \in\{\sigma^{R}\}$
such that $EU^{R}(\sigma^{R*})\geq EU^{R}(\sigma^{R})$ for all $\sigma^{R}\neq\sigma^{R*}.$

We can now establish the following Lemma.
\bigskip

\noindent{\textbf{Lemma A.1: McLennan}}
\textit{The profile of strategies $\sigma^{R*}$ is an equilibrium.}

\begin{proof}
The result is established in two steps, both derived from McLennan (1998). First, since this is a pure common interest game and $\sigma^{R*}=argmax\ EU^{R}(\sigma^{R})$) by construction, it then follows that $\sigma^{R*}$ must be an equilibrium (McLennan, Theorem 1). Second, because the environment is fully symmetric for each class of individuals, experts and non-experts, the game satisfies the requirements of McLennan's Theorem 2. Given $\pi=1/2$ and signal precision symmetric across states, if $\sigma^{R*}$ is maximal with respect to strategies that are symmetric for each group of voters and symmetric with respect to signals content, then it is an equilibrium over all profiles of strategies.\footnote{Note that asymmetric equilibria may exist, and be superior to $\sigma^{R*}$, but $\sigma^{R*}$ is maximal among symmetric strategies. }
\end{proof}

We now proceed to step (2). We begin by studying LD. Later, we will move to DD. 
\vspace{0.5cm}

\noindent{\textbf{Lemma A.2.}}
\textit{Under LD, for any $F(q)$ continuous on $[1/2,p]$, $N=K+M$ finite with $M>1$, MV is not an equilibrium.}

\begin{proof}
Sincere majority voting with voting by all (MV) is feasible within the set of strategies $\sigma^{LD}$ and corresponds to $\widetilde{q}_{LD}=1/2$. Note that if $\widetilde{q}_{LD}=1/2$ is \textit{not} an equilibrium, then we know, by the previous results,
that there must exist an equilibrium of the LD game 
that strictly dominates MV and involves a positive probability of delegation.

Consider the perspective of non-expert voter $i$, with $q_{i}$ in
the neighborhood of $1/2$. Suppose no-one else delegates.
We show in what follows that $i$'s best response is to delegate to
an expert. 

Note first that if no-one delegates, all non-$i$ voters
cast a single vote and have equal weight on the group decision. 
We need to calculate $i$'s interim expected utility from non-delegating
($EU_{i}^{LD}(v_{i},q_{i}))$ or delegating ($EU_{i}^{LD}(d_{i}))$.
We begin by considering the case of $M$ even and $K$ odd (hence $N$
odd) and we use it to clarify the proof strategy. Extending
the analysis to the other possible cases is then straightforward.

(1) Suppose then that $M$ (the number of non-experts) is even and $K$ (the number of experts) is odd ($N$ odd). The expressions
for $EU_{i}^{LD}(v_{i},q_{i})$ and $EU_{i}^{LD}(d_{i})$ are somewhat
cumbersome but conceptually straightforward. With $N=K+M$ odd:

\begin{multline*}
EU_{i}^{LD}(v_{i},q_{i})=\sum_{c_{n}=0}^{M-1}\binom{M-1}{c_{n}}\mu^{c_{n}}(1-\mu)^{M-1-c_{n}}\times\\
\times\left[\sum_{c_{e}=0}^{K}\binom{K}{c_{e}}p{}^{c_{e}}(1-p)^{K-c_{e}}\left(q_{i}I_{c_{n}+c_{e}+1>\frac{(M+K)}{2}}+(1-q_{i})I_{c_{n}+c_{e}>\frac{(M+K)}{2}}\right)\right]
\end{multline*}
\begin{multline*}
EU_{i}^{LD}(d_{i})=\sum_{c_{n}=0}^{M-1}\binom{M-1}{c_{n}}\mu{}^{c_{n}}(1-\mu)^{M-1-c_{n}}\times\\
\times\left[\sum_{c_{e}=0}^{K}\binom{K}{c_{e}}p{}^{c_{e}}(1-p)^{K-c_{e}}\left(\left(\frac{c_{e}}{K}\right)I_{c_{n}+c_{e}+1>\frac{(M+K)}{2}}+\left(\frac{K-c_{e}}{K}\right)I_{c_{n}+c_{e}>\frac{(M+K)}{2}}\right)\right],
\end{multline*}
where $c_{n}$($c_{e}$) indexes
the number of non-experts other than $i$ (the number of experts) whose signals
are correct, $\mu$ is the expected precision of non-experts who choose
to vote, and thus in this conjectured scenario, $\mu=\int_{1/2}^{p}qdF(q)$,
and $I_{C}$ is an indicator function that takes value $1$ if condition
$C$ is satisfied and $0$ otherwise. 

Voter $i$, with $q_{i}$ in the neighborhood of $\underline{q}=1/2$,
strictly prefers delegation if it yields higher expected utility, or:\footnote{It is more transparent to write generic conditions in terms of the parameter $\underline{q}$, and substitute the numerical value $\underline{q}=1/2$ later.}
\[
lim_{q_{i}\rightarrow \underline{q}} \left(EU_{i}^{LD}(v_{i},q_{i})-EU_{i}^{LD}(d_{i})\right)<0
\]

For each realized $c_{n}$ and $c_{e}$, $i$'s expected utility always
equals 1 if $(c_{n}+c_{e})>(M+K)/2$, i.e. if the other voters with
correct signals constitute a majority of the electorate. For $i$,
the choice to delegate or not matters when $(c_{n}+c_{e})$ falls
short of the majority by one vote, that is: $(c_{n}+c_{e})<(M+K)/2$
but $(c_{n}+c_{e}+1)>(M+K)/2$, or $(c_{n}+c_{e})=(M+K-1)/2,$ In
such a case, $EU_{i}^{LD}(v_{i},q_{i})$ equals 1 if $i$'s own signal
is correct (with probability $q_{i}$) and zero otherwise; $EU_{i}^{LD}(d_{i})$
equals 1 if $i$'s vote is delegated to an expert with a correct signal
(with probability $c_{e}/K$) and zero otherwise. 

Denote by $r$ the number of additional correct votes required to
reach a majority, given the votes of the non-experts, excluding $i$,
or $r\equiv(M+K+1)/2-c_{n}$. As just remarked, only the case $c_{e}=(M+K+1)/2-c_{n}-1$,
or $c_{e}=r-1$ is relevant. We can write: 

\begin{multline}
lim_{q_{i}\rightarrow1/2}\left(EU_{i}^{LD}(v_{i},q_{i})-EU_{i}^{LD}(d_{i})\right)=\\
=\sum_{r=1}^{K+1}\binom{M-1}{\frac{M+K+1}{2}-r}\mu{}^{\frac{M+K+1}{2}-r}(1-\mu)^{(M-1)-(\frac{M+K+1}{2}-r)}\binom{K}{r-1}p{}^{r-1}(1-p)^{K-(r-1)}\left(\underline{q}-\frac{r-1}{K}\right)\label{eq:limKsim1-1-1}
\end{multline}
Signing this expression is not immediate because the sign depends
on the last term, $\left(\underline{q} -\frac{r-1}{K}\right)$. However,
the problem can be simplified by noticing its strong symmetry:
In particular, $\left(\underline{q}-\frac{r-1}{K}\right)$ is positive for $r<K\underline{q}+1$ and negative for $r>K\underline{q}+1$. With $\underline{q}=1/2$, 
the expression is positive for $r=\{1,2,..,(K+1)/2\}$
and negative for $r=\{(K+3)/2,...,\}$ (recall that $K$ is odd,
and thus $(K+1)/2$ and $(K+3)/2$ are integers, while $K/2+1$ is
not). Note that the two sets contain the same number of terms and
that to any $r'$ in the first set corresponds $r''=(K+2)-r'$ such
that $\left(1/2-\frac{r''-1}{K}\right)=\left(\frac{r'-1}{K}-1/2\right).$
In addition: 

\[
\binom{M-1}{\frac{M+K+1}{2}-r}=\binom{M-1}{(M-1)-(\frac{M+K+1}{2}-r)};\ \ \binom{K}{r-1}=\binom{K}{K-(r-1)}
\]
Defining a new variable $x$ ranging over the first set of $r$ values, substituting $\underline{q}=1/2$, and exploiting these observations,
equation (\ref{eq:limKsim1-1-1}) can then be written as: 
\begin{multline*}
lim_{q_{i}\rightarrow1/2}\left(EU_{i}^{LD}(v_{i},q_{i})-EU_{i}^{LD}(d_{i})\right)=\sum_{x=1}^{(K+1)/2}\binom{M-1}{\frac{M+K+1}{2}-x}\left(\mu(1-\mu)\right)^{\frac{M+K+1}{2}-(K+2-x)}\times\\
\binom{K}{x}p{}^{x}(1-p)^{x}\left\{ \left(\mu(1-p)\right){}^{K+2-2x}\left[1/2-\frac{x-1}{K}\right]+\left((1-\mu)p\right)^{K+2-2x}\left[\frac{x-1}{K}-1/2\right]\right\} 
\end{multline*}
or$:$ 
\begin{multline*}
lim_{q_{i}\rightarrow1/2}\left(EU_{i}^{LD}(v_{i},q_{i})-EU_{i}^{LD}(d_{i})\right)=\sum_{x=1}^{(K+1)/2}\binom{M-1}{\frac{M+K+1}{2}-x}\left(\mu(1-\mu)\right)^{\frac{M+K+1}{2}-(K+2-x)}\times\\
\binom{K}{x}p{}^{x}(1-p)^{x}\left(1/2-\frac{x-1}{K}\right)\left\{ \left(\mu(1-p)\right){}^{K+2-2x}-\left((1-\mu)p\right)^{K+2-2x}\right\} 
\end{multline*}
Note that $\left(1/2-(x-1)/K\right)>0$ for all $x<(K+2)/2$, and
thus for all relevant $x$ values. It follows that:
\[
\left(\mu(1-p)\right){}^{K+2-2x}<\left((1-\mu)p\right)^{K+2-2x}\Rightarrow lim_{q_{i}\rightarrow1/2}\left(EU_{i}^{LD}(v_{i},q_{i})-EU_{i}^{LD}(d_{i})\right)<0
\]
With all $x<(K+2)/2$, the exponent on both sides is positive, and
we can compare the roots:
\[
\mu(1-p)<(1-\mu)p\Rightarrow lim_{q_{i}\rightarrow1/2}\left(EU_{i}^{LD}(v_{i},q_{i})-EU_{i}^{LD}(d_{i})\right)<0,
\]
a condition that reduces to:
\[
\mu<p
\]
and is always satisfied. Hence $lim_{q_{i}\rightarrow1/2}\left(EU_{i}^{LD}(v_{i},q_{i})-EU_{i}^{LD}(d_{i})\right)<0$:
delegation is the best response. A profile of strategies such that
all non-experts cast their vote with probability 1 cannot be an equilibrium.
\bigskip

(2) Consider now the case $M$ odd and $K$ even (hence again $N$
odd, and recall $M>1$). Equation (\ref{eq:limKsim1-1-1}) is unchanged. Now, 
with $\underline{q}=1/2$, $\left(\underline{q}-\frac{r-1}{K}\right)$ 
equals zero at $r=K/2+1$
(an integer number), is positive for for $r=\{1,2,..,K/2\},$ and
negative for $r=\{(K+4)/2,...,(K+1)\}$. As before, the two sets contain
the same number of terms, and to any $r'$ in the first set corresponds
$r''=(K+2)-r'$ such that $\left(1/2-\frac{r''-1}{K}\right)=\left(\frac{r'-1}{K}-1/2\right).$
As before, we can exploit the symmetry of the binomial terms (here
symmetric around $K/2+1$), define a variable $x$ that spans the first
set of $r$ values, and obtain:
\begin{multline*}
lim_{q_{i}\rightarrow1/2}\left(EU_{i}^{LD}(v_{i},q_{i})-EU_{i}^{LD}(d_{i})\right)=\sum_{x=1}^{K/2}\binom{M-1}{\frac{M+K+1}{2}-x}\left(\mu(1-\mu)\right)^{\frac{M+K+1}{2}-(K+2-x)}\times\\
\binom{K}{x}p{}^{x}(1-p)^{x}\left\{ \left(\mu(1-p)\right){}^{K+2-2x}\left[1/2-\frac{x-1}{K}\right]+\left((1-\mu)p\right)^{K+2-2x}\left[\frac{x-1}{K}-1/2\right]\right\} 
\end{multline*}

The proof then proceeds exactly as in the previous case. 
\bigskip

(3) If $N$ is even, the equations corresponding to $i$'s interim
expected utility from non-delegating ($EU_{i}^{LD}(v_{i},q_{i}))$
or delegating ($EU_{i}^{LD}(d_{i}))$ must be amended because the total
number of votes cast, ignoring \textit{i}, is now odd, and \textit{i}'s
vote can result in a tie. We have:
\begin{multline*}
EU_{i}^{LD}(v_{i},q_{i})=\sum_{c_{n}=0}^{M-1}\binom{M-1}{c_{n}}\mu^{c_{n}}(1-\mu)^{M-1-c_{n}}\times\\
\left[\sum_{c_{e}=0}^{K}\binom{K}{c_{e}}p{}^{c_{e}}(1-p)^{K-c_{e}}\left(q_{i}\left(I_{c_{n}+c_{e}+1>\frac{(M+K)}{2}}+(1/2)I_{c_{n}+c_{e}+1=\frac{(M+K)}{2}}\right)\right.\right.+\\
+\left.\left.(1-q_{i})\left(I_{c_{n}+c_{e}>\frac{(M+K)}{2}}+(1/2)I_{c_{n}+c_{e}=\frac{(M+K)}{2}}\right)\right)\right]
\end{multline*}

\begin{multline*}
EU_{i}^{LD}(d_{i})=\sum_{c_{n}=0}^{M-1}\binom{M-1}{c_{n}}\mu^{c_{n}}(1-\mu)^{M-1-c_{n}}\times\\
\left[\sum_{c_{e}=0}^{K}\binom{K}{c_{e}}p{}^{c_{e}}(1-p)^{K-c_{e}}\left(\left(\frac{c_{e}}{K}\right)\left(I_{c_{n}+c_{e}+1>\frac{(M+K)}{2}}+(1/2)I_{c_{n}+c_{e}+1=\frac{(M+K)}{2}}\right)\right.\right.+\\
+\left.\left.\left(\frac{K-c_{e}}{K}\right)\left(I_{c_{n}+c_{e}>\frac{(M+K)}{2}}+(1/2)I_{c_{n}+c_{e}=\frac{(M+K)}{2}}\right)\right)\right]
\end{multline*}

There are now two events that make \textit{i} pivotal and on which
\textit{i} conditions her actions: $c_{e}+c_{n}+1=(M+K)/2$, and $c_{e}+c_{n}=(M+K)/2$.
Defining $s\equiv(M+K)/2-c_{n}$, we can write:
\begin{multline*}
lim_{q_{i}\rightarrow \underline{q}}\left(EU_{i}^{LD}(v_{i},q_{i})-EU_{i}^{LD}(d_{i})\right)=\\
=\sum_{s=0}^{K}\binom{M-1}{\frac{M+K}{2}-s}\mu{}^{\frac{M+K}{2}-s}(1-\mu)^{(M-1)-(\frac{M+K}{2}-s)}(1/2)\left[\binom{K}{s}p{}^{s}(1-p)^{K-s}\left(\underline{q}-\frac{s}{K}\right)\right]+\\
+\sum_{s=1}^{K+1}\binom{M-1}{\frac{M+K}{2}-s}\mu{}^{\frac{M+K}{2}-s}(1-\mu)^{(M-1)-(\frac{M+K}{2}-s)}(1/2)\left[\binom{K}{s-1}p{}^{s-1}(1-p)^{K-(s-1)}\left(\underline{q}-\frac{s-1}{K}\right)\right].
\end{multline*}

As in the previous cases, signing the expression is complicated by
the need to sign the terms in parentheses---$\left(\underline{q}-\frac{s}{K}\right)$
and $\left(\underline{q}-\frac{s-1}{K}\right)$---but can be simplified
exploiting the symmetry of the different expressions. It helps to
divide the expression into two parts:
\[
lim_{q_{i}\rightarrow \underline{q}}\left(EU_{i}^{LD}(v_{i},q_{i})-EU_{i}^{LD}(d_{i})\right)=f_{1}+f_{2}
\]
where:
\begin{align*}
f_{1} & =\sum_{s=0}^{K}\binom{M-1}{\frac{M+K}{2}-s}\mu{}^{\frac{M+K}{2}-s}(1-\mu)^{(M-1)-(\frac{M+K}{2}-s)}(1/2)\left[\binom{K}{s}p{}^{s}(1-p)^{K-s}\left(\underline{q}-\frac{s}{K}\right)\right]\\
f_{2} & =\sum_{s=1}^{K+1}\binom{M-1}{\frac{M+K}{2}-s}\mu{}^{\frac{M+K}{2}-s}(1-\mu)^{(M-1)-(\frac{M+K}{2}-s)}(1/2)\left[\binom{K}{s-1}p{}^{s-1}(1-p)^{K-(s-1)}\left(\underline{q}-\frac{s-1}{K}\right)\right]
\end{align*}

Consider first $f_{1}$. If $K$ is odd (and thus $M$ is odd as well),
with $\underline{q}=1/2,$ we have: $\left(1/2-\frac{s}{K}\right)>0$
for $s=\{0,1,..,(K-1)/2\}$ and $\left(1/2-\frac{s}{K}\right)<0$
for $s=\{(K+1)/2,..,K\}.$ As before, the two sets contain the same
number of terms and to any $s'$ in the first set corresponds a $s''=K-s'$ such that $\left(1/2-\frac{s''}{K}\right)=\left(\frac{s'}{K}-1/2\right).$
As before, we can divide the summation in two parts, exploit the symmetry
of the binomial terms, and proceed to prove $f_{1}<0$ following exactly
the same logic used in the previous cases. If $K$ is even (and thus
$M$ is even as well), then, with $\underline{q}=1/2$, 
$\left(1/2-\frac{s}{K}\right)=0$
at $s=K/2$. Thus $\left(1/2-\frac{s}{K}\right)>0$ for
$s=\{0,1,..,K/2-1\}$ and $\left(1/2-\frac{s}{K}\right)<0$
for $s=\{(K/2+1,..,K\}.$ Here again, the two sets contain the same
number of terms and to any $s'$ in the first set corresponds a $s''=K-s'$
such that $\left(1/2-\frac{s''}{K}\right)=\left(\frac{s'}{K}-1/2\right).$
We can then proceed as usual: divide the summation in two parts, exploit
the symmetry of the binomial terms, and, given $p>\mu$, use by now
familiar arguments to prove $f_{1}<0$.

Exactly the same strategy can be applied to $f_{2}$, for both $K$
odd and $K$ even, noting that $\left(1/2-\frac{s-1}{K}\right)$
changes sign at $K/2+1$. The steps of the proof proceed identically
as described for the other cases, and we do not repeat them here.
With $p>\mu$, they establish that $f_{2}<0$. Hence $f_{1}+f_{2}<0$,
or $lim_{q_{i}\rightarrow1/2}\left(EU_{i}^{LD}(v_{i},q_{i})-EU_{i}^{LD}(d_{i})\right)<0$, concluding the proof. 
\end{proof}
Note that assuming $\underline{q}=1/2$ is necessary to establish the result
in its full generality, but is also a natural assumption: with a binary choice and own precision known, the lower bound of $F(q)$'s support cannot be inferior to $1/2$. The probability
of realizations near this lower bound needs to be positive, but can
be arbitrarily small.
\vspace{0.5 cm}

Given Lemmas A.1, these results prove Lemma A.2, and thus the first
part of claim (i) in the theorem. When delegation to better informed experts
is possible, there exists an equilibrium in semi-symmetric strategies
and sincere voting that strictly dominates sincere majority voting.
Such an equilibrium must then include a strictly positive probability of delegation.
\vspace{0.5 cm}

We can now analyze DD and prove the second part of claim (i) in the theorem.  We begin with the following Lemma:
\vspace{0.5 cm}

\noindent{\textbf{Lemma A.3}}
\textit{Under DD, for any $F(q)$ continuous on $[1/2,p]$, $N=K+M$ finite with $M>1$, an equilibrium with no abstention exists if and only if $N$ is odd.}
\vspace{0.5 cm}

Lemma A.3 is useful because it tells us that if $N$ is odd, an equilibrium with no abstention can be supported, and thus we cannot invoke Lemma A.1 and McLennan's result to derive any general statement on the relative ranking
of equilibria with and without abstention. Note, however, that, by the same argument, the lemma also tells us that an equilibrium with strictly positive probability of abstention must dominate MV if $N$ is even. 

\begin{proof}
Consider again the perspective of non-expert voter $i$, with $q_{i}$
in the neighborhood of $\underline{q}$. Suppose no-one else abstains.
We want to characterize\textit{ i'}s best response. We denote by $EU_{i}^{DD}(v_{i},q_{i})$
and by $EU_{i}^{DD}(a_{i})$ $i$'s interim expected utility from voting and from abstaining, respectively. Note
that, with everyone else voting, $EU_{i}^{DD}(v_{i},q_{i})=EU_{i}^{LD}(v_{i},q_{i}).$
\bigskip

(1) Suppose first that $N$ is odd. In calculating $EU_{i}^{DD}(a_{i})$,
we must take into account that if voter $i$ abstains and $N$ is odd,
the total number of votes cast is even. Hence:

\begin{multline*}
EU_{i}^{DD}(v_{i},q_{i})=\sum_{c_{n}=0}^{M-1}\binom{M-1}{c_{n}}\mu^{c_{n}}(1-\mu)^{M-1-c_{n}}\times\\
\times\left[\sum_{c_{e}=0}^{K}\binom{K}{c_{e}}p{}^{c_{e}}(1-p)^{K-c_{e}}\left(q_{i}I_{c_{n}+c_{e}+1>\frac{(M+K)}{2}}+(1-q_{i})I_{c_{n}+c_{e}>\frac{(M+K)}{2}}\right)\right]
\end{multline*}
\begin{multline*}
EU_{i}^{DD}(a_{i})=\sum_{c_{n}=0}^{M-1}\binom{M-1}{c_{n}}\mu{}^{c_{n}}(1-\mu)^{M-1-c_{n}}\times\\
\times\left[\sum_{c_{e}=0}^{K}\binom{K}{c_{e}}p{}^{c_{e}}(1-p)^{K-c_{e}}\left(I_{c_{n}+c_{e}>\frac{(M+K-1)}{2}}+(1/2)I_{c_{n}+c_{e}=\frac{(M+K-1)}{2}}\right)\right].
\end{multline*}

Voter $i$, with $q_{i}$ in the neighborhood of $\underline{q}$,
strictly prefers abstention if it yields higher expected utility,
or:
\[
lim_{q_{i}\rightarrow \underline{q}}\left(EU_{i}^{DD}(v_{i},q_{i})-EU_{i}^{DD}(a_{i})\right)<0
\]

It is not difficult to verify that the only feasible event such that\textit{
i}'s abstention affects the outcome is $c_{n}+c_{e}=\frac{(M+K-1)}{2}$.
Hence:
\begin{multline*}
lim_{q_{i}\rightarrow\underline{q}}\left(EU_{i}^{DD}(v_{i},q_{i})-EU_{i}^{DD}(a_{i})\right)=\sum_{c_{n}=0}^{M-1}\binom{M-1}{c_{n}}\mu{}^{c_{n}}(1-\mu)^{M-1-c_{n}}\times\\
\times\left[\sum_{c_{e}=0}^{K}\binom{K}{c_{e}}p{}^{c_{e}}(1-p)^{K-c_{e}}(\underline{q}-1/2)I_{c_{n}+c_{e}=\frac{(M+K-1)}{2}}\right],
\end{multline*}
an expression that equals 0 at $\underline{q}=1/2$.
An equilibrium with no abstention, replicating majority voting, can
be supported. Note that whether $M$ is even and $K$ odd, or $M$
is odd and $K$ even is irrelevant. The conclusion holds identically
for both cases and proves the ``if'' part of the lemma. We now prove
the ``only if'' part by showing that majority voting cannot be supported if $N$ is even.
\bigskip

(2) As in the case of delegation, if $N$ is even, $i$'s interim
expected utility equations must be amended: 

\begin{multline*}
EU_{i}^{DD}(v_{i},q_{i})=\sum_{c_{n}=0}^{M-1}\binom{M-1}{c_{n}}\mu^{c_{n}}(1-\mu)^{M-1-c_{n}}\times\\
\left[\sum_{c_{e}=0}^{K}\binom{K}{c_{e}}p{}^{c_{e}}(1-p)^{K-c_{e}}\left(q_{i}\left(I_{c_{n}+c_{e}+1>\frac{(M+K)}{2}}+(1/2)I_{c_{n}+c_{e}+1=\frac{(M+K)}{2}}\right)\right.\right.+\\
+\left.\left.(1-q_{i})\left(I_{c_{n}+c_{e}>\frac{(M+K)}{2}}+(1/2)I_{c_{n}+c_{e}=\frac{(M+K)}{2}}\right)\right)\right]
\end{multline*}
\begin{multline*}
EU_{i}^{DD}(a_{i})=\sum_{c_{n}=0}^{M-1}\binom{M-1}{c_{n}}\mu{}^{c_{n}}(1-\mu)^{M-1-c_{n}}\times\\
\times\left[\sum_{c_{e}=0}^{K}\binom{K}{c_{e}}p{}^{c_{e}}(1-p)^{K-c_{e}}\left(I_{c_{n}+c_{e}>\frac{(M+K-1)}{2}}\right)\right].
\end{multline*}

If \textit{i} votes, the majority threshold is $(M+K)/2$, but if\textit{
i} abstains, the threshold becomes $(M+K-1)/2$. With $M+K$ even,
there is no possibility of a tie if \textit{i} abstains. It is not
difficult to verify that the only event that makes $i$'s decision pivotal
is $c_{n}+c_{e}=(M+K)/2$, which implies $c_{n}+c_{e}+1>(M+K)/2$
and $c_{n}+c_{e}>(M+K-1)/2$. Hence:
\begin{multline*}
lim_{q_{i}\rightarrow \underline{q}}\left(EU_{i}^{DD}(v_{i},q_{i})-EU_{i}^{DD}(a_{i})\right)=\sum_{c_{n}=0}^{M-1}\binom{M-1}{c_{n}}\mu{}^{c_{n}}(1-\mu)^{M-1-c_{n}}\times\\
\times\left[\sum_{c_{e}=0}^{K}\binom{K}{c_{e}}p{}^{c_{e}}(1-p)^{K-c_{e}}\left((\underline{q}-1)/2\right)\left(I_{c_{n}+c_{e}=\frac{(M+K)}{2}}\right)\right].
\end{multline*}

It follows that $lim_{q_{i}\rightarrow \underline{q}}\left(EU_{i}^{DD}(v_{i},q_{i})-EU_{i}^{DD}(a_{i})\right)<0$  
for all $\underline{q}<1$, thus including $\underline{q}=1/2.$ 
An equilibrium where everyone votes cannot be sustained for $N$ even (whether $M$ and $K$
are both even or both odd is irrelevant). Thus for $N$ even the logic
of the proof applied earlier to LD extends to DD.
Since an equilibrium without abstention cannot be supported, and by
the previous lemma an equilibrium exists, an equilibrium with strictly
positive probability of abstention must exist, and by McLennan's argument
such an equilibrium must dominate voting by all.  
\end{proof}

If $N$ is odd, Lemma A.3 rules out the simple proof strategy
applied to LD. Per se, this does not exclude the possibility
that equilibria with strictly positive probability of abstention,
if they exist, may be systematically (for all relevant $M$, $K$, and $F$)
superior to MV. A counterexample, however, is sufficient
to establish that this cannot be the case, and the lemma teaches us that
any counter-example must have $N$ odd.
\vspace{0.5 cm}

{\textbf{Lemma A.4}}
\textit{Suppose $N=K+M$ finite and odd, with $M>1$. Then there exists $F$ continuous over $[1/2,p]$ such that MV dominates any DD equilibrium with positive probability of abstention.}

\vspace{0.5 cm}

\begin{proof}
Consider the simplest example: $K=1,$ $M=2$ (and thus $N=3$, odd).
By Lemma A.3 there is an equilibrium where all vote. For
arbitrary $N$ and $K$ odd, ex ante expected utility under all voting
is given by:

\[
EU_{MV}=\sum_{c_{n}=0}^{M}\binom{M}{c_{n}}\mu{}^{c_{n}}(1-\mu)^{z-c_{n}}\sum_{c_{e}=0}^{K}\binom{K}{c_{e}}p{}^{c_{e}}(1-p)^{K-c_{e}}I_{c_{n}+c_{e}>\frac{(M+K)}{2}}
\]

With $K=1$ and $M=2$, the expression simplifies to:

\[
EU_{MV}=p[\mu^{2}+2\mu(1-\mu)]+(1-p)\mu^{2}=\mu(\mu+2p-2\mu p)
\]

Consider now possible equilibria with abstention. Recall, from Proposition 1, that experts never abstain. For all $K$ odd (including $K=1$), there is always an equilibrium where all experts vote and all non-experts abstain.\footnote{With $K=1$ and $M=2$, the equilibrium is trivial. Given the other non-expert abstaining, non-expert \textit{i'}s vote only matters if it counters
the expert's vote. But in that case, it would induce a tie, and thus $i$'s best response is to abstain. And with both non-experts abstaining, the expert's best response must be to vote. For general $K$ odd, the basic logic remains unchanged.} 

Consider then such an equilibrium, where ex ante expected utility
$EU_{DD}$ must equal $p$, the probability the expert's signal is
correct. Note that $EU_{MV}(\mu,p)=p$ at $\mu=\mu'=(p-\sqrt{p-p^{2}})/(2p-1)<p$
for all $p\in(1/2,1)$ and $\partial EU_{MV}/\partial\mu >0$.
Whether $EU_{MV}\gtrless p$ depends on $\mu$, and thus on $F$,
and because $\mu'<p$, it is in principle possible that there exists
$F$ defined over $[1/2,p]$ such that $EU_{MV}(\mu,p)>p=EU_{DD}.$
Intuitively, the larger the probability mass close to $p$ (the larger
the probability of signals with precision close to the expert's precision), the
more likely it is that voting by all dominates abstention by non-experts.
For all $p\in(1/2,1)$ it is not difficult to find an $F$ such that
$\mu_{F}>\mu'$ (adding here the subscript $F$ to emphasize the distribution).
For example, if $f(q)=c(q-1/2)^{k}$ (where $c$ is a normalizing constant
ensuring that $\intop_{1/2}^{p}f(q)dq=1$), then $\mu_{F}=(1+2p+2kp)/(4+2k)$,
increasing in $k.$ Since $\mu'=\mu_{F}$ at $k'(p)=\left(1-2\sqrt{p(1-p)}\right)/\left(2\sqrt{p(1-p)}\right)$,
for any $p<1$ there exists a finite $k'(p)$ such that $EU_{MV}=p$.
And since $\mu_{F}$ is increasing in $k$, and thus $EU_{MV}$ is
increasing in $k$, there exists a finite $k>k'(p)$ such that $EU_{MV}>p$.\footnote{In the limit, as $k$ goes to infinity, $lim_{k\rightarrow\infty}EU_{MV}=3p^{2}-2p^{3}>p$ for all $p \in(1/2,1)$.} The power $k$ need not be particularly high: for $k=2$, $EU_{MV}>p$
for all $p<0.947$; for $k=4$, $EU_{MV}>p$ for all $p<0.99$.

These observations show that, for any $p<1$, there exist $F$, $M$, and
$K$ such that there always exists an equilibrium with abstention
that is dominated by MV. The claim in the theorem, however,
applies to \textit{any} equilibrium with abstention. In the case we
are considering, are there other equilibria with strictly positive
probability of abstention? Recall that we focus on equilibria symmetric
across non-experts and that, by Proposition 1, non-expert equilibrium strategies
must be monotone threshold strategies. Hence if other equilibria with positive probability of abstention exist,
then there exists $\widetilde{q}_{DD}\in (1/2,p)$ such that abstaining
if $q_{i}<\widetilde{q}_{DD}$ and voting according to
signal otherwise is a best response when the other non-expert follows the same strategy and the expert votes. 

Denote by $EU_{i}^{DD}(v_{i},\widetilde{q}_{DD})$ ($EU_{i}^{DD}(a_{i},\widetilde{q}_{DD}))$
interim expected utility from voting (abstaining) for a non-expert
with precision $q_{i}$ when all others follow the strategy summarized
by $\widetilde{q}_{DD}$. With $M=2$ and $K=1$:

\[
EU_{i}^{DD}(v_{i},\widetilde{q}_{DD})=F(\widetilde{q}_{DD})\left(\frac{p+q_{i}}{2}\right)+[1-F(\widetilde{q}_{DD})][\mu_{v}(p+q_{i})+pq_{i}(1-2\mu_{v})]
\]
 and
\[
EU_{i}^{DD}(a_{i},\widetilde{q}_{DD})=F(\widetilde{q}_{DD})p+[1-F(\widetilde{q}_{DD})]\left[\mu_{v}p+\frac{\mu_{v}(1-p)}{2}+(1-\mu_{v})\frac{p}{2}\right]
\]
where $\mu_{v}$ is the expected precision of a non-expert voting:
\[
\mu_{v}=\int_{\widetilde{q}_{DD}}^{p}qf(q)dq/(1-F(\widetilde{q}_{DD})
\]

The threshold $\widetilde{q}_{DD}$ must be such that non-expert $i$
with $q_{i}=\widetilde{q}_{DD}$ is indifferent between voting and
abstaining. Imposing $EU_{i}^{DD}(v_{i},\widetilde{q}_{DD})=EU_{i}^{DD}(a_{i},\widetilde{q}_{DD})$, setting $q_i=\widetilde{q}_{DD}$, and
collecting terms, $\widetilde{q}_{DD}$ must solve:
\[
\left(1-F(\widetilde{q}_{DD})\right)\left(\mu_{v}(1-p)+p(1-\mu_{v})(\widetilde{q}_{DD}-1/2)\right)-F(\widetilde{q}_{DD})\left(\frac{p-\widetilde{q}_{DD}}{2}\right)\equiv G(\widetilde{q}_{DD})=0
\]

The equilibria discussed earlier correspond to $\widetilde{q}_{DD}=1/2$
and $F(\widetilde{q}_{DD})=0$ (no abstention), and to $\widetilde{q}_{DD}=p$
and $F(\widetilde{q}_{DD})=1$ (only the expert votes). Both satisfy
the equation. The question is whether there are other interior roots
at some $\widetilde{q}_{DD}\in(1/2,p)$. Note that:
\[
\frac{\partial G(\widetilde{q}_{DD})^{2}}{\partial\widetilde{q}_{DD}^{2}}=-\frac{\partial f(\widetilde{q}_{DD})}{\partial\widetilde{q}_{DD}}\left(\frac{\widetilde{q}_{DD}+p-1}{2}\right)-f(\widetilde{q}_{DD})
\]

As before, consider $f(q)=c(q-1/2)^{k}$. Then $\partial f(\widetilde{q}_{DD})/\partial\widetilde{q}_{DD}=ck(\widetilde{q}_{DD}-1/2)^{k-1}>0.$
It follows that $G(\widetilde{q}_{DD})$ is everywhere concave, for
all $\widetilde{q}_{DD}\in$$(1/2,p)$. But then, if $G(1/2)=G(p)=0$,
there cannot be any $\widetilde{q}_{DD}\in$$(1/2,p)$ such that $G(\widetilde{q}_{DD})=0.$
The only two equilibria are $\widetilde{q}_{DD}=1/2$ (no abstention)
and $\widetilde{q}_{DD}=p$ (only the expert votes), and we have already
shown that for any $p$ there exists $k>1$ such that if $f(q)=c(q-1/2)^{k}$,
the first dominates the second. This concludes the proof.
\end{proof}

We have used as counterexample $M=2$ and $K=1$ because the low numbers
simplify the expressions and the analytical proof. But the result
does not need such extreme case. As long as $N$ is odd, simulations
show that the conclusion holds broadly. 

\bigskip

We can now address claim (ii) in the theorem. Note that claim (i), the first part of the theorem, immediately implies:

\bigskip

{\textbf{Corollary.}}
\textit{There exist $N=K+M$ finite, with $M>1$, and $F$ continuous over $[1/2,p]$ such that there exists an equilibrium under LD
that strictly dominates any equilibrium under DD.}

\vspace{0.3cm}

If there always exists an equilibrium under LD
that strictly dominates
MV, while, for some parameter values, MV strictly dominates any equilibrium under DD, the Corollary must follow. The parametrization $K=1$, $M=2$, $f(q)=c(1/2-q)^2$ and $p \leq 0.9$ is an example.

As the theorem states, the opposite ranking can occur as well. Here too an example is sufficient to make the case. Suppose for example $K=1$, $M=8$, $F$ Uniform and $p=0.7$. DD has an interior equilibrium with $\widetilde{q}_{DD}=0.574$, $EU_{DD}=0.77$; LD has two equilibria, a boundary equilibrium with all non-experts delegating ($\widetilde{q}_{LD}=p$) and $EU_{LD}=0.7$, and an interior equilibrium with $\widetilde{q}_{LD}=0.524$ and $EU_{LD}=0.768$. Expected utility under MV is $EU_{MV}=0.757$. Hence there exists an equilibrium under DD that dominates both MV and any equilibrium under LD.     

This concludes the proof of the theorem. 
\end{proof}

\subsubsection{Equilibrium conditions: \texorpdfstring{$N$ odd and $K$ odd}{N odd and K odd}}\label{subsec:eqm_conditions}

We report here the conditions that characterize equilibrium voting thresholds for arbitrary $N$ odd and $K$ odd. The equations yield the results reported in the numerical exercises of Sections \ref{sec: numerical regularities} and \ref{sec:Experiment-1: Design}.

\vspace{0.5cm}

\textbf{Liquid Democracy}
\vspace{0.2cm}

$EU^{LD}_i(v_i,q_{i},\widetilde{q}_{LD})$ is interim expected utility for a voter with
precision $q_{i}$. $EU^{LD}(d_i, \widetilde{q}_{LD})$ is interim
expected utility from delegating. If there is an interior equilibrium, 
$\widetilde{q}_{LD}$ must solve $EU^{LD}_i(v_i, q_{i},\widetilde{q}_{LD})=EU^{LD}(d_i,\widetilde{q}_{LD})$. With
$\mu_{v}(\widetilde{q}_{LD})\equiv\mathbb{E}_{F}[q_{j}|q_{j}>\widetilde{q}_{LD}]$, we find:

\begin{multline*}
EU^{LD}_i(v_i,q_{i},\widetilde{q}_{LD})=\sum_{z=0}^{M-1}\binom{M-1}{z}\left(1-F(\widetilde{q}_{LD})\right)^{z}F(\widetilde{q}_{LD})^{M-1-z}\sum_{c_{n}=0}^{z}\binom{z}{c_{n}}(\mu_{v}(\widetilde{q}_{LD})){}^{c_{n}}(1-\mu{}_{v}(\widetilde{q}_{LD}))^{z-c_{n}}\\
\times\left\{ \left(\sum_{c_{e}=0}^{K}\binom{K}{c_{e}}p{}^{c_{e}}(1-p)^{K-c_{e}}\sum_{x_{1}=0}^{M-z-1}\sum_{x_{2}=0}^{M-z-1-x_{1}}...\sum_{x_{K-1}=0}^{M-z-1-\sum_{k=1}^{K-2}x_{k}}\frac{(M-z-1)!}{\prod_{k=1}^{K}x_{k}!}\right.\right)\\
\times\left(q_{i}\left((1/K)^{M-z-1}I_{c_{n}+1+c_{e}+\sum_{k=1}^{c_{e}}x_{k}>(M+K)/2}\right)+\right.\\
\left.\left.(1-q_{i})\left((1/K)^{M-z-1}I_{c_{n}+c_{e}+\sum_{k=1}^{c_{e}}x_{k}>(M+K)/2}\right)\right)\right\} \\
,
\end{multline*}
where $x_{K}\equiv M-z-1-\sum_{k=1}^{K-1}x_{k}$, and $I_{C}$ is
an indicator function that equals 1 if condition $C$ is realized
and 0 otherwise. 
Similarly:

\begin{multline*}
EU^{LD}(d_i, \widetilde{q}_{LD})=\sum_{z=0}^{M-1}\binom{M-1}{z}\left(1-F(\widetilde{q}_{LD})\right)^{z}F(\widetilde{q}_{LD})^{M-1-z}\sum_{c_{n}=0}^{z}\binom{z}{c_{n}}(\mu_{v}(\widetilde{q}_{LD})){}^{c_{n}}(1-\mu{}_{v}(\widetilde{q}_{LD}))^{z-c_{n}}\times\\
\times\left\{ \left(\sum_{c_{e}=0}^{K}\binom{K}{c_{e}}p{}^{c_{e}}(1-p)^{K-c_{e}}\sum_{y_{1}=0}^{M-z}\sum_{y_{2}=0}^{M-z-y_{1}}...\sum_{y_{K-1}=0}^{M-z-\sum_{k=1}^{K-2}y_{k}}\frac{(M-z)!}{\prod_{k=1}^{K}y_{k}!}\right.\right)\times\\
\times\left.\left((1/K)^{M-z}I_{c_{n}+c_{e}+\sum_{k=1}^{c_{e}}y_{k}>(M+K)/2}\right)\right\} 
\end{multline*}
where $y_{K}\equiv M-z-\sum_{k=1}^{K-1}y_{k}$.

\vspace{0.2cm}

Ex ante expected utility, $EU_{LD}(\widetilde{q}_{LD})$ equals:
\[
EU_{LD}(\widetilde{q}_{LD})=F(\widetilde{q}_{LD}) EU^{LD}_i(d, \widetilde{q}_{LD})+\int_{\widetilde{q}_{LD}}^{p}EU^{LD}_i(v_i,q)f(q)dq
\]
\vspace{0.2cm}

\textbf{Direct Democracy}
\vspace{0.2cm}

$EU^{DD}_i(v_{i},q_i,\widetilde{q}_{DD})$ is interim expected utility
from voting, given $q_{i}$; $EU^{DD}(a_i, \widetilde{q}_{DD})$ is interim
expected utility from abstaining. If there is an interior equilibrium, 
$\widetilde{q}_{DD}$ must solve $EU^{DD}_i(v_i,q_{i},\widetilde{q}_{DD})=EU^{DD}_i(a_i,\widetilde{q}_{DD})$. With
$\mu_{v}(\widetilde{q}_{DD})\equiv\mathbb{E}_{F}[q_{j}|q_{j}>\widetilde{q}_{DD}]$, we find:

\begin{multline*}
EU^{DD}_i(v_i,q_{i},\widetilde{q}_{DD})=\sum_{v=0}^{M-1}\binom{M-1}{v}\left(1-F(\widetilde{q}_{DD})\right)^{v}F(\widetilde{q}_{DD})^{M-1-v}\sum_{c_{n}=0}^{v}\binom{v}{c_{n}}(\mu_{v}(\widetilde{q}_{DD})){}^{c_{n}}(1-\mu{}_{v}(\widetilde{q}_{DD}))^{v-c_{n}}\\
\times\left\{ \left(\sum_{c_{e}=0}^{K}\binom{K}{c_{e}}p{}^{c_{e}}(1-p)^{K-c_{e}}\right.\right)\\
\times\left(q_{i}\left(I_{c_{n}+1+c_{e}>(v+1+K)/2}+(1/2)I_{c_{n}+1+c_{e}=(v+1+K)/2}\right)+\right.\\
\left.\left.(1-q_{i})\left(I_{c_{n}+c_{e}>(v+1+K)/2}+(1/2)I_{c_{n}+c_{e}=(v+1+K)/2}\right)\right)\right\} ,
\end{multline*}
and:
\begin{multline*}
EU^{DD}_i(a_i,\widetilde{q}_{DD})=\sum_{v=0}^{M-1}\binom{M-1}{v}\left(1-F(\widetilde{q}_{DD})\right)^{v}F(\widetilde{q}_{DD})^{M-1-v}\sum_{c_{n}=0}^{v}\binom{v}{c_{n}}(\mu_{v}(\widetilde{q}_{DD})){}^{c_{n}}(1-\mu{}_{v}(\widetilde{q}_{DD}))^{v-c_{n}}\\
\times\left(\sum_{c_{e}=0}^{K}\binom{K}{c_{e}}p{}^{c_{e}}(1-p)^{K-c_{e}}\right)\left(I_{c_{n}+c_{e}>(v+K)/2}+(1/2)I_{c_{n}+c_{e}=(v+K)/2}\right).
\end{multline*}

\vspace{0.2cm}

Ex ante expected utility, before the realization of $q_{i}$ but under
the correct expectation of $\widetilde{q}_{DD}$, is given by: 
\[
EU_{DD}(\widetilde{q}_{DD})=F(\widetilde{q}_{DD})EU^{DD}(d, \widetilde{q}_{DD})+\int_{\widetilde{q}_{DD}}^{p}EU^{DD}_i(v_i,q,\widetilde{q}_{DD})f(q)dq
\]

\vspace{0.2cm}

\textbf{Majority Voting}
\vspace{0.2cm}

For convenience, we report here again ex ante expected utility under MV, $EU_{MV}$: 
\[
EU_{MV}=\sum_{c_{n}=0}^{M}\binom{M}{c_{n}}\mu{}^{c_{n}}(1-\mu)^{z-c_{n}}\sum_{c_{e}=0}^{K}\binom{K}{c_{e}}p{}^{c_{e}}(1-p)^{K-c_{e}}I_{c_{n}+c_{e}>\frac{(M+K)}{2}}
\]
where, as in earlier use, $\mu\equiv\mathbb{E}_{F}(q_{i})$.


\subsection{Additional numerical results} \label{simulations}

\begin{figure}[!t]
\begin{centering}
\includegraphics[width=\textwidth]{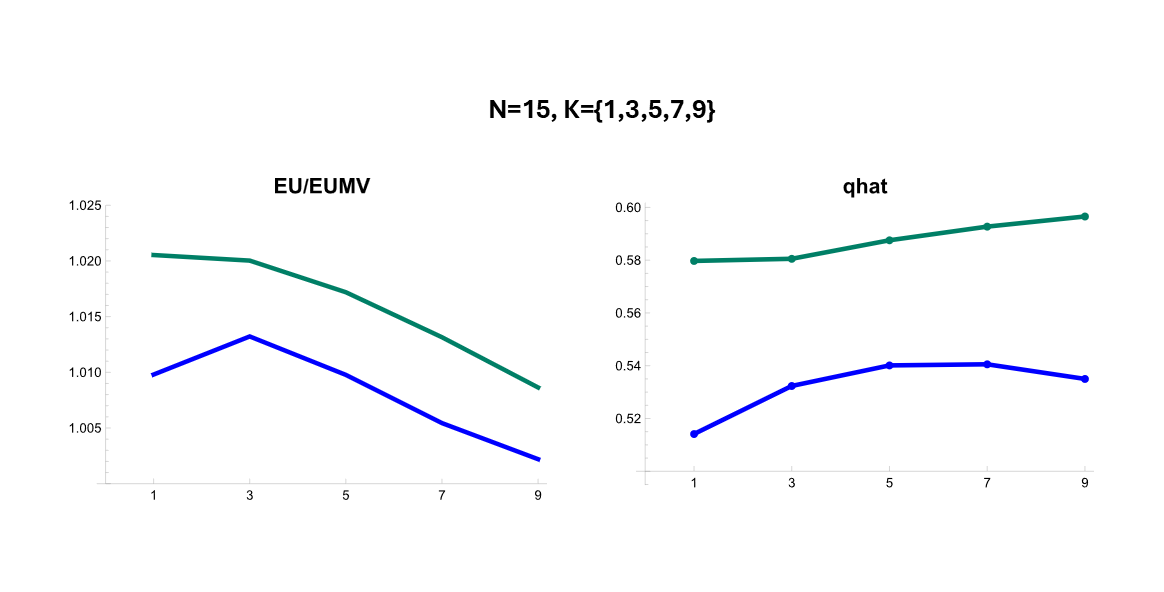}
\caption{\emph{Numerical examples. F Uniform.} Green is DD; Blue is LD. $p=0.7$.}
\label{Fig.Simulations.Uniform.N15}
\end{centering}
\end{figure}

\begin{figure}
\begin{centering}    
\includegraphics[width=\textwidth]{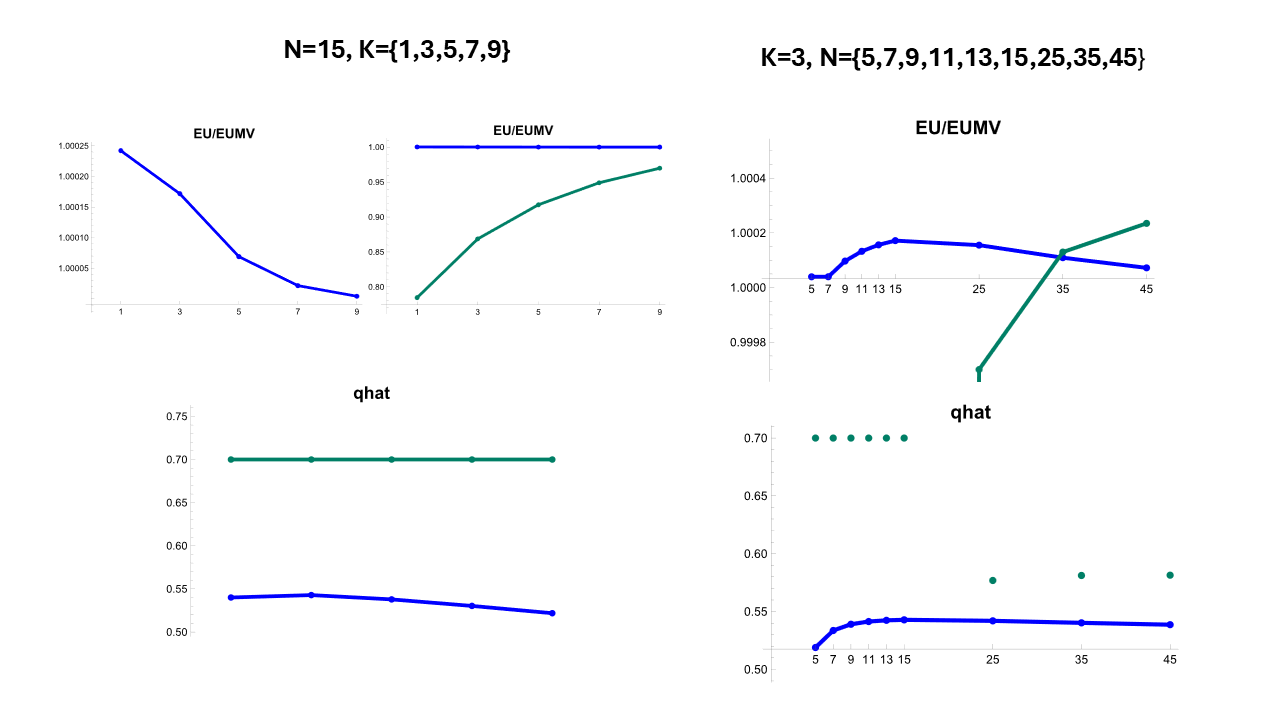}
\caption{\emph{Numerical examples. $f(q)=c(1/2-q)^2$.} Green is DD; Blue is LD. $p=0.7$. For $K=3$ and small $N$, under DD, the only equilibrium with positive abstention is a boundary equilibrium at $\widetilde{q}_{DD}=0.7$. Its expected losses are not represented for reasons of scale.}
\label{Fig.Simulations.Power}
\end{centering}
\end{figure}

We report here some additional numerical results on how LD and DD fare relative to MV and to each other, in response to changes in $N$, $K$, and $F$. Results are sensitive to $F$, and we study two specifications: $F$ Uniform, in Figure \ref{Fig.Simulations.Uniform.N15}, and $F$ with probability mass concentrated on high values ($f(q)=c(1/2-q)^2$) in Figure \ref{Fig.Simulations.Power}. In all panels, Blue curves correspond to LD; Green to DD. When multiple equilibria exist, the plots depict the equilibrium with highest $EU_R$ among equilibria with positive delegation/abstention.\footnote{In all cases, that corresponds to an interior equilibrium, if it exists. Under DD, there are parameter values for which no interior equilibrium exists, and the best equilibrium with positive abstention corresponds to the boundary equilibrium in which all non-experts abstain ($\widetilde{q}_{DD}=p)$.} 

We draw several lessons. First, as expected, gains over MV are larger when the difference in precision between experts and non-experts is larger. Both LD and DD can deliver non-negligible gains in Figure \ref{Fig.Simulations.Uniform.N15}, when $F$ is uniform. But when $F$ is a power distribution and non-experts precision is concentrated at higher values (Figure \ref{Fig.Simulations.Power}), such gains become minuscule under LD and turn into losses under DD.\footnote{All examples have $p=0.7$. With $F$ uniform, $E_F(q)=0.6$; with $f=c(1/2-q)^2$, $E_F(q)=0.65$.} Second, increases in either $K$ or $N$ increase the efficiency of both rules. However, MV's efficiency increases too, and while gains over MV rise initially, at larger electoral sizes expected gains over MV eventually decline, reflecting MV's asymptotic efficiency.\footnote{In Figure \ref{Fig.Simulations.Power}, with $K=3$, abstention gains over MV decline for $N$ larger than $60$, not reported in the figure.} Under both distributions of precisions, DD benefits from larger electorate sizes more than LD. Third, given MV's asymptotic efficiency, we expect $\widetilde{q}$ to approach $1/2$ asymptotically under both rules, thus allowing the best equilibria to replicate MV. The decline however is not monotonic and depends on $F$.\footnote{Here too, the decline in $\widetilde{q}_{DD}$ under the power distribution appears at very large $N$---i.e. $N>60$.} 

Less expected is the relative constancy of the voting thresholds $\widetilde{q}$ in response to changes in either $K$ or $N$. Under both distributions and both rules, if an interior equilibrium exists, thresholds vary only over a very small band of values. The really sizable change occurs when an interior equilibrium becomes sustainable.

We can compare LD and DD directly. Recall that the figures plot the highest efficiency equilibria with positive delegation/abstention, which correspond to interior equilibria, whenever they exist. In all cases, we find $\widetilde{q}_{DD} > \widetilde{q}_{LD}$, or higher equilibrium abstention than delegation. We conjecture that the regularity reflects the higher cost of delegation: in addition to depriving the system of some information, delegation also concentrates votes in the hands of a smaller group. To be advantageous, delegation must be rare. As for efficiency, under the parameters studied here, the most important determinant of the relative performance of the two systems is $F$, the distribution of precisions. When $F$ is uniform, Figure \ref{Fig.Simulations.Uniform.N15}, together with Figure \ref{Fig.SimulationsK1K3.Uniform} in the text, show that at higher $N$, DD dominates LD, by a margin that is sensitive to $N$: at small $N$ the two systems are relatively close (and if $K=1$, LD dominates DD) but the gap increases, in favor of DD, as $N$ increases, before finally very slowly declining as $N$ continues to grow. Under the power distribution, in Figure \ref{Fig.Simulations.Power}, DD is instead inferior to LD for all parameter values at which an interior equilibrium cannot be supported, and in fact remains so as an interior equilibrium appears at $N=25$ (and $K=3$). It is only as $N$ increases further that DD begins to dominate LD by a margin that first increases and then declines as $N$ becomes very large. It is important to note, however, that with precisions concentrated towards high values, all gains over MV are negligible---never reaching 0.1 percent for the parameters in the figure, whether under DD or under LD---while losses under DD can be sizable, reaching 15 percent for $N=15$ when $K=3$. 


\subsection{Additional empirical results}\label{subsec:appendix_additional_empirical}

\subsubsection{Frequency of delegation and abstention, including order and session treatments controls}\label{subsec:main_reg_probit}

\begin{table}[p]
\begin{singlespace}
\centering{}%
\resizebox{\textwidth}{!}{
\begin{tabular}{l>{\centering}p{2.8cm}>{\centering}p{2.8cm}>{\centering}p{2.8cm}>{\centering}p{2.8cm}}
\multicolumn{5}{c}{Experiment 1: Frequency of Delegation or Abstention.}\tabularnewline
\multicolumn{5}{c}{Probit Model.}\tabularnewline
\midrule 
 & \multicolumn{2}{c}{\rule{0pt}{1em}Design 1} & \multicolumn{2}{c}{Design 2} \\
 \cline{2-3} \cline{4-5}
 & \rule{0pt}{1em}(1) & (2) & (3) & (4)\tabularnewline
 & N=5 & N=15 & N=5 & N=15\tabularnewline
\midrule 
LD & 0.938{*}{*} & 0.677{*}{*} & 0.934{*}{*}{*} & 0.744{*}{*} \tabularnewline
 & (0.274) & (0.249) & (0.230) & (0.298) \tabularnewline
\vspace{6pt} & {[}0.001{]}  & {[}0.007{]} & {[}0.000{]} & {[}0.013{]} \tabularnewline
Signal Precision & -2.624{*}{*}{*} & -2.691{*}{*}{*} & -2.388{*}{*}{*} & -2.185{*}{*}{*} \tabularnewline
 & (0.307) & (0.208) & (0.191) & (0.138) \tabularnewline
\vspace{6pt} & {[}0.000{]} & {[}0.000{]} & {[}0.000{]} & {[}0.000{]} \tabularnewline
LD {*} Signal Precision & 0.065 & 0.102 & 0.499{*}{*} & -0.055 \tabularnewline
 & (0.225) & (0.144) & (0.197) & (0.234) \tabularnewline
\vspace{6pt} & {[}0.772{]} & {[}0.477{]} & {[}0.011{]} & {[}0.814{]} \tabularnewline
Round & 0.085 & 0.274 & 0.292 & 0.237{*}{*} \tabularnewline
 & (0.209) & (0.197) & (0.264) & (0.102) \tabularnewline
\vspace{6pt} & {[}0.684{]} & {[}0.164{]} & {[}0.268{]} & {[}0.021{]} \tabularnewline
LD {*} Round & -0.345 & -0.174 & -0.023 & -0.281 \tabularnewline
 & (0.371) & (0.239) & (0.341) & (0.259) \tabularnewline
\vspace{6pt} & {[}0.352{]} & {[}0.466{]} & {[}0.946{]} & {[}0.278{]} \tabularnewline
Second & 0.534{*}{*}{*} & -0.341{*}{*} & & \tabularnewline
 & (0.026) & (0.126) & & \tabularnewline
\vspace{6pt} & {[}0.000{]} & {[}0.007{]} & & \tabularnewline
LD {*} Second & -0.332{*} & 0.125 & & \tabularnewline
 & (0.160) & (0.132) & & \tabularnewline
\vspace{6pt} & {[}0.038{]} & {[}0.343{]} & & \tabularnewline
Second {*} Mixed & -0.451{*}{*}{*} & 0.295{*}{*}{*} & & \tabularnewline
 & (0.025) & (0.030) & & \tabularnewline
\vspace{6pt} & {[}0.000{]} & {[}0.000{]} & & \tabularnewline
LD {*} Second {*} Mixed & -0.033 & -0.577{*}{*} & & \tabularnewline
 & (0.022) & (0.183) & & \tabularnewline
\vspace{6pt} & {[}0.140{]} & {[}0.002{]} & & \tabularnewline
Constant & 0.582{*}{*}{*} & 0.992{*}{*}{*} & 0.836{*}{*}{*} & 0.975{*}{*}{*} \tabularnewline
 & (0.128) & (0.237) & (0.069) & (0.283) \tabularnewline
\vspace{6pt} & {[}0.000{]} & {[}0.000{]} & {[}0.000{]} & {[}0.001{]} \tabularnewline
\midrule 
Observations & 1,920 & 2,880 & 1,920 & 2,880 \tabularnewline
\midrule
\multicolumn{5}{l}{{*}{*}{*} p\textless 0.01, {*}{*} p\textless 0.05, {*} p\textless 0.1}\tabularnewline
\end{tabular}
}
\end{singlespace}
\centering{}\caption{\emph{Determinants of delegation and abstention. }Probit models. Standard
errors in parentheses, clustered at the session level. P-values in
brackets. Delegation/abstention is measured as a binary 0-1 subject
decision. The values for signal precision and round have been scaled
to be between 0 and 1. \label{tab:Determinants-of-delegation-n5}}
\end{table}

Table \ref{tab:Determinants-of-delegation-n5} reports the probit regressions
on the determinants of delegation and abstention in Experiment 1 with
the set of controls for order and treatment composition of the relevant
session. The excluded case is DD; and in Design 1, DD played as first treatment in DD-only sessions. In Design 1, delegation is lower in sessions where
LD is experienced after DD, regardless of group size, but the net
effect of delegation remains positive. Abstention responds to order
in DD-only sessions, and the effect depends on group size.

Table \ref{Table:Exp2.Del.or.Abs.Probit} presents the corresponding
fractional probit estimates for Experiment 2. Here again the excluded case is DD. These estimates confirm
the results of the linear probability model in the text. Accuracy
is not a significant predictor of participation in voting.

\begin{table}[ph]
\begin{centering}
\begin{tabular}{>{\raggedright}p{3.1cm}>{\centering}p{3.1cm}>{\centering}p{3.1cm}}
\multicolumn{3}{c}{Experiment 2: Frequency of Delegation or Abstention. }\tabularnewline
\multicolumn{3}{c}{Fractional Probit Model.}\tabularnewline
\midrule 
 & (1) & (2)\tabularnewline
 & N=5 \& N=15 & N=125\tabularnewline
\midrule 
Accuracy & -0.339 & 0.013\tabularnewline
 & (0.228) & (0.269)\tabularnewline
 & {[}0.138{]} & {[}0.962{]}\tabularnewline
 &  & \tabularnewline
LD & 0.604{*}{*}{*} & 0.577{*}{*}{*}\tabularnewline
 & (0.103) & (0.110)\tabularnewline
 & {[}0.000{]} & {[}0.000{]}\tabularnewline
 &  & \tabularnewline
N=15 & 0.017 & \tabularnewline
 & (0.105) & \tabularnewline
 & {[}0.871{]} & \tabularnewline
 &  & \tabularnewline
Keys: {[}E{]}{[}Y{]} & -0.025 & 0.075\tabularnewline
 & (0.104) & (0.110)\tabularnewline
 & {[}0.809{]} & {[}0.496{]}\tabularnewline
 &  & \tabularnewline
Block & 0.001 & 0.025\tabularnewline
 & (0.032) & (0.034)\tabularnewline
 & {[}0.981{]} & {[}0.460{]}\tabularnewline
 &  & \tabularnewline
Constant & -0.414{*}{*} & -0.534{*}{*}{*}\tabularnewline
 & (0.180) & (0.180)\tabularnewline
 & {[}0.022{]} & {[}0.003{]}\tabularnewline
 &  & \tabularnewline
\midrule 
Observations & 1,800 & 1,500\tabularnewline
\midrule
\multicolumn{3}{l}{{*}{*}{*} p\textless 0.01, {*}{*} p\textless 0.05, {*} p\textless 0.1}\tabularnewline
\end{tabular}
\par\end{centering}
\centering{}\caption{\textit{Frequency of delegation or abstention}. Fractional probit
models. Standard errors are clustered at the individual level. P-values
in brackets. Delegation/abstention is measured as the share of rounds
in a given block in which a subject chose to delegate/abstain (with
a range from 0 to 1). Accuracy is the share of rounds in the block
that subject answered correctly. Subjects randomly use either keys
{[}V{]} and {[}N{]} or {[}E{]} and {[}Y{]} to decide whether to vote;
a dummy for being assigned {[}E{]}{[}Y{]} is included. The values
for block have been scaled to be between 0 and 1; the coefficient for
\textquotedblleft block\textquotedblright{} thus indicates the effect
of going from the first to last block. \label{Table:Exp2.Del.or.Abs.Probit}}
\end{table}

\FloatBarrier

\subsubsection{Frequency of delegation and abstention. First treatments only}

We report here (Tables \ref{tab:Determinants-of-delegation-firsttreat-n5}
and \ref{tab:Determinants-of-delegation-firsttreat-n15}) linear probability
and probit regressions of the individual decision to delegate or abstain,
selecting data from the first treatment played in each session only (and thus, only for Design 1).
That is, corresponding to an in-between subjects design with 20 rounds
only.

\begin{table}[p]
\begin{centering}
\begin{tabular}{>{\raggedright}p{6cm}>{\centering}p{3.8cm}>{\centering}p{3.8cm}}
\multicolumn{3}{c}{Experiment 1: Frequency of Delegation or Abstention, First Treatments.
N=5.}\tabularnewline
\midrule 
 & (1) & (2)\tabularnewline
 & Linear Probability & Probit\tabularnewline
\midrule 
LD & 0.459{*}{*} & 1.450{*}{*}{*}\tabularnewline
 & (0.081) & (0.288)\tabularnewline
 & {[}0.011{]} & {[}0.000{]}\tabularnewline
 &  & \tabularnewline
Round & -0.005 & -0.022\tabularnewline
 & (0.020) & (0.079)\tabularnewline
 & {[}0.816{]} & {[}0.778{]}\tabularnewline
 &  & \tabularnewline
Signal Precision & -0.668{*}{*}{*} & -2.295{*}{*}{*}\tabularnewline
 & (0.107) & (0.423)\tabularnewline
 & {[}0.008{]} & {[}0.000{]}\tabularnewline
 &  & \tabularnewline
LD {*} Round & -0.127{*}{*} & -0.459{*}{*}{*}\tabularnewline
 & (0.024) & (0.117)\tabularnewline
 & {[}0.013{]} & {[}0.000{]}\tabularnewline
 &  & \tabularnewline
LD {*} Signal Precision & -0.322{*} & -0.890{*}\tabularnewline
 & (0.111) & (0.456)\tabularnewline
 & {[}0.062{]} & {[}0.051{]}\tabularnewline
 &  & \tabularnewline
Constant & 0.639{*}{*}{*} & 0.506{*}{*}{*}\tabularnewline
 & (0.056) & (0.172)\tabularnewline
 & {[}0.001{]} & {[}0.003{]}\tabularnewline
 &  & \tabularnewline
\midrule 
Observations & 960 & 960\tabularnewline
R-squared & 0.330 & \tabularnewline
\midrule
\multicolumn{3}{l}{{*}{*}{*} p\textless 0.01, {*}{*} p\textless 0.05, {*} p\textless 0.1}\tabularnewline
\end{tabular}
\par\end{centering}
\centering{}\caption{\emph{Determinants of delegation and abstention; N=5. Design 1, first treatments
only}. Standard errors are clustered at the session level. \label{tab:Determinants-of-delegation-firsttreat-n5}}
\end{table}

\begin{table}[p]
\begin{centering}
\begin{tabular}{>{\raggedright}p{6cm}>{\centering}p{3.8cm}>{\centering}p{3.8cm}}
\multicolumn{3}{c}{Experiment 1: Frequency of Delegation or Abstention, First Treatments.
N=15.}\tabularnewline
\midrule 
 & (1) & (2)\tabularnewline
 & Linear Probability & Probit\tabularnewline
\midrule 
LD & 0.174 & 0.610\tabularnewline
 & (0.105) & (0.375)\tabularnewline
 & {[}0.160{]} & {[}0.104{]}\tabularnewline
 &  & \tabularnewline
Round & 0.078 & 0.263\tabularnewline
 & (0.075) & (0.261)\tabularnewline
 & {[}0.348{]} & {[}0.314{]}\tabularnewline
 &  & \tabularnewline
Signal Precision & -0.868{*}{*}{*} & -2.648{*}{*}{*}\tabularnewline
 & (0.045) & (0.129)\tabularnewline
 & {[}0.000{]} & {[}0.000{]}\tabularnewline
 &  & \tabularnewline
LD {*} Round & -0.048 & -0.162\tabularnewline
 & (0.077) & (0.270)\tabularnewline
 & {[}0.557{]} & {[}0.548{]}\tabularnewline
 &  & \tabularnewline
LD {*} Signal Precision & 0.077 & 0.202\tabularnewline
 & (0.074) & (0.336)\tabularnewline
 & {[}0.351{]} & {[}0.547{]}\tabularnewline
 &  & \tabularnewline
Constant & 0.835{*}{*}{*} & 0.977{*}{*}{*}\tabularnewline
 & (0.094) & (0.297)\tabularnewline
 & {[}0.000{]} & {[}0.001{]}\tabularnewline
 &  & \tabularnewline
\midrule 
Observations & 1,440 & 1,440\tabularnewline
R-squared & 0.301 & \tabularnewline
\midrule
\multicolumn{3}{l}{{*}{*}{*} p\textless 0.01, {*}{*} p\textless 0.05, {*} p\textless 0.1}\tabularnewline
\end{tabular}
\par\end{centering}
\caption{\emph{Determinants of delegation and abstention; N=15. Design 1, first treatments
only}. Standard errors are clustered at the session level. \label{tab:Determinants-of-delegation-firsttreat-n15}}
\end{table}

Standard errors are clustered at the session level. As discussed in
the text, the results are effectively unchanged for the small groups
($N=5$), but there is a loss of precision in the $N=15$ regressions.

\subsubsection{Monotonicity violations and individual thresholds}\label{subsec:appendix_monotonicity}

\begin{figure}[!htbp]
\begin{centering}
\includegraphics[scale=.55]{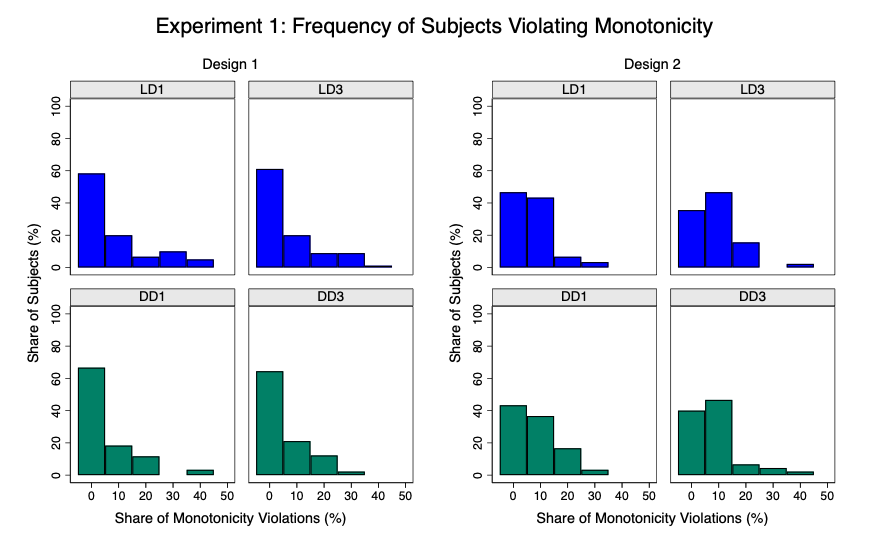}
\includegraphics[scale=.55]{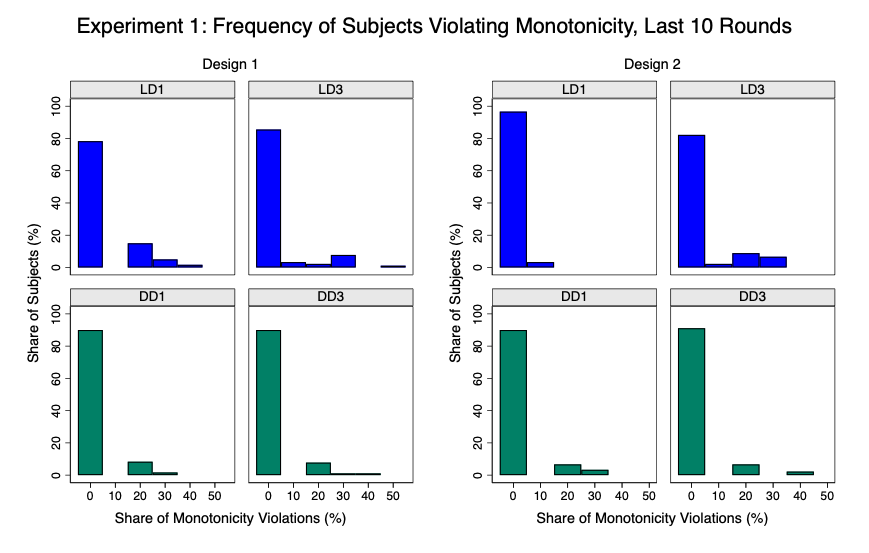}
\par\end{centering}
\caption{\emph{Monotonicity violations. Histograms}. For each subject, we calculate
the frequency of violations given the number of rounds played as non-expert.
The maximum possible frequency is 50\%. \label{fig:monotonicity}}
\end{figure}

As described in the text, for both group sizes, in Design 1, just below 60\% of
subjects have no violations at all under LD; just above 60\% under
DD. In Design 2, 40\% of subjects have no violations under LD, and just above 40\% under DD. (Subjects play twice as many rounds of the same treatment in Design 2, so there are more opportunities for a monotonicity violation to occur.)
Figure \ref{fig:monotonicity} reports histograms of the frequency
of monotonicity violations, by subject.

\begin{figure}[!htbp]
\centering{}\includegraphics[scale=.65]{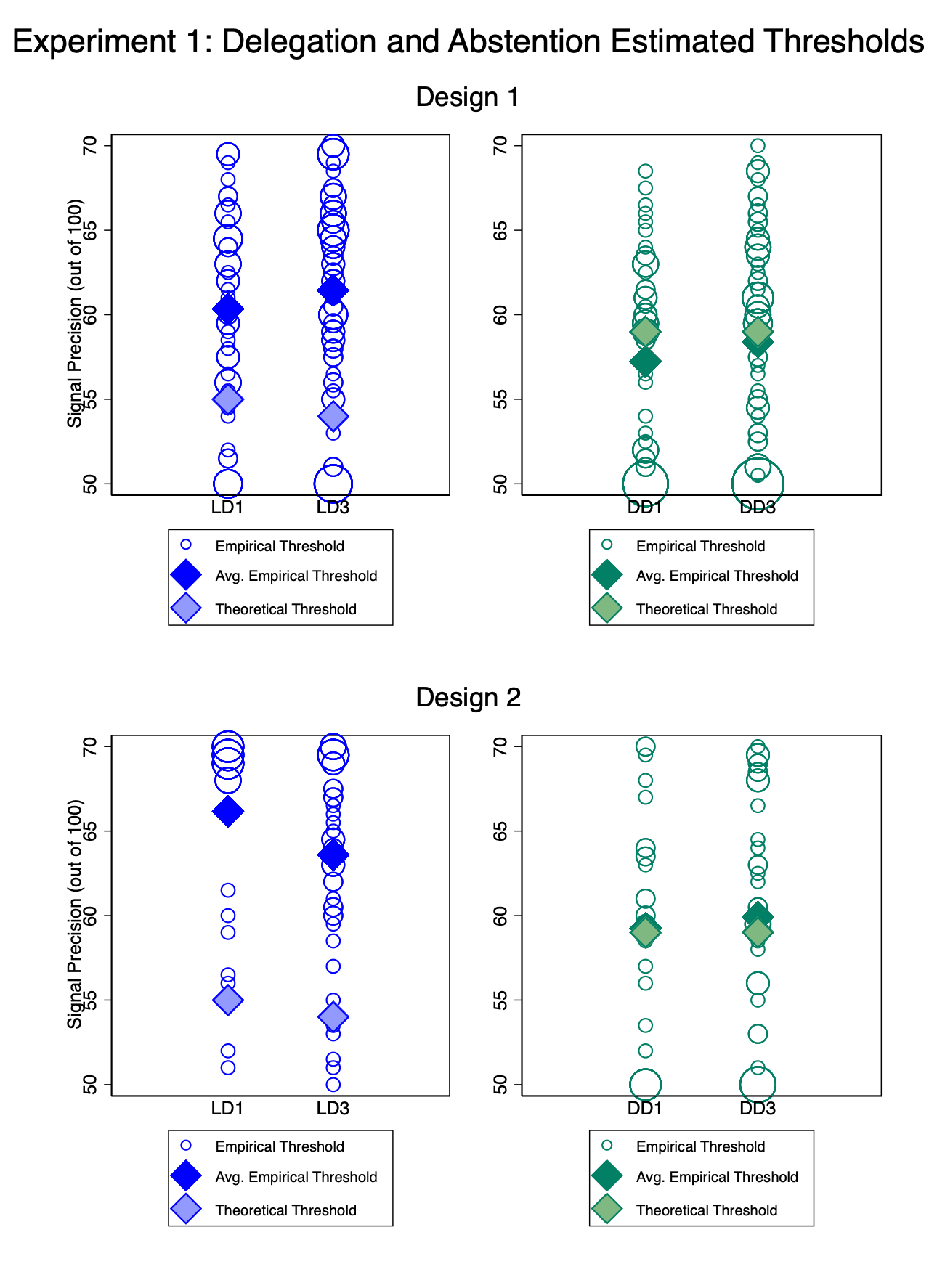}\caption{\textit{Individual delegation and abstention thresholds}. \label{fig:thresholds}}
\end{figure}

We can use monotonicity to estimate individual precision thresholds
for delegation and abstention---the thresholds below which each participant
delegates or abstains. Figure \ref{fig:thresholds} reports, for each
participant, the mean of the range of thresholds that are consistent
with minimal monotonicity violations; the size of the dots is proportional
to the number of participants at that threshold. The dark blue (for
LD) and dark green (for DD) diamonds correspond to the average empirical
thresholds, and the respective light ones to the theory. The figure
confirms the over-delegation that characterizes LD, while again average
values for abstention are close to the theoretical predictions. The
dispersion in estimated thresholds is typical of similar experiments
(for example, Levine and Palfrey, 2007; Morton and Tyran, 2011), but
is in clear tension with the focus on symmetric equilibria.\textbf{ }

\subsubsection{Voting against signal and signal quality}

Table \ref{tab:Frequency-of-voting-against} reports linear probability
and probit regressions studying the determinants of voting against
signal. Across both designs, the only significant variable is signal quality (precision).
Neither the voting system nor the size of the group matter, but a
decrease in signal quality is strongly correlated with an increase
in voting against signal. Although voting against signal is an inferior
action, the loss from doing so is indeed increasing in signal quality.
There could be a small learning effect---in Design 1, voting against signal is less
frequent in the treatments run second---but it is quite noisy.

\begin{table}[p]
\begin{singlespace}
\centering{}%
\resizebox{\textwidth}{!}{
\begin{tabular}{l>{\centering}p{2.8cm}>{\centering}p{2.8cm}>{\centering}p{2.8cm}>{\centering}p{2.8cm}}
\multicolumn{5}{c}{Experiment 1: Frequency of Voting Against Signal.}\tabularnewline
\midrule 
 & \multicolumn{2}{c}{\rule{0pt}{1em}Design 1} & \multicolumn{2}{c}{Design 2} \\
 \cline{2-3} \cline{4-5}
 & \rule{0pt}{1em}(1) & (2) & (3) & (4)\tabularnewline
 & Linear Probability & Probit & Linear Probability & Probit \tabularnewline
\midrule 
Signal Precision & -0.405{*}{*}{*} & -1.993{*}{*}{*} & -0.202{*}{*}{*} & -1.288{*}{*}{*} \tabularnewline
 & (0.035) & (0.174) & (0.028) & (0.123) \tabularnewline
\vspace{6pt} & {[}0.000{]} & {[}0.000{]} & {[}0.000{]} & {[}0.000{]} \tabularnewline
Round & 0.000 & -0.007 & -0.041 & -0.240 \tabularnewline
 & (0.011) & (0.067) & (0.026) & (0.159) \tabularnewline
\vspace{6pt} & {[}0.988{]} & {[}0.918{]} & {[}0.143{]} & {[}0.132{]} \tabularnewline
LD & -0.009 & -0.061 & 0.022 & 0.123 \tabularnewline
 & (0.012) & (0.069) & (0.024) & (0.146) \tabularnewline
\vspace{6pt} & {[}0.466{]} & {[}0.381{]} & {[}0.394{]} & {[}0.399{]} \tabularnewline
N = 15 & -0.009 & -0.080 & -0.022 & -0.132 \tabularnewline
 & (0.029) & (0.178) & (0.024) & (0.146) \tabularnewline
\vspace{6pt} & {[}0.771{]} & {[}0.652{]} & {[}0.399{]} & {[}0.366{]} \tabularnewline
Second & -0.020 & -0.107 & & \tabularnewline
 & (0.016) & (0.085) & & \tabularnewline
\vspace{6pt} & {[}0.237{]} & {[}0.210{]} & & \tabularnewline
Second {*} Mixed & -0.037 & -0.214 & & \tabularnewline
 & (0.025) & (0.179) & & \tabularnewline
\vspace{6pt} & {[}0.181{]} & {[}0.232{]} & & \tabularnewline
N = 15 {*} Second & 0.034 & 0.189 & & \tabularnewline
 & (0.030) & (0.185) & & \tabularnewline
\vspace{6pt} & {[}0.284{]} & {[}0.308{]} & & \tabularnewline
N = 15 {*} Mixed & -0.004 & 0.009 & & \tabularnewline
 & (0.041) & (0.264) & & \tabularnewline
\vspace{6pt} & {[}0.925{]} & {[}0.974{]} & & \tabularnewline
Constant & 0.407{*}{*}{*} & 0.081 & 0.227{*}{*}{*} & -0.582{*}{*}{*} \tabularnewline
 & (0.031) & (0.094) & (0.027) & (0.135) \tabularnewline
\vspace{6pt} & {[}0.000{]} & {[}0.388{]} & {[}0.000{]} & {[}0.009{]} \tabularnewline
\midrule 
Observations & 2,552 & 2,552 & 4,800 & 4,800 \tabularnewline
R-squared & 0.121 & & 0.045 & \tabularnewline
\midrule
\multicolumn{5}{l}{{*}{*}{*} p\textless 0.01, {*}{*} p\textless 0.05, {*} p\textless 0.1}\tabularnewline
\end{tabular}
}
\end{singlespace}
\caption{\emph{Frequency of voting against signal. }Standard errors are clustered
at the session level. When precision is 50, all votes are considered sincere. Experts excluded. In Design 1, delegators/abstainers are excluded. In Design 2, if a non-expert delegates/abstains, their MV choice is used. \label{tab:Frequency-of-voting-against}}
\end{table}

\FloatBarrier
\pagebreak{}
\setcounter{page}{1}
\section{Online Appendix\label{sec:online-Appendix-B}}

\setcounter{figure}{0} \renewcommand{\thefigure}{B.\arabic{figure}}
\setcounter{table}{0} \renewcommand{\thetable}{B.\arabic{table}}

\subsection{Robustness}\label{subsec:robustness}

\begin{figure}[h]
\begin{centering}
\includegraphics[scale=0.5]{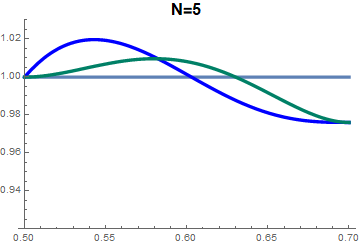}
\includegraphics[scale=0.5]{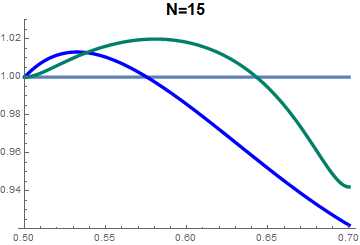}
\par\end{centering}
\centering{}\caption{\textit{Robustness to strategic errors}. The horizontal axis is $\widetilde{q}$, the vertical axis is the probability of reaching
the correct decision, relative to MV. Blue is LD, green is DD, and
gray is MV. $F$ is Uniform over $[1/2,p]$, and experts' precision $p$ is fixed at $0.7$ in both panels\label{fig.robustness}}
\end{figure}

In the lab, as in life, some deviation from optimal strategies is
highly likely. For both LD and DD, how robust are potential improvements
over MV to strategic mistakes? We consider here a particularly simple
parametrization of strategic mistakes: we suppose that behavior remains
symmetric within the group of non-experts, but the precision threshold for delegation or abstention
is chosen incorrectly. In Figure \ref{fig.robustness}, the horizontal
axis is the threshold , and the vertical axis reports gains
and losses in expected utilities relative to MV (fixed at 1). Thus
the plots depict the percentage changes in the probability of the
group making the correct choice, relative to MV, at different delegation
or abstention thresholds. We plot the results for both rules in the same graph (LD in blue; DD in green), but there is no presumption that $\widetilde{q}$ be equal across LD and DD; for any $\widetilde{q}$, the blue (green) curve reports the gain or loss relative to MV for LD (DD). The
first panel corresponds to $N=5$, $K=1$; the second to $N=15$,
$K=3$. The highest points on the blue and green curves coincide with
the respective equilibrium thresholds. The declines away from the
highest points capture the cost of strategic mistakes: delegating
or abstaining too much (at too high thresholds of precisions, relative
to equilibrium, or to the right of the highest points), or too little
(voting at too low precision, or to the left of the highest points). 

At $\widetilde{q}=1/2$, no-one delegates
their vote or abstains, and all curves equal MV and coincide. At $\widetilde{q}=0.7$, all non-experts delegate or abstain,
and only the expert(s) decide(s).\footnote{When $K=3$, the blue and green curves do not coincide at $\widetilde{q}=0.7$
because under DD3 each expert has the same weight, while under LD3
the number of votes each of them commands is stochastic.} In the first panel, with a small group, the maximum potential improvement
over MV from delegation (from LD) is higher than from abstention (DD).
However, this is not true in the second panel, with the larger group.
The two results were already shown in Tables 1 and 2 in the text; here the figure
depicts them graphically. It adds to the earlier tables the range
of thresholds for which each voting rule dominates MV. Here the message
is consistent across the two group sizes: in both cases, the range
of thresholds that deliver improvements over MV is limited, and particularly
limited for LD. When the group is larger, LD's potential for losses
is evident in the figure, as is its increased fragility, relative
to DD: the range of thresholds that improve over MV is half as large
under LD3 than under DD3. With both voting schemes, but with LD in
particular, while potential gains are small, there is the real danger
of reaching worse decisions: under LD3, maximal potential losses are
more than six times maximal potential gains.

\FloatBarrier

\subsection{The bootstrapping procedure: allowing for individual correlation across rounds}\label{subsec:online_app_bootstrapping}

Replicating what happens in an individual session, we draw with replacement
15 subjects from the relevant treatment, each with all choices made
over the 20 rounds (or 40 rounds in Design 2). Among these 15 subjects, we draw, with replacement,
3 subjects, assigning to each of them one choice they made as expert,\footnote{If the subject was never an expert, the subject is dropped and another
one is drawn.} and 12 subjects, assigning to each one choice made as non-expert.
In $N=15$, that constitutes the group and yields one group decision;
in $N=5$ treatments, we divide the 12 subjects randomly into three
groups of 4 and assign to each group one of the experts drawn earlier,
generating three group decisions. We repeat new draws of 3 experts
and 12 non-experts as above 20 (40) times, generating 20 (40) decisions from
the same sample of 15 subjects if the treatment has $N=15$, and 60 (120)
if the treatment has $N=5$, thus simulating one experimental session.
We then draw, always with replacement, a new group of 15 subjects,
and repeat the procedure, each time generating 20 (40) rounds from
the same group of 15 subjects. 
For Design 1, we repeat the whole procedure 4 times for $N=5$ or 6 times for $N=15$; for Design 2, we repeat the procedure 2 times for $N=5$ or 3 times for $N=15$. 
For both designs, we generate 240 decisions for $N=5$ and 120 decisions for $N=15$, as in our data for each of the treatments.
We then calculate the frequency of correct decisions, and consider that one data point for
that treatment. 
We repeat the whole process 100,000 times and generate a distribution of the frequency with which the correct decision was reached.

\subsection{A short note on beliefs}

\begin{figure}[h]
\begin{centering}
\includegraphics[scale=0.3]{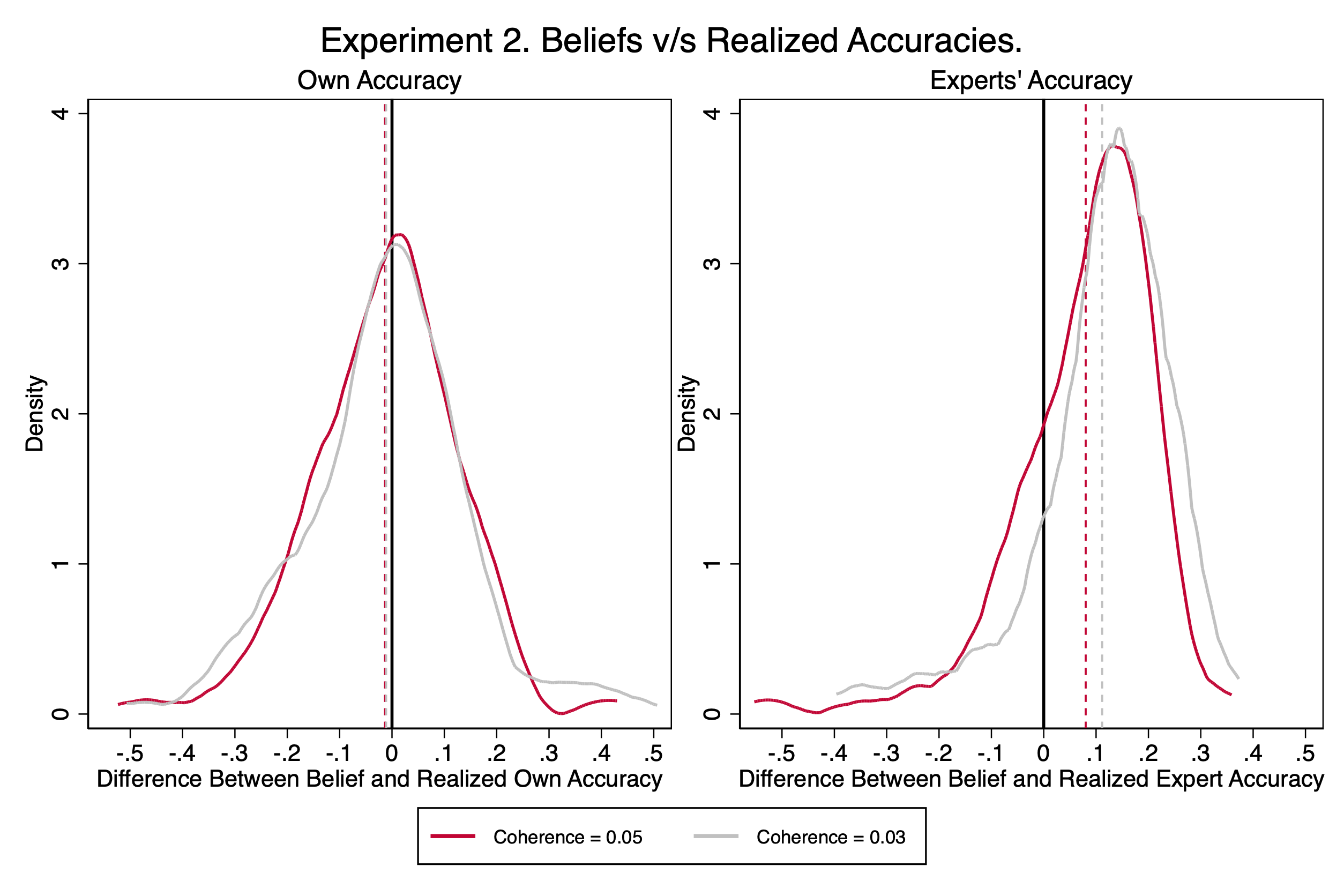}
\par\end{centering}
\caption{\emph{Difference between actual and reported accuracies}. The left
panel refers to own accuracy; the right panel to the average of the
experts' accuracy. The dashed vertical lines correspond to the means.\label{fig:Difference-between-actual} }
\end{figure}

As reported in the text, at the end of the RDK experiment, we asked:
``On average what percentage of trials in the second part do you
think you got right?'' and ``On average what percentage of trials
in the second part do you think the experts got right?'' Participants
earned 25 cents for replies the fell within 5\% of the observed percentage
in the group to which the participant was assigned. Figure \ref{fig:Difference-between-actual}
plots distributions of the difference between actual own accuracy
and the corresponding reported belief, in the left panel, and between
actual average experts' accuracy and the corresponding reported belief,
in the panel on the right, for the two different coherence levels.
Both panels rely on a single measure of beliefs per subject, and thus
the unit of analysis is the subject. 

As the figure shows, accuracy of beliefs varies little with coherence.
On average, beliefs about own accuracy are remarkably correct, with
barely detectable underestimation. The distribution is instead shifted
to the right in the case of experts, with about 15\% average overestimation,
imputable, as we say in the text, to stochasticity in accuracies and
reversion to the mean, neglected in the formation of beliefs. Overestimation
is somewhat larger at smaller coherence, when the task is more difficult. 

Table \ref{tab:Beliefs-and-the} reports variants of the regressions
in Table \ref{Table:Exp2.Del.or.Abs.LPM} in the text, now including
measures of beliefs. The dependent variable is the frequency of delegation
or abstention at the subject level (120 tasks per subject), with data
aggregated by coherence level.

\begin{table}[ph]
\begin{centering}
\begin{tabular}{l>{\centering}p{3.5cm}>{\centering}p{3.5cm}}
\multicolumn{3}{c}{Experiment 2: Frequency of Delegation or Abstention.}\tabularnewline
\midrule 
 & (1) & (2)\tabularnewline
 & Coherence = 0.05 & Coherence = 0.03\tabularnewline
\midrule 
Own Accuracy & -0.250 & -0.014\tabularnewline
 & (0.288) & (0.334)\tabularnewline
 & {[}0.387{]} & {[}0.968{]}\tabularnewline
 &  & \tabularnewline
Experts' Accuracy & -0.282 & \tabularnewline
 & (0.389) & \tabularnewline
 & {[}0.469{]} & \tabularnewline
 &  & \tabularnewline
Belief About Own Accuracy & -0.714{*}{*}{*} & -0.607{*}{*}{*}\tabularnewline
 & (0.144) & (0.161)\tabularnewline
 & {[}0.000{]} & {[}0.000{]}\tabularnewline
 &  & \tabularnewline
Belief About Experts' Accuracy & 0.337{*}{*} & 0.615{*}{*}{*}\tabularnewline
 & (0.160) & (0.181)\tabularnewline
 & {[}0.037{]} & {[}0.001{]}\tabularnewline
 &  & \tabularnewline
LD & 0.249{*}{*}{*} & 0.230{*}{*}{*}\tabularnewline
 & (0.037) & (0.041)\tabularnewline
 & {[}0.000{]} & {[}0.000{]}\tabularnewline
 &  & \tabularnewline
Constant & 0.757{*}{*}{*} & 0.223\tabularnewline
 & (0.269) & (0.194)\tabularnewline
 & {[}0.005{]} & {[}0.252{]}\tabularnewline
 &  & \tabularnewline
\midrule 
Observations & 300 & 250\tabularnewline
R-squared & 0.184 & 0.163\tabularnewline
\midrule
\multicolumn{3}{l}{{*}{*}{*} p\textless 0.01, {*}{*} p\textless 0.05, {*} p\textless 0.1}\tabularnewline
\end{tabular}
\par\end{centering}
\centering{}\caption{\emph{Beliefs and the decision to delegate/abstain}. Linear probability
models. Expert accuracy is excluded from coherence 0.03 because, with
a single group per treatment, it is collinear with LD. \label{tab:Beliefs-and-the} }
\end{table}

As described in the text, beliefs are strongly significant and have
the expected signs but cannot explain the higher tendency towards
delegation rather than abstention. 

\subsection{The Random Dot Kinematogram}\label{subsec:online_app_rdk}

In a Random Dot Kinematogram (RDK), the perceptual stimulus consists
of a number of dots being displayed on a screen. A proportion of these
dots are determined to be signal dots, while the remaining are noise
dots. Signal dots all move in a determined direction, while noise
dots move at random according to an algorithm. The task consists in
reporting the direction in which the signal dots are moving. This
direction is called the coherent direction and the proportion of signal
dots, the coherence, is the main factor in determining the difficulty
of the task. 

The task can be programmed in various ways, using a variety of parameters
(e.g. color, duration, algorithm, number of dots). Research has been
done to study the various effects of using different combinations
of these parameters (Pilly and Seitz, 2009; Schütz et al., 2010).
We take advantage of the recent development of a customizable version
of the RDK (Rajananda et al., 2018) which can be implemented as a
plugin in jsPsych. This version allows for the configuration of various
parameters in order to adjust the task as desired by the researchers.
We report in Table \ref{tab:rdk-parameters} the parameters that we used in our experiment.
The reader can find details about how they affect the task in Rajananda
et al., 2018 and in the following link: \href{https://www.jspsych.org/6.3/plugins/jspsych-rdk/}{https://www.jspsych.org/6.3/plugins/jspsych-rdk/}.
It is important to emphasize again that our objective in using the
RDK was not to study perception in itself, but rather to create a
common task that is reasonably well controlled and calibrated and
where, nevertheless, the information about the accuracy of the signals
remains ambiguous. 

\begin{table}[h]
\begin{centering}
\begin{tabular}{|l|l|}
\hline 
Duration: & 1 second\tabularnewline
\hline 
Directions: & Left/Right\tabularnewline
\hline 
Number of dots: & 300\tabularnewline
\hline 
Background color: & Black\tabularnewline
\hline 
Color of dots: & White\tabularnewline
\hline 
Dot radius: & 2 pixels\tabularnewline
\hline 
Dot movement per frame: & 1 pixel\tabularnewline
\hline 
Aperture width: & 600 pixels\tabularnewline
\hline 
Aperture height: & 400 pixels\tabularnewline
\hline 
Signal selection: & Same\tabularnewline
\hline 
Minimal screen resolutions & 1000x600 pixels\tabularnewline
\hline 
Noise type: & Random direction\tabularnewline
\hline 
Aperture shape: & Ellipse\tabularnewline
\hline 
Reinsertion: & Dots reappear randomly when hitting edge\tabularnewline
\hline 
Fixation cross: & No\tabularnewline
\hline 
Aperture border: & No\tabularnewline
\hline 
Coherence: & 20\% to 3\% (according to treatment)\tabularnewline
\hline 
\end{tabular}
\par\end{centering}
\caption{\emph{Experiment 2: RDK parameters}}
\label{tab:rdk-parameters}
\end{table}

\FloatBarrier

\subsection{Discussion: Some additional regressions}\label{sec:online_app_discussion}

We report below regressions discussed in Section \ref{subsec:disc-del-vs-abs}. The first set (Tables \ref{tab:response_to_prev_round1}--\ref{tab:response_to_prev_round2})
tests the sensitivity of delegation/abstention to one-period lagged
delegation/abstention by others. As discussed in the text, if participants
believe that others are voting at too low precisions, delegation should
respond negatively to lagged delegations by others. We see no convincing
evidence in the data.\footnote{We do not estimate probit regressions to avoid biases due to lagged
dependent variables. } 

\begin{table}[p]
\begin{centering}
\begin{tabular}{lcccc}
\multicolumn{5}{c}{Experiment 1, Design 1: Frequency of Delegation or Abstention. N=5 and N=15.}\tabularnewline
\midrule 
 & (1) & (2) & (1) & (2)\tabularnewline
 & LD \& N=5 & LD \& N=15 & DD \& N=5 & DD \& N=15\tabularnewline
\midrule 
\# Delegations/Abstentions & -0.027 & -0.013 & 0.008 & -0.012\tabularnewline
by Others in Prev. Round & (0.033) & (0.015) & (0.021) & (0.010)\tabularnewline
 & {[}0.481{]} & {[}0.411{]} & {[}0.737{]} & {[}0.267{]}\tabularnewline
 &  &  &  & \tabularnewline
Signal Precision & -0.866{*}{*}{*} & -0.849{*}{*}{*} & -0.782{*}{*}{*} & -0.865{*}{*}{*}\tabularnewline
 & (0.079) & (0.046) & (0.085) & (0.049)\tabularnewline
 & {[}0.002{]} & {[}0.000{]} & {[}0.003{]} & {[}0.000{]}\tabularnewline
 &  &  &  & \tabularnewline
Second & 0.071 & -0.065{*}{*} & 0.144{*}{*}{*} & -0.108{*}\tabularnewline
 & (0.068) & (0.019) & (0.010) & (0.045)\tabularnewline
 & {[}0.375{]} & {[}0.018{]} & {[}0.001{]} & {[}0.063{]}\tabularnewline
 &  &  &  & \tabularnewline
Second {*} Mixed & -0.159{*}{*}{*} & -0.100 & -0.118{*}{*}{*} & 0.090{*}{*}{*}\tabularnewline
 & (0.026) & (0.068) & (0.010) & (0.011)\tabularnewline
 & {[}0.009{]} & {[}0.200{]} & {[}0.001{]} & {[}0.000{]}\tabularnewline
 &  &  &  & \tabularnewline
Round & -0.087 & 0.022 & 0.018 & 0.092\tabularnewline
 & (0.051) & (0.026) & (0.065) & (0.068)\tabularnewline
 & {[}0.183{]} & {[}0.433{]} & {[}0.805{]} & {[}0.235{]}\tabularnewline
 &  &  &  & \tabularnewline
Constant & 1.055{*}{*}{*} & 1.131{*}{*}{*} & 0.681{*}{*}{*} & 0.883{*}{*}{*}\tabularnewline
 & (0.039) & (0.089) & (0.056) & (0.100)\tabularnewline
 & {[}0.000{]} & {[}0.000{]} & {[}0.001{]} & {[}0.000{]}\tabularnewline
 &  &  &  & \tabularnewline
\midrule 
Observations & 912 & 1,368 & 912 & 1,368\tabularnewline
R-squared & 0.296 & 0.290 & 0.267 & 0.302\tabularnewline
\midrule
\multicolumn{3}{l}{{*}{*}{*} p\textless 0.01, {*}{*} p\textless 0.05, {*} p\textless 0.1} &  & \tabularnewline
\end{tabular}
\par\end{centering}
\centering{}\caption{\emph{Response of delegation/abstention to number of others who delegated/abstained
in group in previous round; N=5 and N=15. }Design 1. Linear probability models.
Standard errors in parentheses, clustered at the session level. \label{tab:response_to_prev_round1}}
\end{table}

\begin{table}[p]
\begin{centering}
\begin{tabular}{lcccc}
\multicolumn{5}{c}{Experiment 1, Design 2: Frequency of Delegation or Abstention. N=5 and N=15.}\tabularnewline
\midrule 
 & (1) & (2) & (1) & (2)\tabularnewline
 & LD \& N=5 & LD \& N=15 & DD \& N=5 & DD \& N=15\tabularnewline
\midrule 
\# Delegations/Abstentions & -0.004 & -0.010 & 0.0357{*}{*} & 0.0173\tabularnewline
by Others in Prev. Round & (0.006) & (0.005) & (0.003) & (0.020)\tabularnewline
 & {[}0.648{]} & {[}0.191{]} & {[}0.050{]} & {[}0.471{]}\tabularnewline
 &  &  &  & \tabularnewline
Signal Precision & -0.478{*} & -0.708{*}{*} & -0.810{*}{*} & -0.775{*}{*}{*} \tabularnewline
 & (0.057) & (0.084) & (0.037) & (0.048)\tabularnewline
 & {[}0.076{]} & {[}0.014{]} & {[}0.029{]} & {[}0.004{]}\tabularnewline
 &  &  &  & \tabularnewline
Round & 0.092 & -0.002 & 0.090 & 0.058 \tabularnewline
 & (0.079) & (0.089) & (0.098) & (0.038)\tabularnewline
 & {[}0.452{]} & {[}0.980{]} & {[}0.525{]} & {[}0.270{]}\tabularnewline
 &  &  &  & \tabularnewline
Constant & 0.992{*}{*} & 1.116{*}{*}{*} & 0.747{*}{*} & 0.764{*} \tabularnewline
 & (0.041) & (0.038) & (0.014) & (0.181)\tabularnewline
 & {[}0.026{]} & {[}0.001{]} & {[}0.012{]} & {[}0.052{]}\tabularnewline
 &  &  &  & \tabularnewline
\midrule 
Observations & 936 & 1,404 & 936 & 1,404\tabularnewline
R-squared & 0.134 & 0.215 & 0.247 & 0.240\tabularnewline
\midrule
\multicolumn{3}{l}{{*}{*}{*} p\textless 0.01, {*}{*} p\textless 0.05, {*} p\textless 0.1} &  & \tabularnewline
\end{tabular}
\par\end{centering}
\centering{}\caption{\emph{Response of delegation/abstention to number of others who delegated/abstained
in group in previous round; N=5 and N=15. }Design 2. Linear probability models.
Standard errors in parentheses, clustered at the session level. \label{tab:response_to_prev_round2}}
\end{table}

The second set (Tables \ref{tab:high_precision_reg_n5}--\ref{tab:high_precision_reg_n15}) tests whether delegation is more frequent than abstention at high signal precisions (65\% or above). As discussed in the text, if participants focus only on their own delegation to the expert, delegation is an optimal choice at any precision less than $p$ (given that the number of experts is odd). On the other hand, this focus on only the expert may be less likely in DD, as abstention induces one to consider all those who do not abstain. Accordingly, if such a distortion is at work, at high signal precisions we should see higher rates of delegation than of abstention. Indeed, the tables show that across both designs of Experiment 1, delegation is more frequent than abstention at high precisions.

\begin{table}[p]
\begin{singlespace}
\begin{centering}
\resizebox{\textwidth}{!}{
\begin{tabular}{l>{\centering}p{2.8cm}>{\centering}p{2.8cm}>{\centering}p{2.8cm}>{\centering}p{2.8cm}}
\multicolumn{5}{c}{Experiment 1: Frequency of Delegation or Abstention at High Signal Precisions. N=5.}\tabularnewline
\midrule 
 & \multicolumn{2}{c}{\rule{0pt}{1em}Design 1} & \multicolumn{2}{c}{Design 2} \\
 \cline{2-3} \cline{4-5}
 & \rule{0pt}{1em}(1) & (2) & (3) & (4)\tabularnewline
\midrule 	
LD & 0.099** & 0.102*** & 0.468*** & 0.483*** \tabularnewline
 & (0.027) & (0.018) & (0.066) & (0.065) \tabularnewline
 \vspace{6pt} & {[}0.014{]} & {[}0.002{]} & {[}0.006{]} & {[}0.005{]} \tabularnewline			
Signal Precision & & -0.615** & & -0.929* \tabularnewline
 & & (0.207) & & (0.333) \tabularnewline
 \vspace{6pt} & & {[}0.031{]} & & {[}0.069{]} \tabularnewline
Second & & 0.070** & & \tabularnewline
 & & (0.027) & & \tabularnewline
 \vspace{6pt} & & {[}0.046{]} & & \tabularnewline
Second {*} Mixed & & -0.067** & & \tabularnewline
 & & (0.022) & & \tabularnewline
 \vspace{6pt} & & {[}0.028{]} & & \tabularnewline
Round & & -0.044 & & 0.040 \tabularnewline
 & & (0.052) & & (0.083) \tabularnewline
 \vspace{6pt} & & {[}0.429{]} & & {[}0.668{]} \tabularnewline
Constant & 0.051 & 0.596** & 0.133 & 0.921* \tabularnewline
 & (0.026) & (0.193) & (0.057) & (0.321) \tabularnewline
 \vspace{6pt} & {[}0.102{]} & {[}0.027{]} & {[}0.100{]} & {[}0.064{]} \tabularnewline
\midrule 
Observations & 548 & 548 & 556 & 556 \tabularnewline
R-squared & 0.028 & 0.070 & 0.240 & 0.269 \tabularnewline
\midrule
\multicolumn{5}{l}{{*}{*}{*} p\textless 0.01, {*}{*} p\textless 0.05, {*} p\textless 0.1}\tabularnewline
\end{tabular}
}
\par\end{centering}
\end{singlespace}
\centering{}\caption{\emph{Determinants of delegation or abstention at precisions of 65\% or above; N=5. } Linear probability models. Standard errors in parentheses, clustered at the session level. \label{tab:high_precision_reg_n5} }
\end{table}

\begin{table}[p]
\begin{singlespace}
\begin{centering}
\resizebox{\textwidth}{!}{
\begin{tabular}{l>{\centering}p{2.8cm}>{\centering}p{2.8cm}>{\centering}p{2.8cm}>{\centering}p{2.8cm}}
\multicolumn{5}{c}{Experiment 1: Frequency of Delegation or Abstention at High Signal Precisions. N=15.}\tabularnewline
\midrule 
 & \multicolumn{2}{c}{\rule{0pt}{1em}Design 1} & \multicolumn{2}{c}{Design 2} \\
 \cline{2-3} \cline{4-5}
 & \rule{0pt}{1em}(1) & (2) & (3) & (4)\tabularnewline
\midrule 	
LD & 0.106** & 0.104*** & 0.185** & 0.190** \tabularnewline
 & (0.033) & (0.005) & (0.067) & (0.067) \tabularnewline
 \vspace{6pt} & {[}0.014{]} & {[}0.000{]} & {[}0.039{]} & {[}0.036{]} \tabularnewline			
Signal Precision & & -0.658*** & & -1.339*** \tabularnewline
 & & (0.130) & & (0.327) \tabularnewline
 \vspace{6pt} & & {[}0.001{]} & & {[}0.009{]} \tabularnewline
Second & & -0.110*** & & \tabularnewline
 & & (0.024) & & \tabularnewline
 \vspace{6pt} & & {[}0.003{]} & & \tabularnewline
Second {*} Mixed & & 0.036 & & \tabularnewline
 & & (0.035) & & \tabularnewline
 \vspace{6pt} & & {[}0.345{]} & & \tabularnewline
Round & & 0.063 & & 0.028 \tabularnewline
 & & (0.059) & & (0.056) \tabularnewline
 \vspace{6pt} & & {[}0.325{]} & & {[}0.641{]} \tabularnewline
Constant & 0.083*** & 0.675*** & 0.171*** & 1.337*** \tabularnewline
 & (0.020) & (0.122) & (0.039) & (0.306) \tabularnewline
 \vspace{6pt} & {[}0.005{]} & {[}0.001{]} & {[}0.007{]} & {[}0.007{]} \tabularnewline
\midrule 
Observations & 850 & 850 & 816 & 816 \tabularnewline
R-squared & 0.024 & 0.070 & 0.044 & 0.111 \tabularnewline
\midrule
\multicolumn{5}{l}{{*}{*}{*} p\textless 0.01, {*}{*} p\textless 0.05, {*} p\textless 0.1}\tabularnewline
\end{tabular}
}
\par\end{centering}
\end{singlespace}
\centering{}\caption{\emph{Determinants of delegation or abstention at precisions of 65\% or above; N=15. } Linear probability models. Standard errors in parentheses, clustered at the session level. \label{tab:high_precision_reg_n15} }
\end{table}

\FloatBarrier

\subsection{Discussion: Three simple models}\label{sec:three_simple_models}

\subsubsection{Introducing private values}

As in our main model, we focus on symmetric pure equilibria with sincere
voting where the decision to delegate or abstain is symmetric across
signals. Because the preferences of the expert are publicly known,
such a decision typically conveys information about the voter's preferences.
But note that we have introduced private values while preserving the
symmetry of the model. Symmetric equilibria with sincere voting and
delegation/abstention decisions symmetric across signals continue
to exist. 

We say that two individuals are \emph{congruent} if they
share the same private preferences, and denote by $s$ the state of
being congruent with (``similar'' to) the expert. The set of strategies
is $\{x,x(s),x(-s),d/a\}$, where $x$ stands for voting unconditionally
on congruence, $x(s)$ for voting if congruent and delegating or abstaining
if not; $x(-s)$ stands for voting if non-congruent and delegating
or abstaining otherwise, and $d/a$ stands for delegating or abstaining
always. We begin by analyzing LD.

\paragraph{LD}

As always, for any $\{p,q\}$, there is an equilibrium in which both
non-experts delegate. Is there an equilibrium with voting always?

Consider the expected utility of non-expert voter $i$ who is congruent
with the expert. Suppose first that non-expert $j$ always votes,
whether congruent or not. Then:
\begin{align*}
EU_{i}(x|i_{s},x_{j}) & =pq+[(1-p)q+(1-q)p][q/2+(1-q)/2]=(p+q)/2\\
EU_{i}(d) & =p
\end{align*}
Consider $EU_{i}(x|i_{s},x_{j})$. Because $i$ is $s$, from $i's$
perspective, the correct decision is reached if both $i$ and the
expert have the correct signal (with probability $pq$), or if one
of them has the wrong signal (with probability $p(1-q)+q(1-p)$) but
the desired decision wins thanks to $j$'s vote---which happens if
$j$ is $s$ and has the correct signal (with probability $(1/2)q$),
or if $j$ is $\not s$ and has the wrong signal (with probability
$(1/2)(1-q)$). (All derivations below follow a similar logic).

With $p>q$, delegation dominates voting for the congruent non-expert.
Hence there is no pure symmetric equilibrium such that the two non-experts
always vote, whether congruent or not.

Suppose then that $j$ delegates if congruent, and votes if not. It
follows that $i$ can be pivotal only if $j$ votes, and thus if $j$
is $\not s$. Conditioning on pivotality, if $i$ is congruent:
\begin{align*}
EU_{i}(x|i_{s},x(-s)_{j}) & =pq+p(1-q)^{2}+q(1-p)(1-q)=p-q(1-q)(2p-1)\\
EU(d|i_{s}) & =p
\end{align*}

Given $p>1/2,$ if $i$ is congruent, $i'$s best response to $j$
voting when non-congruent is to delegate.

If $i$ is not congruent, again conditioning on pivotality: 
\begin{align*}
EU_{i}(x|i_{\not s},x(-s)_{j}) & =q^{2}+2(1-q)q(1-p)\\
EU(d|\text{\ensuremath{i}}_{\not s}) & =1-p
\end{align*}

For all $p>q$, conditioning on pivotality, $EU(d|\text{\ensuremath{i}}_{\not s})<$$EU_{i}(x|\text{\ensuremath{i_{\not s}}\ensuremath{x(-s\ensuremath{)_{j}}}})$.
Thus indeed there is an equilibrium in which a non-expert delegates
if congruent, and votes if non congruent.

Finally, as expected, there is no equilibrium such that a non-expert
votes if congruent and delegates if non-congruent. Suppose $j$ follows
such a strategy, and consider $i$'s best response when $i$ is not
congruent. Again, $i$ can be pivotal only if $j$ does not delegate,
and thus if $j$ votes and is congruent. Then, if $j$ votes and is
congruent:
\begin{align*}
EU_{i}(x|\text{\ensuremath{i_{\not s}},\ensuremath{(xs)_{j}}}) & =q(1-q)(1-p)+q[(1-q)p+q(1-p)]+(1-q)^{2}(1-p)\\
EU(d|\text{\ensuremath{i}}_{\not s}) & =1-p
\end{align*}

For all $p>q$, conditioning on pivotality, $EU(d|\text{\ensuremath{i}}_{\not s})<$$EU_{i}(x|\text{\ensuremath{i_{\not s}}\ensuremath{,x(s\ensuremath{)_{j}}}})$:
if $i$ is not congruent, voting dominates delegating when $j$ follows
strategy $x(s)$. Voting when congruent and delegating when not is
not an equilibrium. 

Summarizing, then, under LD there are two pure equilibria with symmetric
strategies: in one equilibrium, non-experts always delegate; in the
second equilibrium, the equilibrium strategy is to delegate if congruent,
and to vote if not congruent. 

\paragraph{DD}

Consider now the problem when the choice for non-experts is between
voting and abstaining. We begin by establishing that there is no equilibrium
in which both non-experts always abstain. Suppose $i$ is not congruent
and $j$ abstains. Then:
\begin{align*}
EU_{i}(x|\text{\ensuremath{i_{\not s}},\ensuremath{a_{j}}}) & =q(1-p)+(1/2)qp+(1/2)(1-p)(1-q)\\
EU(a|\text{\ensuremath{i}}_{\not s},a_{j}) & =1-p
\end{align*}
where we denote by $a_{j}$ $j$'s choice to abstain.

For all $p>q$, $EU(x|\text{\ensuremath{i}}_{\not s},a_{j})>$$EU_{i}(a|\text{\ensuremath{i_{\not s}},\ensuremath{a_{j})}:}$
if $j$ abstains, non-congruent $i$ prefers to vote. Hence indeed
there is no equilibrium where both non-experts always abstain.

There is always, however, an equilibrium where both non-experts always
vote. Suppose $j$ always votes, and suppose first that $i$ is congruent.
Then: 
\begin{multline*}
EU_{i}(x|\text{\ensuremath{i_{s}},\ensuremath{x_{j}}})=pq(q/2+(1-q)/2)+pq((1-q)/2+q/2)+\\
+(1-q)p(q/2+(1-q)/2)+(1-p)q(q/2+(1-q)/2)=\\
=pq+(1/2)(p(1-q)+q(1-p))\\
EU(a|\text{\ensuremath{i}}_{s},x_{j})=p(q/2+(1-q)/2)+p((1-q)/2+\\
+q/2)(1/2)+(1-p)(q/2+(1-q)/2)(1/2)=p/2+1/4
\end{multline*}

For all $q\in(1/2,p)$ and all $p<1$, $EU_{i}(x|\text{\ensuremath{i_{s}},\ensuremath{x_{j}}})>EU_{i}(a|\text{\ensuremath{i_{s}},\ensuremath{x_{j}}}).$
Suppose now that $i$ is not congruent. Then:
\begin{align*}
EU_{i}(x|\text{\ensuremath{i_{\not s}},\ensuremath{x_{j}}}) & =(1-p)q((1-q)/2+q/2)+(1-p)q(q/2+(1-q)/2)+\\
+ & (1-p)(1-q)((1-q)/2+q/2)+pq((1-q)/2+q/2)=\\
 & =(1+q-p)/2\\
EU(a|\text{\ensuremath{i}}_{\not s},x_{j}) & =(1-p)((1-q)/2+q/2)+(1-p)(q/2+(1-q)/2)+\\
+ & p((1-q)/2+q/2)(1/2)=\\
 & =(1-p)/2+1/4=(1+1/2-p)/2
\end{align*}

Indeed, for all $q\in(1/2,p)$ and all $p<1$, $EU_{i}(x|\text{\ensuremath{i_{\not s}},\ensuremath{x_{j}}})>EU_{i}(a|\text{\ensuremath{i_{\not s}},\ensuremath{x_{j}}}).$
For all relevant parameter values, an equilibrium exists in which
both non-experts always vote. 

Consider now a candidate equilibrium in which non-experts abstain
if congruent and vote if not congruent. Suppose $j$ follows such
a strategy. What is $i$'s best response? 

Suppose first that $i$ is congruent. Then:
\begin{align*}
EU_{i}(x|\text{\ensuremath{i_{s}},\ensuremath{x(-s)_{j}}}) & =(1/2)(pq+p(1-q)/2+q(1-p)/2)+\\
 & +(1/2)(pq(1-q)+pq^{2}+p(1-q)^{2}+(1-p)q(1-q)=\\
 & =(p+q)/4+(1/2)[pq+(1-q)\left(p(1-q)+q(1-p)\right)]\\
EU(a|\text{\ensuremath{i}}_{s},x(-s)_{j}) & =(1/2)p+(1/2)(pq/2+p(1-q)+\\
+ & (1-p)(1-q)/2=\\
= & p(2-q)/2+(pq+(1-p)(1-q))/2
\end{align*}

If instead $i$ is non-congruent, expected utilities are:
\begin{align*}
EU_{i}(x|\text{\ensuremath{\text{\ensuremath{i}}_{\not s},},\ensuremath{x(-s)_{j}}}) & =(1/2)((1-p)q+pq/2+(1-q)(1-p)/2)+\\
 & +(1/2)((1-p)q^{2}+2(1-p)q(1-q)+pq{}^{2}\\
EU(a|\text{\ensuremath{\text{\ensuremath{i}}_{\not s},}},x(-s)_{j}) & =(1/2)(1-p)+(1/2)((1-p)q+pq/2+(1-p)(1-q)/2)=\\
= & (1-p)(1+q)/2+(pq+(1-p)(1-q))/4
\end{align*}

For any $q\in(1/2,p)$ and all $p<1$, $EU_{i}(x|\text{\ensuremath{i_{\not s}},\ensuremath{x(-s)_{j}}})>EU_{i}(a|\text{\ensuremath{i_{\not s}},\ensuremath{x(-s)_{j}}}):$
a non-congruent $i$ indeed prefers to vote. As for a congruent $i$,
$i$'s best response is less straightforward and depends on parameters:
for any $p\in(1/2,1)$, there exists $\overline{q}(p)\in(1/2,p)$
such that $EU(a|i_{s},x(-s)_{j})>$$EU_{i}(x|i_{s},\text{\ensuremath{x(-s)_{j}})}$
if $q<\overline{q}(p),$ and not otherwise.\footnote{The precise expression is
\[
\overline{q}(p)=\frac{-1+p+\sqrt{1/2-p+p^{2}}}{-1+2p}.
\]

For example, if $p=0.7,$then $\overline{q}(0.7)=0.596$.} Thus an equilibrium such that non-experts prefer to vote if non-congruent
and abstain if congruent exists for $q\leq\overline{q}(p)$.

Finally, we want to verify that voting when congruent and abstaining
when non-congruent cannot be an equilibrium strategy. Suppose $j$
follows such a strategy, and consider $i$'s best response when $i$
is non-congruent. Then:
\begin{align*}
EU_{i}(x|\text{\ensuremath{\text{\ensuremath{i}}_{\not s},},\ensuremath{x(s)_{j}}}) & =(1/2)\left((1-p)(1-q)q+(1-p)q(1-q)+\right.\\
+ & \left.(1-p)(1-q)^{2}+p(1-q)q\right)+\\
+ & (1/2)\left((1-p)q+pq/2+(1-p)(1-q)/2\right)\\
EU(a|\text{\ensuremath{\text{\ensuremath{i}}_{\not s},}},x(s)_{j}) & =(1/2)(1-p)+(1/2)\left((1-p)(1-q)+(1-p)q/2+p(1-q)/2\right)
\end{align*}

It is readily verified that non-congruent $i$ prefers to deviate
and vote.

Summarizing then, under DD there are two pure symmetric sincere equilibria:
for all relevant parameter values, there is an equilibrium in which
both non-experts always vote; given $p\in(1/2,1],$ there exists a
second equilibrium in which non-experts vote if non-congruent and
abstain if congruent if $q$ is not too large (i.e. there exists $\overline{q}(p)\in(1/2,p)$
such that the equilibrium exists if $q\leq\overline{q}(p)$). 

\paragraph{Comparing LD and DD}

Can we rank the different equilibria and the two voting systems? We
begin by focusing on non-experts only. We compare equilibria by comparing
ex ante expected utility (before the realization of the voter's types).

Under LD, we denote by $EU_{D,all}$ ex ante expected utility in the
delegation equilibrium, and by $EU_{D,s}$ ex ante expected utility
in the equilibrium where delegation is only chosen if congruent. Thus,
keeping in mind that the ex ante probability of agreeing with the
expert is $1/2$:
\begin{align*}
EU_{D,all} & =(1/2)p+(1/2)(1-p)=1/2\\
EU_{D,s} & =(1/4)p+(1/4)p+(1/4)(1-p)+(1/4)[q^{2}(1-p)+2(1-p)q(1-q)+q^{2}p]
\end{align*}

As we state in the text, the equilibrium with full delegation does
poorly because of the ex ante uncertainty on the preferences of the
expert. With $EU_{D,all}=1/2$, it is dominated by the equilibrium
with congruent delegation for all $q\in(1/2,p)$: $EU_{D,s}>EU_{D,all}$. 

Under DD, denoting by $EU_{A,none}$ ex ante expected utility when
non-experts always vote, and $EU_{A,s}$ ex ante expected utility
in the equilibrium with congruent abstention (and $q\leq\overline{q}(p)$),
we find:
\begin{align*}
EU_{A,none} & =(1/4)\left(pq^{2}+2pq(1-q)+(1-p)q^{2}\right)+\\
+ & (1/4)\left(pq(1-q)+pq^{2}+p(1-q)^{2}+(1-p)q(1-q)\right)+\\
+ & (1/4)\left((1-p)q(1-q)+(1-p)q^{2}+(1-p)(1-q)^{2}+q(1-q)p\right)+\\
+ & (1/4)\left((1-p)q^{2}+2(1-p)q(1-q)+pq^{2}\right)=\\
= & (1/4)(1+2q)\\
EU_{A,s} & =(1/4)p+(1/4)\left(p(1-q)+pq/2+(1-p)(1-q)/2\right)+\\
+ & (1/4)\left((1-p)q+(1-p)(1-q)/2+pq/2\right)+\\
+ & (1/4)\left((1-p)q^{2}+2(1-p)q(1-q)+pq^{2}\right)
\end{align*}
 When the equilibrium with congruent abstention exists, i.e. if $q\leq\overline{q}(p),$
it dominates the no abstention equilibrium: $EU_{A,s}>EU_{A,none}$. 

Comparing ex ante expected utilities across the two systems, LD and
DD, $EU_{A,none}>EU_{D,all}$ for all $q\in[1/2,p]$; $EU_{A,s}>EU_{D,all}$
whenever the equilibrium with congruent abstention exists, i.e. for
all $q\in(1/2,\overline{q}(p)]$; finally, there exists $q'(p)$\textgreater$\overline{q}(p)$
such that $EU_{A,none}>EU_{D,s}$ if $q>q'(p)$, but the reverse holds
otherwise.\footnote{$q'(p)=(p-\sqrt{(p-p^{2)}})/(2p-1).$ For example, if $p=0.7$, $\overline{q}(0.7)=0.596$,
and $q'(0.7)=0.604$. If $q=0.6$, the equilibrium with congruent
abstention does not exist, and we find $EU_{D,all}=0.5$, $EU_{D,s}=0.551>$$EUA_{none}=0.55$.
If $p=0.8$, $\overline{q}(0.8)=0.638$ and $q'(0.8)=0.667$. If $q=0.6$,
$EU_{D,all}=0.5$, $EU_{D,s}=0.564=$$EU_{A,s}=0.564>EUA_{none}=0.55$;
if $q=0.7$, $EU_{D,all}=0.5$ $EU_{D,s}=0.593<$$EUA_{none}=0.6$. } 

Particularly instructive is the comparison between $EU_{D,s}$ and
$EU_{A,s},$ that is, between congruent delegation and congruent abstention.
In this model, the two expressions are identical. Comparing $EU_{D,s}$
and $EU_{A,s}$ above, note that they must be identical if both non-experts
are congruent (with probability 1/4)---in which case the expert alone
decides and each member obtains utility 1 with probability $p$; or
if both non-experts are not-congruent (with probability 1/4)---in
which case all vote and each obtains utility 1 if all three voters
vote in the same direction (with probability $q^{2}(1-p)$), or if
two of the three voters vote in the direction preferred by the non-experts
(with probability $2(1-p)q(1-q)+pq^{2}$). Where the expressions differ
is when the two non-experts disagree with each other. Suppose first
$\{i=s,j=\not s\}$(an event with probability (1/4)). Under congruent
delegation, $i$ delegates and obtains utility 1 with probability
$p$; under congruent abstention, $i$ abstains and obtains utility
1 with probability $(1+p-q)/2<p$. Conditional on being the voter
who is congruent with the expert, congruent delegation is preferred
to congruent abstention because it makes it possible for the voter
to shift all decision power to the expert, something abstention cannot
achieve. But suppose now $\{i=\text{\ensuremath{\not}s},j=s\}$(again
with probability (1/4)). Under congruent delegation, $j$ delegates
and thus $i$ achieves her preferred outcome only with probability
$(1-p)$. Under congruent abstention, $j$ cannot make the expert
dictator, and $i$'s probability of the preferred outcome is $(1+q-p)/2>1-p$:
conditional on being the voter who is not congruent with the expert,
congruent abstention is better exactly because it leaves the voter
with some decision power. In this simple model, the two effects, based
on equally probable events, exactly cancel each other: $(1/4)(p+1-p)=(1/4)(1+p-q+1+q-p)=(1/4).$ 

Contrary to what one may have expected, when private values are included,
LD need not dominate DD: if we focus on the worst equilibrium, it
corresponds to LD with full delegation; if we focus instead on the
best equilibrium, at $q\leq\overline{q}(p)$ it corresponds to congruent
delegation or abstention, and LD and DD are equivalent; at $q>q'(p)$,
the best equilibrium requires voting by all and can only be supported
under DD. LD dominates only over the range $(\overline{q}(p),q'(p))$,
when congruent abstention is not an equilibrium and voting by all
is dominated. The range $(q'(p)-\overline{q}(p))$ is increasing in
$p,$ but typically not large; for all $p\leq0.8,$ for example, it
covers less than 10\% of the possible range of $q$ values. 

How about the expert then? Here the answer is unambiguous and expected:
the expert always prefers LD with full delegation because in such
an equilibrium the expert has full control and achieves her preferred
outcome with probability $p$. For completeness, using the notation
$EUe$ to indicate the expert's ex ante expected utility (with indices
specifying the voting regime and the equilibrium), we report here
the expert's ex ante utilities under the different equilibria: 
\begin{align*}
EUe_{D,all} & =p\\
EUe_{D,s} & =(1/4)p+2(1/4)p+(1/4)[(1-q)^{2}p+2q(1-q)p+(1-q)^{2}(1-p)]\\
EUe_{A,none} & =(1/4)[q^{2}p+2(1-q)qp+(1-p)q^{2}]+\\
+ & (1/4)[(1-q)^{2}p+2(1-q)qp+(1-p)(1-q)^{2}]+\\
+ & (1/2)[(1-q)qp+(1-q)^{2}p+q^{2}p+q(1-q)(1-p)]=\\
= & (1/4)(1+2p)\\
EUe_{A,s} & =(1/4)p+(1/2)[(1-q)p+pq/2+(1-p)(1-q)/2]+\\
+ & (1/4)[(1-q)^{2}p+2q(1-q)p+(1-q)^{2}(1-p)]
\end{align*}

It is not difficult to verify that $EUe_{D,s}$, $EUe_{A,none},$
and $EUe_{A,s},$ all are inferior to $EUe_{D,all}=p$. If we construct
a measure of welfare that includes the expert's, introducing private
values as we did in this model can lead to LD being preferred. But
the reason is the expert's preference for full control, not the non-experts'
option of delegating to someone aligned with their own preferences.
And that is because, as we point out in the text, if we deviate from
a common interest scenario and recognize the existence of heterogeneous
private values, we also need to acknowledge that some voters will
feel misrepresented by an expert with whose preferences they disagree. 

\subsubsection{Introducing costly information acquisition}

We focus on pure equilibria with sincere voting but now it is natural
to allow for asymmetric equilibria: with pure common interest, free-riding
on other voters' investment in information can be an attractive and
intuitive strategy. The expert has no choice to make; a non-expert
voter must choose whether to invest in information or not, and whether
to vote or not. A non-expert who delegates (under LD) or abstains
(under DD) will not invest. We denote the three relevant strategies
$\{xc,x,\{d\text{ or \ensuremath{a}\}\}}$where $xc$ stands for the
decision to invest and vote; $x$ for voting without investing, and
$d$ or $a$ for not voting and delegating (under LD) or abstaining
(under DD), and not investing. Consider LD first.

\paragraph{LD}

As always, for any $\{p,q,c\}$, there is an equilibrium where both
non-experts delegate. Are there equilibria with voting?

Consider the expected utility of non-expert voter $i$. Suppose first
that non-expert $j$ invests and votes. Then:
\begin{align*}
EU_{i}(xc|xc) & =p^{3}+3p^{2}(1-p)-c\\
EU_{i}(x|xc) & =p^{2}+2pq(1-p)\\
EU_{i}(d|xc) & =p
\end{align*}

Note that $EU_{i}(x|xc)>EU_{i}(d|xc)$ always. Hence $i's$ best response
to $j$ investing and voting is always to vote, with investment in
information if $c$$\leq$$\widetilde{c}\equiv2p(1-p)(p-q)$. For
any $p$, there is an equilibrium such that both non-experts invest
in information and vote if $\ensuremath{c\leq\widetilde{c}}.$

Suppose now that non-expert $j$ votes without investing. Then:

\begin{align*}
EU_{i}(xc|x) & =p^{2}+2pq(1-p)-c\\
EU_{i}(x|x) & =q^{2}+2pq(1-q)\\
EU_{i}( & d|x)=p
\end{align*}

There are two possibilities. If $p\leq p'=q^{2}/[1-2q(1-q)]$, $i$'s
best response is to vote, with investment in information if $c\leq c'_{vote}\equiv(p-q)(p+q-2pq)$.
If instead $p>p'$, $i$'s best response is to delegate if $c>c'_{del}\equiv p(1-p)(2q-1)$,
and invest in information and vote otherwise. 

Note that if $p\leq p'$, then $\widetilde{c}\leq c'_{vote}$; if
$p>p'$, the ranking between $\widetilde{c}$ and $c'_{del}$ depends
on $\{p,q\}$. Consider first $p\leq p'$. If $c\in(\widetilde{c},c'_{vote})$,
there is an equilibrium in which one non-expert invests and one does
not. If $c\geq c'_{vote}$, both non-experts vote but neither invests.
Suppose now $p>p'.$ Then: (i) if $c\geq c'_{del}$ neither non-expert
invests, and the only equilibrium is the delegation equilibrium. (ii)
If $c_{del}>\widetilde{c}$ and $c\in(\widetilde{c},c'_{del})$, then
there is an equilibrium in which one non-expert invests and one does
not. As an example, suppose $p=0.7$ and $q=0.6.$ Then $p>p'$, $\widetilde{c}=c'_{del}=0.042$.
There exists an equilibrium where both non-experts vote; both invest
in information if $c\leq0.042$, and neither invests otherwise. The
probability of a correct outcome is 0.784 when both experts invest
and vote, 0.696 when both vote without investing, and 0.7 when both
delegate to the expert.

\paragraph{DD}

Under DD, a non-expert who does not vote abstains. Suppose first
that non-expert $j$ invests and votes. Then:
\begin{align*}
EU_{i}(xc|xc) & =p^{3}+3p^{2}(1-p)-c\\
EU_{i}(x|xc) & =p^{2}+2pq(1-p)\\
EU_{i}(a|xc) & =p^{2}+2p(1-p)/2
\end{align*}

Since $q>1/2$, voting without investing again dominates not voting
whenever the other non-expert invests and votes. Thus, as in the case
of delegation, $i$'s best response is to invest and vote if $c\leq\widetilde{c}$,
and vote without investing otherwise. Thus, for any $p$, under DD
as well there is an equilibrium such that both non-experts invest
in information and vote if $\ensuremath{c\leq\widetilde{c}}.$

Suppose now that non-expert $j$ votes without investing. Then:

\begin{align*}
EU_{i}(xc|x) & =p^{2}+2pq(1-p)-c\\
EU_{i}(x|x) & =q^{2}+2pq(1-q)\\
EU_{i}( & a|x)=pq+p(1-q)/2+q(1-p)/2=(p+q)/2
\end{align*}

For all $p<1$, voting without investing dominates abstaining. Hence
$i$'s best response is to invest and vote if $c\leq c'_{vote}$,
vote without investing otherwise. 

Finally, suppose that $j$ abstains. Then:
\begin{align*}
EU_{i}(xc|a) & =p^{2}+2p(1-p)/2-c\\
EU_{i}(x|a) & =pq+p(1-q)/2+q(1-p)/2=(p+q)/2\\
EU_{i}( & a|a)=p
\end{align*}

With $q<p$ and $c>0$, abstaining is the best response to $j$ abstaining.

It then follows that: (1) for any $\{p,q,c\}$ in the relevant range
($q\in[1/2,p],$$p<1,$$c>0$) there is an equilibrium where both
non-experts abstain. As in the case of LD, there is also an equilibrium
where both non-experts vote, investing or not investing in information
depending on $c$. If $c\leq\widetilde{c}$, both non-experts invest
and vote; if $c\in(\widetilde{c},c'_{vote})$, in equilibrium one
non-expert invests and one does not; if $c\geq c'_{vote}$, neither
invests.

\paragraph{Comparing LD and DD}

Under both LD and DD, for all parameter values, there is an equilibrium
in which only the expert votes, and no-one invests in information.
More interesting is the comparison of the equilibria with voting.
For all $\{p,q\}$, under both LD and DD there is an equilibrium
in which both non-experts invest and vote if $c\leq\widetilde{c}$.
Under both LD and DD, there is an asymmetric equilibrium in which
both non-experts vote but only one invests if $c\in(\widetilde{c},c'_{vote})$,
but this equilibrium exists only for $p<p'$ under LD, while it exists
for all $p$ under DD. If $p>p'$, under LD the asymmetric equilibrium
requires $c'_{del}>\widetilde{c}$ and $c\in(\widetilde{c},c'_{del})$.
However, $c'_{del}<c'_{vote}$ for all $p>p'$:\footnote{The threshold $c'_{del}$ is decreasing in $p$, while $c'_{vote}$
is increasing in $p$. It then follows that $(c'_{vote}-c'_{del})$
is minimal at minimum $p$, that is, at $p=p'$. But $c'_{vote}=c'_{del}=((1-q)^{2}q^{2}(2q-1))/(1-2(1-q)q)^{2}$
at $p=p'.$ The statement then follows.} the asymmetric equilibrium exists for a strictly larger range of
parameter values under DD than under LD. This is the statement reported
in the text: whenever parameter values are such that investment in
information (either by both non-experts or by one only) can be supported
in equilibrium under LD, then it can be supported under DD. But there
exists parameters ($p>p'$ and $c\in(c'_{del},c'_{vote})$) such that
there exists an equilibrium in which one non-expert invests in information
under DD but none does under LD. If $p=0.7$ and $q=0.6$, for example,
there is an equilibrium where both non-experts invest and vote if
$c\leq0.042$ under both LD and DD. However, if $c\in(0.042,$0.046),
under DD there is an equilibrium in which both non-experts vote and
one of them invests in information; under LD in the only equilibrium
the expert is dictator and none of the non-experts invests. Expected
welfare (the sum of expected utilities) is higher under DD. 

Note also, however, that if $p>p'$ and $c>c'_{del},$ under LD, delegation
is the only equilibrium. Under DD, the equilibrium in which both
non-experts abstain exists as well, but so does a second equilibrium
in which both vote without investing. The range of parameter values
supporting the equilibrium with voting without information acquisition
is strictly larger under DD than under LD.

\subsubsection{Introducing correlated signals}

Once again, we study the pure common interest model with $N=3$. We
now suppose that all signals are conditionally independent with probability
$\alpha<1;$ with probability $(1-\alpha)$, the two non-experts receive
the same signal, of precision $q$, while the expert alone has a conditionally
independent signal, of precision $p\in(q,1).$ The expert thus benefits
both from higher signal precision, and from the signal's (conditional)
independence. The probability $\alpha$ is known, but it is not known
whether the realized non-experts' signals are correlated (fully, in
this example). We focus on pure symmetric equilibria with no communication
and with sincere voting, where all votes are always cast in line with
the received signal, and non-experts' strategies are $\{x,d/a\}$---that
is, whether to vote or to delegate/abstain. Consider first LD.

\paragraph{LD}

As always, for all parameter values there is an equilibrium in which
both non-experts delegate, and $EU_{i}(d|d)=p$. Is there an equilibrium
in which the non-experts do not delegate? Suppose $j$ does not delegate.
Then:
\begin{align*}
EU_{i}(d|x_{j}) & =p\\
EU_{i}(x|x_{j}) & =(1-\alpha)q+\alpha[pq^{2}+2pq(1-q)+q^{2}(1-p)]
\end{align*}
Note that if the non-experts' signals are fully correlated (with probability
$(1-\alpha))$ and votes are sincere, the non-experts always vote
in the same direction and always prevail; the correct decision is
taken if the non-experts' common signal is correct, that is, with
probability $q$. 

From the expressions above, non-expert $i$'s best response to $j$
voting is to vote if:
\[
\alpha(q^{2}+2pq(1-q)-p)\geq(1-\alpha)(p-q)
\]
and delegate otherwise. It then follows that there exists $\widetilde{q}(p,\alpha)$
such that if $q$\textless$\widetilde{q}$ LD admits a unique equilibrium
where both non-experts always delegate; if $q\geq\widetilde{q}$,
then under LD there are two equilibria: the equilibrium with full
delegation and a second equilibrium in which the two non-experts cast
their votes.\footnote{The threshold $\widetilde{q}(p,\alpha)$ is increasing in $p$ and
decreasing in $\alpha$. It is given by:
\[
\widetilde{q}(p,\alpha)=\frac{1+\alpha(2p-1)-\sqrt{1-(2-\alpha)\alpha(2p-1)^{2}}}{2\alpha(2p-1)}.
\]
} 

\paragraph{DD}

Consider now DD. Suppose first that $j$ abstains. Then:
\begin{align*}
EU_{i}(a|a_{j}) & =p\\
EU_{i}(x|a_{j}) & =[pq+1/2[(1-p)q+p(1-q)]=(1/2)(p+q)
\end{align*}

With $p>q$, abstaining is the superior response. Thus, for all relevant
parameter values, there is always an equilibrium in which both non-experts
abstain. 

Suppose now that $j$ votes. Then:
\begin{align*}
EU_{i}(a|x_{j}) & =(1/2)(p+q)\\
EU_{i}(x|x_{j}) & =(1-\alpha)q+\alpha[pq^{2}+2pq(1-q)+q^{2}(1-p)]
\end{align*}

If both non-experts vote, the probability of reaching the correct
decision, and thus $EU_{i}(x|x_{j})$, must be identical under LD
or DD; if only one votes, however, such probability is lower under
DD than under LD because abstention does not benefit as much as delegation
from the expert's superior signal. It follows that supporting the
equilibrium in which both non-experts vote is easier under DD than
under LD---that is, there is a larger range of parameter values for
which $\{x,x\}$ is an equilibrium. Under DD, non-expert $i$'s best
response to $j$ voting is to vote if:
\[
\alpha\left(q^{2}+2pq(1-q)-\frac{(p+q)}{2}\right)\geq(1-\alpha)\left(\frac{p-q}{2}\right)
\]
and abstain otherwise. There exists $q'(p,\alpha)$ such that if $q$\textless$q'(p,\alpha)$
DD admits a unique equilibrium where both non-experts always abstain;
if $q\geq q'$, there are two equilibria: the equilibrium with full
abstention and a second equilibrium in which the two non-experts cast
their votes. Note that for any given $(p,\alpha)$, $q'<\widetilde{q}$.\footnote{The threshold $q'(p,\alpha)$, again increasing in $p$ and decreasing
in $\alpha$, is given by:
\[
q'(p,\alpha)=\frac{1+2\alpha(2p-1)-\sqrt{1-4(1-\alpha)\alpha(2p-1)^{2}}}{4\alpha(2p-1)}.
\]
} 

\paragraph{Comparing LD and DD}

Thus, both LD and DD support, for all parameter values, an equilibrium
in which the expert alone dictates the decision (non-experts do not
vote). Both voting systems also support an equilibrium in which both
experts vote, but only for $q$ high enough, higher than $\widetilde{q}(p,\alpha)$
in the case of LD, and higher than $q'(p,\alpha)$ in the case of
DD, with $q'<\widetilde{q}$. Comparing the two equilibria, the first
equilibrium, with the expert dictating, is superior if $q<\widetilde{q}(p,\alpha)$,
while the second, with the non-experts voting, is superior for $q>\widetilde{q}(p,\alpha)$.
It follows, as stated in the text, that both voting systems can support
the superior equilibrium for any parameter values. However, both systems
also can support delegation/abstention when the probability of reaching
the correct decision would be higher if the non-experts voted. Finally,
there exists a range of parameters ($q\in(q'(p,\alpha),\widetilde{q}(p,\alpha))$
such that DD but not LD can support voting by non-experts when the
expert alone dictating would be superior.



\end{document}